\documentclass[12pt]{article}

\usepackage{graphicx}
\usepackage{graphics}
\usepackage{amssymb}
\usepackage{amsmath}

\usepackage{listings}

\usepackage{wrapfig,lipsum}

\usepackage{epsf,amsfonts,hyperref}

\usepackage{tikz,mathtools}
\usetikzlibrary{knots}
\usetikzlibrary{calc}
\usetikzlibrary{decorations.pathreplacing} 
\usetikzlibrary{decorations.markings}
\usetikzlibrary{arrows}
\tikzset{->-/.style={decoration={
  markings,
  mark=at position .5 with {\arrow{>}}},postaction={decorate}}}

\usepackage[
backend=biber,
style=alphabetic
]{biblatex}
\newcommand\encadremath[1]{\vbox{\hrule\hbox{\vrule\kern8pt 
\vbox{\kern8pt \hbox{$\displaystyle #1$}\kern8pt} 
\kern8pt\vrule}\hrule}}
\def\enca#1{\vbox{\hrule\hbox{
\vrule\kern8pt\vbox{\kern8pt \hbox{$\displaystyle #1$}
\kern8pt} \kern8pt\vrule}\hrule}}

\newcommand\figureframex[3]{
\begin{figure}[bth]
\hrule\hbox{\vrule\kern8pt 
\vbox{\kern8pt \vbox{
\begin{center}
{\mbox{\epsfxsize=#1.truecm\epsfbox{#2}}}
\end{center}
\caption{#3}
}\kern8pt} 
\kern8pt\vrule}\hrule
\end{figure}
}
\newcommand\figureframey[3]{
\begin{figure}[bth]
\hrule\hbox{\vrule\kern8pt 
\vbox{\kern8pt \vbox{
\begin{center}
{\mbox{\epsfysize=#1.truecm\epsfbox{#2}}}
\end{center}
\caption{#3}
}\kern8pt} 
\kern8pt\vrule}\hrule
\end{figure}
}

\makeatletter
\@addtoreset{equation}{section}
\makeatother
\newtheorem{theorem}{Theorem}[section]

\newtheorem{remark}{Remark}[section]
\newtheorem{proposition}{Proposition}[section]
\newtheorem{lemma}{Lemma}[section]
\newtheorem{corollary}{Corollary}[section]
\newtheorem{definition}{Definition}[section]
\newtheorem{example}{Example}[section]
\def\br{\begin{remark}\rm\small}
\def\er{\end{remark}}
\def\bt{\begin{theorem}}
\def\et{\end{theorem}}
\def\bd{\begin{definition}}
\def\ed{\end{definition}}
\def\bp{\begin{proposition}}
\def\ep{\end{proposition}}
\def\bl{\begin{lemma}}
\def\el{\end{lemma}}
\def\bc{\begin{corollary}}
\def\ec{\end{corollary}}
\def\bex{\begin{example}}
\def\eex{\end{example}}
\def\beaq{\begin{eqnarray}}
\def\eeaq{\end{eqnarray}}
\newcommand{\proof}[1]{{\noindent \bf proof:}\par
{#1} $\square$}

\newcommand{\beq}{\begin{equation}}
\newcommand{\eeq}{\end{equation}}
\newcommand{\bea}{\begin{eqnarray}}
\newcommand{\eea}{\end{eqnarray}}

\renewcommand{\and}{{\qquad {\rm and} \qquad}}

\newcommand{\Res}{\mathop{\,\rm Res\,}}
\newcommand{\Discr}{\operatorname{Discr}} 

\newcommand{\CC}{{\mathbb C}}
\newcommand{\ZZ}{{\mathbb Z}}
\newcommand{\RR}{{\mathbb R}}

\newcommand{\td}[1]{{\tilde{#1}}}

\newcommand{\ii}{{\mathrm{i}}}

\newcommand{\Pint}{{\int\kern -1.em -\kern-.25em}}

\renewcommand{\Im}{{\mathrm{Im}}}

\newcommand{\curve}{{\Sigma}}
\newcommand{\Field}{{\mathbb F}}

\newcommand{\genus}{{\mathfrak g}}

\newcommand{\acycle}{{\cal A}}
\newcommand{\bcycle}{{\cal B}}

\newcommand{\order}{\operatorname{order}}

\newcommand{\modsp}{{\mathcal M}}

\newcommand{\Newt}{{\cal N}}
\newcommand{\dNewt}{{\partial\cal N}}
\newcommand{\Newtint}{\displaystyle{\mathop{{\cal N}}^{\circ}}}

\def\grid#1#2{\draw[step=1.0,black,thin,xshift=0.5,yshift=0.5] (-0.5,-0.5) grid ({#1+0.5},{#2+0.5});
}

\newcommand\drawpolytope[2][1=red]{
    \foreach [count=\x] \pt in #2
        {\node[#1] (p\x) at \pt {$\bullet$};}
    }

\begin{document}
\sloppy


\pagestyle{empty}
\addtolength{\baselineskip}{0.20\baselineskip}
\begin{center}
\vspace{26pt}
{\large \bf {Lecture on the combinatorial algebraic method for computing algebraic integrals}}
\newline
\vspace{26pt}

{\sl B.\ Eynard}\hspace*{0.05cm}\footnote{ E-mail: bertrand.eynard@ipht.fr }\\
\vspace{6pt}
Service de Physique Th\'{e}orique de Saclay,\\
F-91191 Gif-sur-Yvette Cedex, France.\\
\end{center}

\vspace{20pt}
\begin{center}
{\bf Abstract}
\end{center}
Consider an algebraic equation $P(x,y)=0$ where $P\in \CC[x,y] $ (or $\Field[x,y]$ with $\Field\subset \CC$ a subfield) is a bivariate polynomial, it defines a plane algebraic curve.
We provide an efficient method for computing integrals of the type
$ \int_\gamma R(x,y)dx $ where $R(x,y)\in \CC(x,y) $ is any rational fraction, and $y$ is solution of $P(x,y)=0$, and $\gamma$ any Jordan arc open or closed on the plane algebraic curve.
The method uses only algebraic and combinatorial manipulations, it rests on the combinatorics of the Newton's polygon.
We illustrate it with many practical examples.

%





\vspace{26pt}
\pagestyle{plain}
\setcounter{page}{1}


\tableofcontents

\section{Notations and introduction}

In all what follows, $P(x,y)\in \Field[x,y]$ is a bivariate polyomial with coefficients in a subfield $\Field\subset\CC $
(e.g. $\Field = \mathbb Q, \RR, \CC, $ number fields, ...etc).
$P_x(x,y)$ and $P_y(x,y)$ denote its partial derivatives
\beq
P_x(x,y) = \frac{\partial}{\partial x}P(x,y)
\quad , \quad
P_y(x,y) = \frac{\partial}{\partial y}P(x,y).
\eeq
Likewise $P_{xx},P_{xy},P_{yy},P_{xxy},\dots$  denote higher derivatives.


We shall write
\beq
P(x,y) = \sum_{i=0}^d y^i P_i(x)
\eeq
(we assume that $P_d\neq 0$.)
And
\beq
P(x,y) = \sum_{(i,j)\in \Newt} P_{i,j} x^i y^j,
\eeq
where $\Newt\subset \ZZ^2$ is a finite polytope (set of points).
If $P$ is a bivariate polynomial, we denote the coefficient of the monomial $x^iy^j$ as
\beq
[x^iy^j] P(x,y) = P_{i,j}.
\eeq

\subsection{Purpose and results}

The goal is to provide a practical way to compute integrals of the type
$$ 
\int_\gamma R(x,y)dx 
$$ 
where $R(x,y)\in \CC(x,y)$ is any rational function, $\gamma$ any Jordan arc or loop on the curve $P(x,y)=0$,  and $y$ is a solution of $P(x,y)=0$, analytic along a tubular neighborhood of $\gamma$.

Some methods to achieve this goal, do already exist in textbooks, and are implemented in some mathematical softwares.

However, here we propose a self-contained survey of recent combinatorial--algebraic methods.

The method described here uses only algebraic and combinatorial manipulations.


\subsubsection{Example 1: Rational case}

This is the case where there exists a rational parametric solution to $P(x,y)=0$, i.e. there exists 2 rational functions $X(z)\in \CC(z)$ and $Y(z)\in \CC(z)$ such that
\beq
P(x,y) =0 \quad \Leftrightarrow \quad \exists z\in \CC  \ , \ x=X(z) \text{ and } y=Y(z).
\eeq
Any rational function $R(x,y)\in \CC(x,y)$ can then be written as a rational function of $z$, and then decomposed into sum of poles:
\beq
R(x,y)dx = 
R(X(z),Y(z)) X'(z) dz = \sum_{p = \text{poles}} \sum_{k=0}^{\deg_p-1} t_{p,k} (z-p)^{-k-1}dz 
\eeq
for which integrals are easy to compute.
The method mostly amounts to finding the coefficients $t_{p,k}$, by taking residues
\beq
t_{p,k} = \Res_{z\to p} (z-p)^{k} R(x,y)dx = \Res_{z\to p} (z-p)^{k} R(X(z),Y(z)) X'(z) dz.
\eeq
We then have
\beq
\int_{\gamma} R(x,y)dx = \sum_{p}\sum_k t_{p,k} \int_\gamma (z-p)^{-k-1}dz.
\eeq

The method presented below provides a generalization to non-rational cases.

\subsubsection{Motivation Example 2: Elliptic integrals}

This is the case where $P(x,y)$ is of degree $\deg_y P=2$ and $\deg_x P= 3$ or $4$:
\beq
P(x,y) = y^2-\hat P(x) \qquad \deg \hat P = 3 \text{ or } 4.
\eeq
The Weierstrass\index{Weierstrass} canonical form is cubic in $x$ (we change a factor 4 compared to usual normalization)
\beq
P(x,y) = y^2 - x^3+g_2 x +g_3,
\eeq
while the Legendre\index{Legendre} canonical form is quartic and even in $x$
\beq
P(x,y) = y^2 - (1-x^2)(1-k^2 x^2).
\eeq
Legendre elliptic integrals are of the type
\beq
\int_\gamma R(x,y) \ \frac{dx}{y} \quad \text{with} \ \  y=\sqrt{(1-x^2)(1-k^2 x^2)}.
\eeq
It has been known since Legendre, Jacobi, Weierstrass,... how to compute such integrals, they have been reduced to algebraic combinations of 6 elementary blocks, called 1st, 2nd and 3rd kind, (see \cite{SpecF,Abram}) and often denoted
$K(k), K'(k), E(k), F(x,k), \Pi(u,k),\Pi(u,x,k)$:
\beq
K(k) = \int_{0}^1 \frac{dx}{\sqrt{(1-x^2)(1-k^2 x^2)}}
\qquad , \quad K'(k)=K(\sqrt{1-k^2}),
\eeq
\beq
E(k) = \int_{0}^1 \sqrt{\frac{1-k^2 x^2}{1-x^2}} \ dx
\eeq
\beq
F(x,k) = \int_0^x \frac{dx'}{\sqrt{(1-x'^2)(1-k^2 x'^2)}},
\eeq
\beq
\Pi(u,k) = \int_{0}^1 \frac{dx}{(1-u\ x^2)\sqrt{(1-x^2)(1-k^2 x^2)}}.
\eeq
\beq
\Pi(u,x,k) = \int_{0}^{x} \frac{dx'}{(1-u\ x'^2)\sqrt{(1-x'^2)(1-k^2 x'^2)}}.
\eeq
All  elliptic integrals can be written as combinations of these 6 elementary blocks, with coefficients rational functions of $k$ and of the coefficients of $R(x,y)$.

There exist many textbooks \cite{SpecF,Abram} and tables of elliptic integrals providing algorithms to reduce every elliptic integrals to these elementary ones, and these algorithms are implemented in formal mathematical computer languages as {\em Mathematica, Maple, Sympy, Sage, CoCalc...}

The Legendre-Jacobi-Weierstrass methods are not so easy to generalize beyond elliptic or hyperelliptic.

The method of Jacobi uses other elementary functions: the Theta-functions (see \cite{SpecF,Abram} or section~\ref{sec:theta}). For example, writing $\tau = \ii \frac{K'(k)}{K(k)}$ one has
\beq
K(k) = \frac{\pi}{2} \theta(\tau)^2
\quad , \quad
\theta(\tau) = \sum_{n\in \ZZ} e^{\ii\pi\tau n^2} .
\eeq
Again algorithms to decompose every algebraic integral to Jacobi Theta-functions are known, and rather easy to use.
However let us mention a subtle point: Many integrals that we find at the end are algebraic combinations, whereas Theta-functions are transcendental functions, and with this method it is not obvious at all that some combinations of transcendantal functions can in the end reduce to algebraic expressions. Also the generalization to non-elliptic curve is not so easy.

\medskip

Our goal is to summarize more combinatorial methods in a ready-to-use formulation.

\subsection{Generalization beyond elliptic integrals}

The method  to compute an integral $\int_\gamma R(x,y)dx$ on a plane curve $P(x,y)=0$  proceeds in 3 steps, described below:

1) identify the elementary blocks. This depends only on $P$ and not on $R$. This is the preparatory step. 

2) Compute a decomposition of $R$, using the kernel computed in step 1).

3) Then compute integrals as combinations of the elementary blocks.

\subsubsection{Preparation step: Elementary blocks}

The elementary blocks that we shall define here below (let  $\genus$ the genus) generalize the 6 Legendre Elliptic functions as follows:
\begin{itemize}

\item The role of $K(k)$ is replaced by a $\genus\times \genus$ invertible matrix $\mathcal K$. In the Legendre case $\mathcal K=\mathcal K_{1,1}  = 2K(k)$.
$\mathcal K$ is not algebraic in terms of the coefficients of $P$, it is genuinely transcendental.

\item The role of $F(x,k)$ is played by a collection of $\genus$ functions called the Abel map, collectively written as a $\genus$-dimensional vector  $F(x,y)=\{F_i(x,y) \}_{i=1,\dots,\genus}$.

\item A fundamental object, rarely considered in the classical Legendre--Jacobi methods (although introduced by Riemann as the "fundamental 2nd kind differential") is the Bergman kernel\footnote{In fact there are 2 famous Bergman kernels, there is another one in Hilbert-spaces Riesz representation theory. The two are in fact closely related, and both were studied by Bergman \cite{BergSchif}. We call this one Bergman kernel mostly because of the works of Bergman and Schiffer \cite{BergSchif,BergSchifrev} and Kokotov and Korotkin \cite{Koko}.}
$$
B(x_1,y_1;x_2,y_2)= B^{\text{comb}}(x_1,y_1;x_2,y_2) + \frac{dx_1\otimes dx_2}{P_y(x_1,y_1)P_y(x_2,y_2)} S(x_1,y_1;x_2,y_2)
$$
where $B^{\text{comb}}$ is algebraic in the coefficients $P_{i,j}$ and given by a combinatorial formula. 

As Riemann envisioned, the kernel $B$ is the key to parametrize every differential 1-form, and to compute any integral, it is really fundamental.

\item The term $S$ in $B$ plays the role of the Legendre  $E(k)$, or in fact it generalizes the 2nd Eisenstein series $G_2(\tau)$ or the 3rd derivative of the Theta function. 
$S(x_1,y_1;x_2,y_2)\in \CC[x_1,y_1,x_2,y_2]$ is a polynomial of 4 variables, with coefficients parametrized by a symmetric $\genus\times \genus$ matrix,  not algebraic in the coefficients of $P$ (transcendantal) and  algebraically independent of $\mathcal K$. In fact $S$ is related to the derivative of $\mathcal K$, or alternatively 
$S$ is also closely related to derivatives of the Theta function at half-integer characteristics.
In the Legendre case $S$ is a constant polynomial worth $S=\frac{G_2}{4 K^2} + \frac23(1+k^2) = \frac{2E}{K} -1+k^2$ and $G_2 = -\frac13 \frac{\Theta'''_\chi}{\Theta'_\chi}$.


\item The role of $K'(k)$ or rather $\tau = \ii K'/K$ is played by a $\genus\times \genus$ Siegel\footnote{A Siegel matrix is a complex symmetric matrix whose imaginary part is positive definite.} matrix $\tau$.
It is also equal to
\beq
\tau_{i,j} = \frac{1}{2\pi i} \oint_{\bcycle_i}\oint_{\bcycle_j} B,
\eeq
where the cycles $\bcycle_i$ span a Lagrangian in the homology space $H_1(\curve,\ZZ)$, called polarization.

\item
 The role of $\Pi(u,k)$ is played by the 3-rd kind differentials
$$
dS_{p_1,p_2}(x,y)= dS^{\text{comb}}_{p_1,p_2}(x,y) + \frac{dx}{P_y(x,y)} (\zeta(p_1)-\zeta(p_2))
$$
where  $dS^{\text{comb}}$ is algebraic and given by a combinatorial formula,
and  $\zeta(p)$ is a vector of dimension $\genus$, whose derivative is algebraic, and depends algebraically on the previously defined elementary functions, in particular $S$.

\item
 The role of $\Pi(u,x,k)$ is played by the logarithm of the prime form that we shall introduce in section \ref{sec:primeform}. It is needed only for incomplete integrals.

\item Moreover, let us mention that there is a relationship between $S$ and Theta-functions.
This relation is actually not needed to compute integrals, but is nice for book-keeping and to relate to other formulas.


\end{itemize}

\subsubsection{Second step: Canonical decomposition of a rational 1-form}

After defining the elementary blocks $\mathcal K,\tau,S,\zeta$, the procedure to compute any integral works as follows:

let $R=R(x,y)dx$ an arbitrary Abelian differential 1-form with $R(x,y)\in \CC(x,y)$.

\begin{itemize}
    \item Find all the poles of $R$, for a pole $p$,  let $a_p=\order_p (x-x_p)$ (with the convention $x_p=0$ if $p$ is a pole of $x$).
    \item Define the "times" of $R$ for each pole $p$ and each $k=0,\dots,-1+\deg_p R$, as
    \beq
        t_{p,k}(R) = \Res_{p} (x-x_p)^{k/a_p} R(x,y)dx
    \eeq
    The times are algebraic in the coefficients of $R$ and of $P$.
    
    \item Compute for $k\geq 1$:
    \bea
    B_{p,k}(x,y) 
    &=& \frac{1}{k}\Res_{(x',y')\to p} (x'-x_p)^{-k/a_p} B(x,y,x',y') \cr
    &=& \frac{1}{k}\Res_{(x',y')\to p} (x'-x_p)^{-k/a_p} B^{\text{comb}}(x,y,x',y') \cr
    && + \frac{1}{k}\Res_{(x',y')\to p} (x'-x_p)^{-k/a_p} \frac{S(x,y,x',y')dx dx'}{P_y(x,y) P_y(x',y')} \cr    
    \eea
    Notice that $B_{p,k}$ is a polynomial of $S$ of degree 1, with coefficients algebraic in the coefficients of $P$.

    \item Compute 
    \bea
    \td R(x,y)dx &=& R(x,y)dx - \sum_{p=\text{poles}} \sum_{k=1}^{-1+\deg_p R} t_{p,k}(R) B_{p,k}(x,y) \cr
    && - \sum_{p=\text{poles}} t_{p,0}(R)dS_{o,p}(x,y)
    \eea
    Since we have subtracted all poles, $\td R(x,y)dx$ is a holomorphic 1-form (and it is independent of the point $o$). $\td R(x,y)$ is polynomial of degree 1 in $S$ and in $\zeta$ with coefficients algebraic of the coefficients of $P$ and $R$. 
    It turns out that the product $P_y(x,y) \td R(x,y)$ is a polynomial of $x,y$.
    
    \item Compute the periods of $R$ for $i=1,\dots,\genus$:
    \beq
        t_i(R) = \sum_{(k,l)} \mathcal K_{(k,l),i} \ [x^k y^l] ( P_y(x,y) \td R(x,y))
    \eeq    
    Notice that the computation of periods is purely algebraic and combinatorial, it doesn't require computing integrals, only residues.
    Moreover, periods are polynomial of $S$, $\zeta$ and  $\mathcal K$ of degree 1, with coefficients algebraic functions of the $P_{i,j}$ and the coefficients of $R$.

    \item Eventually, this gives the canonical decomposition of $R$ as:
    \bea
    R(x,y)dx &=& \sum_{p=\text{poles}} \sum_{k=1}^{-1+\deg_p R} t_{p,k}(R) B_{p,k}(x,y) \cr
    && + \sum_{p=\text{poles}} t_{p,0}(R)dS_{o,p}(x,y) \cr
    && + \sum_{i=1}^{\genus} t_i(R) \omega_i(x,y) .
    \eea

\end{itemize}
 
\subsubsection{Third step: Computing integrals}

It remains to compute the integrals of each term on a given  $\gamma$.

\begin{itemize}

\item[$\circ$] \textbf{Complete integrals:} if $\gamma$ is a closed Jordan loop, it can be decomposed as a linear combination of cycles $\acycle_i, \bcycle_i$, and small circles $\mathcal C_p$ around the poles
\begin{itemize}
    \item[$\bullet$] $\oint_{\acycle_i} B_{p,k} = 0  $, $\oint_{\acycle_i} dS_{o,p} = 0$, $\oint_{\acycle_i} \omega_j = \delta_{i,j}  $.
    \item[$\bullet$] $\oint_{\bcycle_i} B_{p,k} = \frac{1}{k}\Res_{p} (x-x_p)^{-k/a_p} \omega_i$, $\oint_{\bcycle_i} dS_{o,p} = F_i(p)-F_i(o) $, $\oint_{\bcycle_i} \omega_j = \tau_{i,j}  $.
    \item[$\bullet$] $\oint_{\mathcal C_p} B_{q,k} = 0$, $\oint_{\mathcal C_p} dS_{o,q} = 2\pi\ii(\delta_{p,q}-\delta_{p,o}) $, $\oint_{\mathcal C_p} \omega_j = 0  $.
\end{itemize}
 
\item[$\circ$] \textbf{Incomplete integrals:}  if $\gamma = \gamma_{p_1\to p_2}$ is an open Jordan arc, we have
\begin{itemize}
        \item[$\bullet$] $\int_{\gamma} B_{p,k} = \frac{1}{k}\Res_{p} (x-x_p)^{-k/a_p} dS_{p_2,p_1}$
        \item[$\bullet$] $\int_{\gamma} \omega_j = F_j(p_2)-F_j(p_1)  $.
        \item[$\bullet$] $\int_{\gamma} dS_{o,p} = \ln{\left({E(p_1,p)E(p_2,o)}/{E(p_1,o)E(p_2,p)}\right)} $
        where $E(p,q)$ is called the prime form (see section \ref{sec:primeform}), and is an independent fundamental block that generalizes the Legendre $\Pi(u,x,k)$.
\end{itemize}

\end{itemize}

With this method, the computation of any integral can be entirely automatized, and requires only combinatorial and algebraic manipulations, no analysis and no transcendantal functions, no integrations.
Moreover there is no question of how to choose branches (e.g. choosing the sign of the square-root).

With this method, the choice of $\acycle$ cycles, as well as choice of branches, is encoded  in $\mathcal K$ and $S$, and nowhere else, so can be computed once for all when we compute these blocks.
The choice of $\bcycle$ cycles is encoded only in $\tau$.

Every integral is a polynomial of degree 1 in $S$, in ${\mathcal K}$, in $F$, in $\zeta$, in $\tau$, and in $\ln E$, with coefficients algebraic functions of the $P_{i,j}$s and of the coefficients of $R(x,y)$.

\medskip
So let us now see the details.

\section{Discriminant and degeneracy}

The first step is to identify the type of degeneracies that $P$ can have, this is done through the discriminant
(we recall the general definition and formula of discriminants in appendix \ref{App:Discr}).

\bd[Discriminant]
Let the discriminant
\beq
\Delta(x) = \Discr_y (P(x,y)) \ \in \Field[x]
\eeq
and
\beq
\Delta = \Discr_x \Delta(x)\ \in \Field.
\eeq

If we write $P(x,y) = \sum_{j=0}^{d} y^j P_j(x)$, the discriminant $\Delta(x)$ is the determinant of the following matrix $\hat\Delta(x)$ of size $2d-1$
\beq\label{eq:discrmatrix}
\Delta(x) = \det \hat\Delta(x) 
\quad \text{with} \quad \hat\Delta(x) =
\begin{pmatrix}
  P_d    & P_{d-1} & \dots   & P_0 & 0      &       \dots 0         \cr
   0      & \ddots  &         &     & \ddots & \vdots    \cr
   0\dots      &    0    &  P_d    &     & \dots  & P_0       \cr
  dP_d    & (d-1)P_{d-1} & \dots   & P_1 & 0      &      \dots 0          \cr
   0      & \ddots  &         &     & \ddots & \vdots     \cr
   0\dots      &    0    &  d P_d    &     & \dots  & P_1           \cr
\end{pmatrix}
\eeq
\ed

\bd[Generic/Degenerate]
We say that $P$ is generic if $\Delta\neq 0$, and is degenerate if $\Delta=0$.
\ed

\br
Remark that $\Delta(x)$ is always multiple of $P_d(x)$, and thus the zeros of $P_d(x)$ are always zeros of $\Delta(x)$.
\er

\br Remark that $\Delta\neq 0$ means that $\Delta(x)$ has only simple zeros, and $\Delta= 0$ means that $\Delta(x)$ has at least one zero of order $\geq 2$. 
\er

\br[Generic vs Degenerate]
In the following, we shall always present the methods first for the generic case $\Delta\neq 0$, then for the degenerate cases.

The reader interested in the generic case only can easily skip the degenerate sections.

\er

\br
One could think that studying the generic  case would be general enough, but  very often in practical applications, the curves of interest are degenerate, this is what makes them more special.

\er

\bex[Weierstrass curve]
Let $P(x,y)=y^2-x^3+ax+b$,
we have
\beq
\Delta(x) = -4(x^3-ax-b)
\eeq
and
\beq
\Delta= 2^{8}(27b^2-4a^3).
\eeq
It is degenerate iff $4a^3=27 b^2$, and generic otherwise.

\eex

The following is a well known theorem
\bd[Degenerate points]
Degenerate points (called  nodal or branchpoints) are the zeros of $\Delta(x)$.
A degenerate point $\beta$ is a pair $\beta=(x_\beta,y_\beta)$ where $x_\beta$ is a zero of $\Delta(x)$, and $y_\beta$ is such that $(y_\beta^{2d-2},\dots,y_\beta^2,y_\beta,1)$ is a null eigenvector of the matrix $\hat\Delta(x_\beta)$.
They are all the solutions of the system
\beq
\left\{\begin{array}{l}
P(x_\beta,y_\beta)=0 \cr
P_y(x_\beta,y_\beta)=0
\end{array}\right. .
\eeq
\ed

In the case where $\Delta(x)$ has a simple zero at $x_\beta$, $\hat\Delta(x_\beta)$ has a unique null vector, and it can be written with Kramer's formula
\beq\label{eq:KramerseigenvectorDiscr}
y_\beta = - \ \frac{\operatorname{Minor}_{d,2d-2} \hat\Delta(x_\beta)}{\operatorname{Minor}_{d,2d-1}\hat\Delta(x_\beta)}.
\eeq

\bd[Field extension]

\begin{itemize}

\item Let the set of all roots of $\Delta(x)$ in the algebraic closure $\bar\Field$ of $\Field$:
\beq
\mathcal R_x=\{ x \in \bar\Field \ | \ \Delta(x)=0\}.
\eeq
Let the field extension that contains all the roots
\beq
\Field_\Delta=\Field[\mathcal  R_x] \ \subset \bar\Field.
\eeq

\item Let the degenerate points given by
\beq
\mathcal R=\{ (x,y) \ | \ \Delta(x)=0, \ P(x,y)=0, \ P_y(x,y)=0 \}.
\eeq

\item  For a given $x\in \mathcal  R_x $, if $\Delta(x)$ has a simple zero, the value of $y$ is given by Kramers formula \eqref{eq:KramerseigenvectorDiscr}, which shows that $y\in\Field[x]$, and in particular $y\in \Field_\Delta$.
More generally, if $\Delta(x)$ has a zero of higher order, $\hat\Delta(x)$ has a null-subspace, generated by lower dimensional minors, and we still have that $y\in \Field_\Delta$.

 Therefore
\beq 
\forall \beta=(x_\beta,y_\beta)\in \mathcal R , \quad x_\beta\in \Field_\Delta \ \text{ and } y_\beta\in \Field_\Delta.
\eeq

\end{itemize}
\ed

\section{Riemann surface, topology, homology}

See \cite{Shaf,Farkas} for classical textbooks.

{\em This section is only for geometric intuition, it is in fact not needed in computations. For example the immersion map defined below needs not be computed for practical applications. It is just useful to remember that it exists.

Similarly, the choice of basis of homology cycles is only to give a geometric meaning to the building blocks $\mathcal K, S, \zeta,\tau$, it is not actually needed in computations.
}

\medskip

The zero locus of the algebraic equation $P(x,y)=0 $ defines a subset $\td\curve$ of $\CC\times \CC$, which is locally a 1-dimensional complex manifold (a complex curve), or equivalently a real 2-dimensional manifold (a surface)
\beq
\td\curve  = \{(x,y)\in\mathbb C\times \mathbb C \,\, | \,\, P(x,y)=0\},
\eeq

In the generic case $\Delta\neq 0$, $\td\curve$ is a smooth and connected surface, but is not compact (there are punctures, where $x$ and/or $y$ tend to $\infty$).

However in degenerate cases $\Delta=0$, this surface might be not connected (if $P$ is factorizable), and in fact it is not even a surface, as it may have non-smooth points $(x,y)\in\mathcal R$, that can be either "cusps" or  with neighborhoods not homeomorphic to a Euclidian disc (rather union of discs), called nodal points, viewed as "pinchings" in the figure below.
$$
\includegraphics[scale=0.35]{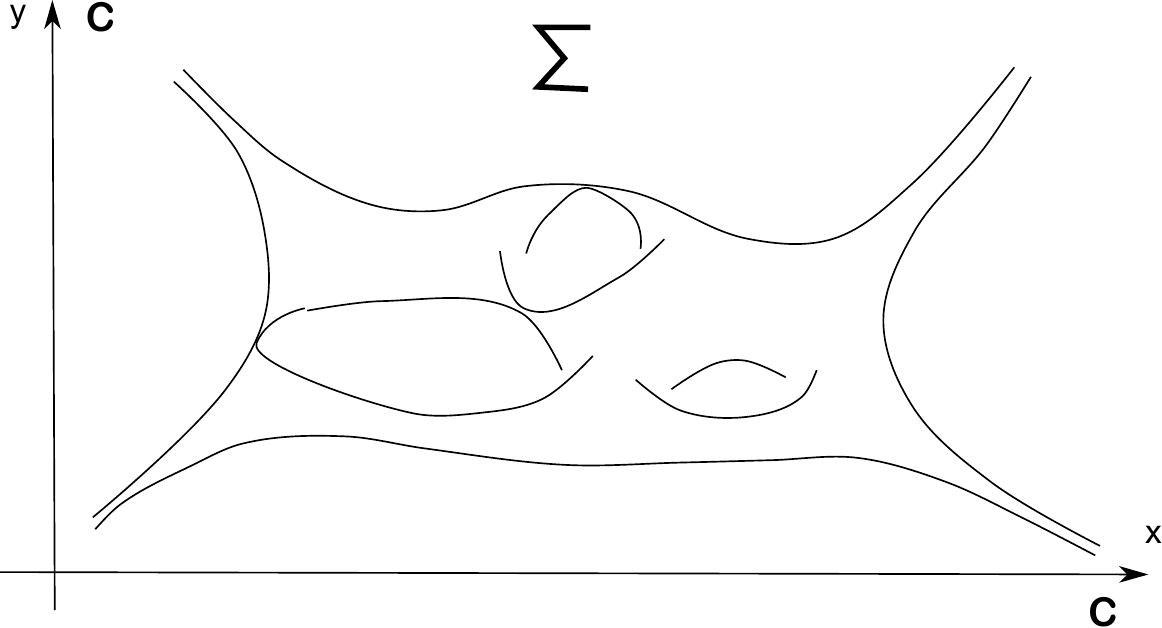}
$$
Let $\Sigma$ the normalization of $\td\Sigma$ (possibly disconnected), a compact smooth Riemann surface, equipped with two meromorphic functions $X:\curve\to \CC, \,Y:\curve\to \CC$, such that
\beq
\td\curve  = \{(x,y)\in\mathbb C\times \mathbb C \,\, | \,\, P(x,y)=0\}
= \{(X(p),Y(p))\,\,|\,\,p\in \curve\setminus\{\text{punctures}\}\}.
\eeq

$\bullet$ The map 
\bea
\ii:\Sigma & \hookrightarrow & \CC\times \CC \cr
p & \mapsto & (X(p),Y(p))
\eea
is a meromorphic immersion, whose image is $\ii(\Sigma)=\td\Sigma$.

$\bullet$ The \textbf{punctures} are the locus where either $x$ or $y$ tends to $\infty$, i.e. the poles of $X$ and/or $Y$.

$\bullet$ at all points $(x,y)\in\td\curve$ where the vector $\nabla P=(P_x(x,y),P_y(x,y))\neq (0,0)$, the surface is smooth, it has a well defined tangent plane $T_{(x,y)}\td\Sigma = (P_y(x,y),-P_x(x,y))\CC$.
At the points where $\nabla P=(0,0)$, there is no tangent plane, the surface is not smooth.

$\bullet$ The meromorphic map
\bea
X:\Sigma & \to & \CC P^1 \cr
p & \mapsto & X(p)
\eea
is a holomorphic ramified covering of $\CC P^1$ by $\curve$.
Its ramification points occur where 2 (or more) branches meet, and thus at $p=$ zeros of $P_y(X(p),Y(p))$, and/or possibly at punctures.

$\bullet$ zeros of $P_y(x,y)=0$ are the zeros of $\Delta(x)$, they can be either regular ramification points, or they can also be nodal points i.e. self-intersection points, and they can be higher ramified.

$\bullet$ For generic $P$ (i.e. $\Delta\neq 0$), the zeros of $P_y(x,y)$ and the zeros of $P_x(x,y)$ are distinct on $\curve$. In that case  $\td\curve$ has everywhere a tangent and is smooth. However for non-generic $P$ these zeros may coincide, and the surface is not smooth. We have a degenerate curve with nodal points of possibly higher degeneracy.

\medskip
\bex[Weierstrass curve]
$P(x,y)=y^2-x^3+g_2 x+g_3$.
\begin{center}
\begin{tikzpicture}

\begin{scope}[scale = 1, xshift=0cm, yshift=0.5cm]
\draw (0,0) ellipse [x radius=1cm,y radius=.5cm];
\draw (-0.5,0.1)  to[out=-45,in=-135] (0.5,0.1) ;
\draw (-0.3,0.05)  to[out=30,in=120] (0.3,0.05) ;
\end{scope}

\begin{scope}[scale = 1, xshift=4cm, yshift=0cm]
\draw (0,0) -- (1.5,0) -- (1.9,1) -- (0.4,1) -- (0,0);
\draw (0,0) node[below,left] {$0$};
\draw (0.4,1) node[left] {$\tau$};
\draw (1.5,0) node[right] {$1$};
\end{scope}

\end{tikzpicture}
\end{center}

$\bullet$  For generic $g_2,g_3$, the curve $\curve$ is a torus, represented as a parallelogram $\CC/(\ZZ+\tau\ZZ)$, and the immersion map $\ii:\curve \to \CC\times \CC$ can be parametrized by the elliptic Weierstrass\footnote{The Weierstrass function $\wp(z,\tau)$ is bi-periodic $\wp(z+1,\tau)=\wp(z+\tau,\tau)=\wp(z)$, and it has a double pole $\wp(z,\tau)\sim z^{-2}+O(z^2)$ at $z=0$ and no other pole. It is the unique bi-periodic function with these properties.} function $\wp(z,\tau)$ and its $z$-derivative:
\bea
X(z) &=& \nu^2 \wp(z,\tau) \cr
Y(z) &=& \frac12\nu^3 \wp'(z,\tau) \cr
\eea
The parameters $(\nu,\tau)$ are functions of $(g_2,g_3)$, the inverse map being:
\beq
g_2 = 15\nu^4 G_4(\tau)
\quad , \quad
g_3 = 60\nu^6 G_6(\tau),
\eeq
with $G_4$ and $G_6$ the modular Eisenstein $G$-series.

There are 3 branchpoints, zeros of $\wp'(z,\tau)$, located at $z=\frac12$, $z=\frac\tau{2}$, $z=\frac12(1+\tau)$.
There is one puncture (pole of $X$ and $Y$) at $z=0$, with $\deg_0 X=2$ and $\deg_0 Y=3$, at which we have the asymptotics  $Y\sim \pm  X^{\frac32}$.
\begin{center}
\begin{tikzpicture}
\begin{scope}[scale = 1, xshift=4cm, yshift=0cm]
\draw (0,0) -- (1.5,0) -- (1.9,1) -- (0.4,1) -- (0,0);
\draw[red] (0,0) node {$\bullet$};
\draw[blue] (0.75,0) node {$\bullet$};
\draw[blue] (0.2,0.5) node {$\bullet$};
\draw[blue] (0.95,0.5) node {$\bullet$};
\draw (0,0) node[below,left] {$0$};
\draw (0.4,1) node[left] {$\tau$};
\draw (1.5,0) node[right] {$1$};
\end{scope}
\end{tikzpicture}
\end{center}

$\bullet$  If $4g_2^3-27g_3^2=0$, then the torus is degenerate, $\curve=\CC P^1$ is then a sphere, and $\td\curve$ has a nodal point.
\begin{center}
\begin{tikzpicture}

\begin{scope}[scale = 1, xshift=0cm, yshift=0.5cm]
\draw (0,0) ellipse [x radius=1cm,y radius=.5cm];
\draw (-0.5,0.1)  to[out=-45,in=-135] (1.,0.) ;
\draw (-0.3,0.05)  to[out=30,in=120] (1.,0.) ;
\draw (1,0) node {$\bullet$};
\end{scope}

\end{tikzpicture}
\end{center}

The immersion map $\ii:\curve\to\CC\times \CC$ can be parametrized by polynomial functions
\bea
X(z) &=& z^2-2u \cr
Y(z) &=& z^3-3uz \cr
\eea
with $u$ given by
\beq
g_2=3 u^2 \quad , \quad g_3 = -2u^3.
\eeq
The nodal point is $\beta=(u,0)$, with $x_\beta = u$ and $y_\beta=0$, corresponding to the immersion  $\beta = \ii(\sqrt{3u}) = \ii(-\sqrt{3u})$ of the 2 points $z=\pm \ii\sqrt{3u} $ in $\curve$.

There is 1 branchpoint, at $z=0$.
There is one puncture (pole of $X$ and $Y$) at $z=\infty$, with $\deg_\infty X=2$ and $\deg_\infty Y=3$, at which $Y\sim \pm X^{\frac32}$.

\eex

\subsection{Genus and cycles}

{\em This section is again only for geometric intuition, it is in fact not needed in computations. In fact a choice of cycles is encoded only in the matrix $\mathcal K$ and in $S$ defined below, and in fact the user never needs to worry about topology and homology if they want to get algebraic expressions with building blocks $\mathcal K$ and $S$. }
\medskip

The compact Riemann surface $\Sigma$ is possibly disconnected $\curve=\curve_1\cup \dots \cup\curve_m$, and each connected component has some genus $\genus_i$.
Let us define the total genus
\beq
\genus=\sum_{i=1}^m \genus_i.
\eeq
It is well known (and we shall recover it below) that the genus is at most the number of interior points to the Newton's polygon:
\beq
0\leq \genus \leq \#\Newtint.
\eeq
The Homology space $H_1(\Sigma,\ZZ)$ has dimension
\beq
\dim H_1(\Sigma,\ZZ) =  2\genus,
\eeq
which means that there exists $2\genus$ independent non-contractible cycles, and it is possible (but not uniquely) to choose a symplectic basis:
\beq
\acycle_1,\dots,\acycle_{\genus}, \ \bcycle_1,\dots,\bcycle_{\genus},
\eeq
such that
\beq
\acycle_i\cap \acycle_j = 0
\ \ , \quad
\bcycle_i\cap \bcycle_j = 0
\ \ , \quad
\acycle_i\cap \bcycle_j = \delta_{i,j}.
\eeq

Such a choice of symplectic basis of cycles is called a Torelli marking of $\curve$. It is not unique.

\begin{center}
\begin{tikzpicture}

\begin{scope}[scale = 1, xshift=0cm, yshift=0.5cm]
  \draw[scale=0.5, domain=0:2, smooth, variable=\x, blue] plot ({2.5*\x}, {sqrt(-\x^4+2+3.5*\x^2)});
  \draw[scale=0.5, domain=0:2, smooth, variable=\x, blue] plot ({2.5*\x}, {-sqrt(-\x^4+2+3.5*\x^2)});
    \draw[scale=0.5, domain=0:2, smooth, variable=\x, blue] plot ({-3*\x}, {sqrt(-\x^4+2+3.5*\x^2)});
    \draw[scale=0.5, domain=0:2, smooth, variable=\x, blue] plot ({-3*\x}, {-sqrt(-\x^4+2+3.5*\x^2)});
\end{scope}

\begin{scope}[scale = 1, xshift=1cm, yshift=0.5cm]
\draw[blue] (-0.5,0.1)  to[out=-45,in=-135] (0.5,0.1) ;
\draw[blue] (-0.3,0.05)  to[out=30,in=120] (0.3,0.05) ;
\end{scope}

\begin{scope}[scale = 1, xshift=-1.5cm, yshift=0.5cm]
\draw[blue] (-0.5,0.1)  to[out=-45,in=-135] (0.5,0.1) ;
\draw[blue] (-0.3,0.05)  to[out=30,in=120] (0.3,0.05) ;
\end{scope}

\draw[red] (-1.5,0.5) ellipse [x radius=1.1cm,y radius=.5cm];
\draw[red] (1,0.5) ellipse [x radius=1.1cm,y radius=.5cm];

\draw[red] (-1.5,0.67)  to[out=100,in=-145] (-1.1,1.45) ;
\draw[red] (1,0.67)  to[out=100,in=-145] (1.2,1.55) ;

\draw[black] (1.8,1.3) node {$\bullet$};
\draw[black] (-1.8,1.2) node {$\bullet$};
\draw[black] (-2,-.4) node {$\bullet$};

\end{tikzpicture}
\end{center}

However, here we need to go a little further.
\bd[Marking of Jordan loops]
We call a marking of $\curve$, a choice of $2\genus$ \textbf{Jordan loops} on $\curve$, satisfying
\beq
\acycle_i\cap \acycle_j = 0
\ \ , \quad
\bcycle_i\cap \bcycle_j = 0
\ \ , \quad
\acycle_i\cap \bcycle_j = \delta_{i,j}.
\eeq
Their homotopy class in $H_1(\Sigma,\ZZ)$ is a Torelli marking of $\curve$ in the usual sense.
We moreover require that they don't contain punctures neither ramification or nodal points.
\ed

\br we insist that $\acycle_i$ and $\bcycle_i$ are not defined as cycles, they are Jordan loops. Cycles are  homotopy classes of Jordan loops, and their linear combinations. Therefore there is a projection from loops to cycles, by taking the homotopy class, but which is not invertible.
\er

\subsection{Holomorphic differentials}

Let $H^1(\curve)$ the space of holomorphic differential 1-forms on $\curve$.

The following is a classical theorem going back to Riemann
\bt[Riemann]
\beq\dim H^1(\curve) =\genus.
\eeq
Having made a choice of Torelli marking of $\Sigma$, there exists a unique basis $\omega_1,\dots,\omega_{\genus}$ of the space $H^1(\Sigma)$  of holomorphic differentials on $\Sigma$, such that
\beq
\oint_{\acycle_i}\omega_j=\delta_{i,j}.
\eeq
From there we define the \index{Riemann matrix of periods}\textbf{Riemann matrix of periods}
\beq
\tau_{i,j} = \oint_{\bcycle_i}\omega_j.
\eeq
$\tau$ is a  ${\genus}\times{\genus}$  \index{Siegel}\textbf{Siegel matrix}, i.e. a symmetric matrix, whose imaginary part is positive definite:
\beq
\tau^T=\tau
\quad , \quad
\Im\ \tau>0.
\eeq
\et
\proof{The method of the proof uses the now famous "Riemann bilinear identity", but we refer the reader to textbooks \cite{Farkas}.}

\section{Newton's polygon and combinatorics}

\begin{center}
\begin{tikzpicture}
\begin{scope}[scale = 0.35]
\grid{8}{7};
\draw[red,thick] (0,0) -- (0,6) --  (1,6)  -- (4,5) -- (5,4) -- (6,3) -- (7,1) -- (7,0) -- cycle;

\def\N{(0,0),(0,6),(1,4),(1,6),(2,5),(4,3),(4,5),(5,4),(6,3),(7,1),(7,0),(3,1),(5,0),(0,3)};
\drawpolytope[black]\N;

\end{scope}
\end{tikzpicture}
\end{center}

{\em As a preliminary we need to introduce the \textbf{combinatorial} setting, a polytope of lattice points, the famous Newton's polytope.

This method is considered introduced by Isaac Newton in 1676, but was more developed by Puiseux \cite{Puiseux} and then followers after 1850 \cite{Baker}.

We suggest the lectures by Casselman \cite{Casselman}.

}

\bd[Newton's polytope, Newton's polygon]

\textbf{Newton's polytope} is the finite set of points $\Newt\subset \ZZ_+\times \ZZ_+$ at which the bivariate polynomial $P(x,y)\in \CC[x,y]$ has non-vanishing coefficients ($P_{i,j}\neq 0$):
\beq
P(x,y) = \sum_{(i,j)\in \Newt} P_{i,j} x^i y^j.
\eeq
The \textbf{Newton's polygon} is its \textbf{convex envelope}.

We define its completion with all integer points enclosed within its convex envelope:
\beq
\bar\Newt = \{(i,j) \in \ZZ\times \ZZ \,|\,(i,j)\in\,\text{inside or on the boundary  of the convex envelope of }\,\Newt\}.
\eeq

We define its \textbf{interior} $\Newtint\subset\mathbb Z\times \mathbb Z$, shifted by $(-1,-1)$:
\beq
\Newtint = \{(i,j)\in \bar\Newt \ |\,(i+1,j+1)\in\,\text{strictly interior of the convex envelope of }\,\Newt\}.
\eeq
and its \textbf{boundary} (the integer points of the convex envelope)
\beq
\dNewt=\bar{\Newt}\setminus(\Newtint+(1,1)),
\eeq
and we define
\beq
\Newt'''=\{(i,j)\in \bar\Newt \ | \ (i+1,j+1) \in \dNewt \} = \text{"3rd kind points"}
\eeq
\beq
\Newt''=\{(i,j)\in \bar\Newt \ | \ (i+1,j+1) \  \notin \bar{\Newt}\}= \text{"2nd kind points"}.
\eeq
\beq
\Newt'=\Newtint=\{(i,j)\in \bar\Newt \ | \ (i+1,j+1) \ \text{strictly interior}  \}= \text{"1st kind points"}.
\eeq
In other words:
\begin{itemize}
\item 1st kind $\Newtint$ = interior : $(i+1,j+1)\in$ strict interior 
\item 3rd kind $\Newt'''$ = boundary : $(i+1,j+1)\in$ boundary
\item 2nd kind $\Newt''$ = exterior : $(i+1,j+1)\in$ exterior
\end{itemize}
\ed
We shall show below in section \ref{sec:forms} why they are called 1st, 2nd or 3rd kind, or refer to textbooks, for instance my lectures notes \cite{eynardlecturesRS}.



\bex{Weierstrass curve}\\ $P(x,y)=y^2-x^3+g_2 x+g_3$ has the following Newton's polygon
\begin{center}
\begin{tikzpicture}
\begin{scope}[scale = 0.55]
\grid{3}{2};
\fill [red, opacity=0.2] (0,0) -- (0,2) -- (3,0)  ;

\draw [red, thick] (0,0)  -- (0,2)  -- (3,0)  -- (1,0)  -- cycle ;
\draw [blue] (1,1) node {$\circ$};
\def\N{(0,0), (1,0),(3,0),(0,2)};

\drawpolytope[black]\N;

\end{scope}
\end{tikzpicture}
\end{center}

where the dots $\bullet$ represent non-zero coefficients $P_{i,j}$, and $\circ$ at position $(1,1)$ is the only interior point to the polygon. Therefore $\Newtint=\{(0,0)\}$,
$\partial \Newt=\{(0,0),(0,1),(0,2),(1,0),(2,0),(3,0)\}$, $\Newt'''=\emptyset$, $\Newt''=\{(0,2),(0,1),(1,0),(2,0),(3,0),(1,1)\}$.
\eex

\bp[properties]
Under $x\to 1/x$, the Newton's polytope is sent to its reflection by a vertical axis.
Under $y\to 1/y$, the Newton's polytope is sent to its reflection by a horizontal axis.

\begin{center}
\begin{tikzpicture}

\begin{scope}[scale = 0.35, xshift=-10cm]
\begin{scope}[xshift=-7cm]
\grid{4}{3};
\fill [red, opacity=0.2] (0,3) -- (0,1) -- (1,0) -- (4,0) -- (3,2) -- (2,3) ;
\draw [red, thick] (0,3) -- (0,1) -- (1,0) -- (4,0) -- (3,2) -- (2,3) -- cycle ;
\end{scope}
\draw (0,2) node {$x\leftrightarrow 1/x$};
\begin{scope}[xshift=3cm]
\grid{4}{3};
\fill [red, opacity=0.2] (0,0) -- (3,0) -- (4,1) -- (4,3) -- (2,3) -- (1,2) ;
\draw [red, thick] (0,0) -- (3,0) -- (4,1) -- (4,3) -- (2,3) -- (1,2) -- cycle ;
\end{scope}
\end{scope}

\begin{scope}[scale = 0.35, xshift=10cm]
\begin{scope}[xshift=-7cm]
\grid{4}{3};
\fill [red, opacity=0.2] (0,3) -- (0,1) -- (1,0) -- (4,0) -- (3,2) -- (2,3) ;
\draw [red, thick] (0,3) -- (0,1) -- (1,0) -- (4,0) -- (3,2) -- (2,3) -- cycle ;
\end{scope}
\draw (0,2) node {$y\leftrightarrow 1/y$};
\begin{scope}[xshift=3cm]
\grid{4}{3};
\fill [red, opacity=0.2] (0,0) -- (2,0) -- (3,1) -- (4,3) -- (1,3) -- (0,2) ;
\draw [red, thick] (0,0) -- (2,0) -- (3,1) -- (4,3) -- (1,3) -- (0,2) -- cycle ;
\end{scope}
\end{scope}

\end{tikzpicture}
\end{center}

\ep

\bd[Shifted Newton's polygon]
\label{def:shiftedNewton}
Let $(x_0,y_0)\in \CC\times \CC$.
Let us define $\Newt^{(x_0,y_0)}$ as the Newton polygon of $P$ shifted by
\beq
P^{(x_0,y_0)}(x,y) = P(x_0+x,y_0+y).
\eeq
The shift $x\to x+x_0$ can modify the left side of the polygon, 
the shift $y\to y+y_0$ can modify the bottom side of the polygon.

If $(x_0,y_0)\in \CC\times \CC $ is a generic point, not solution of $P(x_0,y_0)=0$, then $\Newt^{(x_0,y_0)}$ is a polygon whose  bottom-left corner is the point $(0,0)$.

If $(x_0,y_0)\in \td\curve $ is a generic point solution of $P(x_0,y_0)=0$, then $\Newt^{(x_0,y_0)}$ is a polygon with the bottom-left corner truncated by a $1\times 1$ triangle.
\begin{center}
\begin{tikzpicture}
\begin{scope}[scale = 0.35]
\grid{4}{3};
\fill [red, opacity=0.2] (0,3.2) -- (0,1) -- (1,0) -- (4.2,0) -- (4.2,3.2) ;
\draw [red, thick] (0,3) -- (0,1) -- (1,0) -- (4,0);
\end{scope}
\end{tikzpicture}
\end{center}

If $(x_0,y_0)\in \mathcal R $ is a non--generic point solution of $P(x_0,y_0)=0$, then $\Newt^{(x_0,y_0)}$ is a polygon with the bottom-left corner truncated by a larger area, see section \ref{sec:nodalbp}.
\ed

\subsection{Degree}

{\em It is convenient to grade algebraic curves by their degree, which is at the same time the size $d$  of the smallest $d\times d$ triangle that contains $\Newt$, and the generic number of intersections of the surface $\td\curve$ with a straight line.}

\begin{center}
\begin{tikzpicture}
\begin{scope}[scale = 0.45]
\grid{4}{4};
\fill [blue, opacity=0.1] (0,4) --  (4,0) -- (0,0)  ;
\fill [red, opacity=0.2] (0,3) --  (2,2) --  (3,0) -- (0,0)  ;
\draw [red, thick] (0,3) --  (2,2) -- (3,0) -- (0,0) -- cycle ;
\draw[red] (0,3) node {$\bullet$};
\draw[red] (2,2) node {$\bullet$};
\draw[red] (3,0) node {$\bullet$};
\draw[red] (0,0) node {$\bullet$};

\draw [blue, thick] (0,4) --  (4,0)  ;

\end{scope}
\end{tikzpicture}
\end{center}

\bd[Degree]
Let
\beq
\deg P = \max_{(i,j)\in \Newt} i+j.
\eeq
\begin{itemize}
    \item  A curve of degree 1 is a line $ax+by+c=0$. It has genus $0$.
    \item $\deg P$ is the number of intersections of a generic line $y=ax+b$ with the surface $\td\curve$, i.e. the number of solutions of $P(x,y=ax+b)=0$ for generic $a,b$.
    \item  A curve of degree 2 is a conic (ellipse, hyperbola, parabola, or union of 2 lines). It has genus $0$.
    \item  A curve of degree 3 is an elliptic curve, it has generically genus $\genus= 1$, or genus $\genus=0$ if degenerate.
    \item The degree $d=\deg P$ is the size of the smallest $d\times d$ triangle that contains $\Newt$.
    
\end{itemize}

\ed

\subsection{Nodal/branch points}
\label{sec:nodalbp}

For $\beta=(x_\beta,y_\beta)\in \mathcal R$, in def \ref{def:shiftedNewton} we defined $P^{(\beta)}$ and its Newton's polygon $\Newt^{(\beta)}$ as
\beq
P^{(\beta)}(x,y) = P(x_\beta+x,y_\beta+y) = \sum_{(i,j)\in \Newt_\beta} P^{(\beta)}_{i,j} x^i y^j \quad \in \Field[x_\beta][x,y]\subset  \Field_\Delta[x,y].
\eeq
Notice that the points $(0,0)$ and  $(0,1)$ are necessarily outside of $ \Newt^{(\beta)}$.

\subsubsection{Generic case}

If $\Delta\neq 0$, all $\beta=(x_\beta,y_\beta)\in \mathcal R$ are generic branchpoints, i.e. $P(x_\beta,y_\beta)=0$ and $P_y(x_\beta,y_\beta)=0$, but no other derivative vanishes.
Then $\Newt^{(\beta)}$ is a polygon with the bottom-left corner truncated by a $1\times 2$ triangle as
\begin{center}
\begin{tikzpicture}
\begin{scope}[scale = 0.35]
\grid{4}{3};
\fill [red, opacity=0.2] (0,3.2) -- (0,2) -- (1,0) -- (4.2,0) -- (4.2,3.2) ;
\draw [red, thick] (0,3) -- (0,2) -- (1,0) -- (4,0);
\draw (1,0) node {$\bullet$} ;
\draw (0,2) node {$\bullet$} ;

\end{scope}
\end{tikzpicture}
\end{center}
Notice that removing this corner doesn't affect the number of interior points of $\Newtint$.

\subsubsection{Degenerate case $\Delta= 0$}

Let $\beta=(x_\beta,y_\beta)\in \mathcal R$, such that more derivatives of $P$ vanish at $\beta$.
Then $\Newt^{(\beta)}$ is a polygon with the bottom-left corner truncated by a concave region larger than a $1\times 2$ triangle:
\begin{center}
\begin{tikzpicture}
\begin{scope}[scale = 0.65]
\grid{5}{5};

\fill [red, opacity=0.2] (0,5.2) -- (0,4) -- (1,2) -- (2,1) -- (3,0) -- (5.2,0) -- (5.2,5.2) -- (0,5.2) ;
\draw [red, thick, opacity=0.6] (0,5) -- (0,4) -- (1,2) -- (2,1) -- (3,0) -- (5,0)  ;
\node (A1) at (0,4) {$\bullet$};
\node[blue] (A2) at (1,2) {$\bullet$};
\node[blue] (A3) at (2,1) {$\bullet$};
\node (A4) at (3,0) {$\bullet$};
\node[blue] (B) at (1,1) {$\bullet$};

\draw (3,1) [blue] node {$\circ$};
\draw (2,2) [blue] node {$\circ$};
\draw (1,3) [blue] node {$\circ$};
\draw (3,2) [blue] node {$\circ$};
\draw (1,4) [blue] node {$\circ$};
\draw (2,3) [blue] node {$\circ$};
\draw (3,3) [blue] node {$\circ$};
\draw (2,4) [blue] node {$\circ$};
\draw (4,1) [blue] node {$\circ$};
\draw (4,2) [blue] node {$\circ$};

\fill [blue, opacity=0.1] (0,0) -- (3,0) -- (2,1) -- (1,1) -- (1,2) -- (0,4) -- (0,0)  ;

\fill [blue, opacity=0.2] (2,1) -- (1,2) -- (1,1) -- cycle ;

\end{scope}
\end{tikzpicture}
\end{center}

\begin{itemize}

\item Let $\ell_\beta$ the number of bottom-left irreducible integer segment boundaries of $\Newt^{(\beta)}$ (see figure above).
We have
\beq
\ell_\beta  = \# \ii^{-1}(\beta) = \#\text{ bottom-left integer segment boundaries of } \Newt^{(\beta)}.
\eeq
(in the figure above $\ell_\beta=3$ segments.)

\item If $\ell_\beta>1$, $\beta$ is called a \textbf{nodal point}, and if $\ell_\beta=1$, $\beta$ is called a \textbf{branchpoint}.

\item Let $\check\Newt^{(\beta)}$ the set of lattice points  strictly between the interior of $\Newt^{(\beta)}$ and the coordinate axis $i=0$ and $j=0$, shifted by $(-1,-1)$:
\beq
\check\Newt^{(\beta)} = \{(i-1,j-1)  \ | \ (i,j)=  \text{ points strictly between  }\Newtint^{(\beta)} \text{ and coordinate axis} \}.
\eeq
(in the figure above it $\check\Newt^{(\beta)}$ is the middle $1\times 1$ triangle, with $\#\check\Newt^{(\beta)}=3$ lattice points.)

\item Let $\genus_\beta$ (called the genus of $\beta$) the number of integer points in it
\beq
\genus_\beta = \# \check\Newt^{(\beta)}.
\eeq
Notice that $\genus_\beta\geq \ell_\beta-1$, and in particular, nodal points always have $\genus_\beta>0$.

(in the figure above  $\genus_\beta=3$.)

\item Under a deformation of $P(x,y)$, the points in the bottom left corner may become occupied by some coefficienst of $P$, and may become interior points. Therefore the genus of the whole surface may increase by up to $\genus_\beta$, i.e. the resolution of the nodal points may increase the genus of $\curve$  up to $\genus+\genus_\beta$.

\item for each irreduciple integer segment $s$ of the bottom-left boundary of $\Newt^{(\beta)}$ of normal vector $(a_s,b_s)$, we have
\beq
\order_\beta X = a_s
\qquad , \qquad
\order_\beta Y = b_s.
\eeq
A local model is
\beq
y-y_\beta =  C_s^{\frac{1}{a_s}} (x-x_\beta)^{\frac{b_s}{a_s}}
\eeq
where $C_s$ is a zero of the polynomial:
\beq
\sum_{(i,j) \in \Newt\cap s} P^{(\beta)}_{i,j} C_s^{j/a_s} = 0.
\eeq
If the boundary $s$ is single (not aligned with other boundaries), this polynomial is of degree 1, so that $C\in \Field[x_\beta]\subset \Field_\Delta $.
If the boundary is not single, then $C_s$ belongs to a Field extension of $\Field[x_\beta]$.

(in the figure above there are 3 boundaries of normal vectors $(2,1)$, $(1,1)$, $(1,1)$.
One corresponds to $y-y_\beta \sim \sqrt{C(x-x_\beta)} $ with $P_{0,4}C^2+P_{1,2}C=0$ i.e. $C=-P_{1,2}/P_{0,4}$,
and the other 2 correspond to $y-y_\beta \sim C(x-x_\beta) $ with $C$ the 2 roots of $P_{1,2}C^2+P_{2,1}C+P_{3,0}=0$.
)

\item
Let
\beq
\deg \beta = \sum_{s=1}^{\ell_\beta} a_s = \text{height of  } \check\Newt_\beta.
\eeq
(in the figure above  $\deg \beta=4=2+1+1$.)

\item the neighborhood of $\beta$ in $\ii^{-1}(\curve)$ is a union of $\ell_\beta$ ramified discs of respective degrees $a_i$, and of total degree $\deg\beta$.

\item For a \textbf{generic branchpoint}, we have $P_x(\beta)\neq0 $ and $P_{yy}(\beta)\neq 0$, the bottom left boundary of the Newton's polygon is the segment $[(1,0),(0,2)]$.
We have $\ell_\beta=1$, and $a=\order_\beta x=2$, $b=\order_\beta y=1$. We have $\deg\beta=2$, $\genus_\beta=0$. A neighborhood of the branchpoint is a disc which is a degree 2 cover of a disc in the base.
The local model is $y-y_\beta \sim \sqrt{C(x-x_\beta)}$ with $C=-2P_x(\beta)/P_{yy}(\beta)$.

\begin{center}
\begin{tikzpicture}
\begin{scope}[scale = 0.35]
\grid{4}{3};
\fill [red, opacity=0.2] (0,3.2) -- (0,2) -- (1,0) -- (4.2,0) -- (4.2,3.2) ;
\draw [red, thick] (0,3) -- (0,2) -- (1,0) -- (4,0);
\draw[red] (0,2) node {$\bullet$};
\draw[red] (1,0) node {$\bullet$};

\end{scope}
\end{tikzpicture}
\end{center}

\item For a \textbf{generic nodal point}, we have $P_x(\beta)=0 $ and $P_{xx},P_{xy},P_{yy}\neq 0$, the bottom left boundary of the Newton's polygon is the segments $[(2,0),(1,1)]\cup [(1,1),(0,2)]$.
We have $\ell_\beta=2$, with $a_i=\order_\beta x=1$, $b_i=\order_\beta y=1$. We have  $\deg\beta=2=a_1+a_2$, and $\genus_\beta=1$. The neighborhood of the nodal point is a union of 2 unramified discs, i.e. a pinched cyclinder. A deformation will resolve the nodal point into a cylinder, and may increase the genus by 1.
The local model is $y-y_\beta \sim C_i(x-x_\beta)$ with $C_i$ the 2 solutions of $P_{xx} + 2P_{xy}C+P_{yy}C^2=0$. Or in other words the vectors $(1,C_1)$ and $(1,C_2)$ are the 2 eigenvectors of the Hessian matrix of $P$  at $\beta$.

\begin{center}
\begin{tikzpicture}
\begin{scope}[scale = 0.35]
\grid{4}{3};
\fill [red, opacity=0.2] (0,3.2) -- (0,2) -- (1,1) -- (2,0) -- (4.2,0) -- (4.2,3.2) ;
\draw [red, thick] (0,3) -- (0,2) -- (1,1) -- (2,0) -- (4,0);
\draw[red] (0,2) node {$\bullet$};
\draw[red] (2,0) node {$\bullet$};
\draw[blue] (1,1) node {$\bullet$};

\end{scope}
\end{tikzpicture}
\end{center}

\end{itemize}

\bex{Degenerate Weierstrass curve}
$P(x,y)=y^2-x^3+g_2 x+g_3$, with $4g_2^3-27g_3^2=0$.
$\curve=\CC P^1 $ is a surface of genus $\genus=0$,  equipped with the immersion map $\ii:\curve\to\CC\times \CC, \ z\mapsto (X(z),Y(x))$:
\bea
X(z) &=& z^2-2u \cr
Y(z) &=& z^3-3uz \cr
\eea
with 
$g_2=3 u^2 \quad , \quad g_3 = -2u^3$.
The nodal point is $\beta=(u,0)$ and $\ii^{-1}(\beta) =\{\sqrt{3u},-\sqrt{3u}\}$.
We have
\beq
P^{(\beta)}(x,y) = P(x+u,y) = y^2-x^3+3ux^2.
\eeq
$\beta$ is a nodal point, it has $\ell_\beta=2$, $\deg\beta=2$, $\genus_\beta=1$.

\begin{center}
\begin{tikzpicture}
\begin{scope}[scale = 0.45]
\grid{4}{3};
\fill [red, opacity=0.2] (0,2) --  (2,0) -- (3,0)  ;
\draw [red, thick] (0,2) --  (2,0) -- (3,0) -- cycle ;
\draw[red] (0,2) node {$\bullet$};
\draw[red] (2,0) node {$\bullet$};
\draw[red] (3,0) node {$\bullet$};
\draw[blue] (1,1) node {$\bullet$};
\end{scope}
\end{tikzpicture}
\end{center}

Under a deformation of $P$, the resolution of the nodal point, will be a surface of genus 
\beq
\genus+\genus_\beta=0+1=1=\#\Newtint.
\eeq
\eex

\subsection{Moduli space}

If $K$ is a finite polytope $K\subset\ZZ_{+}\times \ZZ_{+}$, we denote the vector sub-space of $\Field[x,y]$ of bivariate polynomials with monomial powers in $K$:
\beq
\Field[K] = \operatorname{span}_{(i,j)\in K} \left\langle  x^{i}y^{j} \right\rangle \subset \Field[x,y].
\eeq

\bd[Moduli space]
We define the moduli space associated to $P$ as
\beq
\modsp(P)
= \cap_{p\in \CC\times \CC} \ \Field[\Newtint^{(p)}].
\eeq
\ed

\br
For generic $p$ we have $\Field[\Newtint^{(p)}] = \Field[\Newtint]$.
It is only at special points (nodal points and punctures) that the number of interior points could be lower and the intersection might restrict the space, and thus the intersection is in fact finite, and we have $\modsp(P) \subset \Field[\Newtint]$.
\er

\br For generic $P$, we have
$$ \modsp(P) = \Field[\Newtint].$$
For non generic $P$ we only have
$\modsp(P) \subset \Field[\Newtint]$.
\er

The following classical theorem is really what makes the Newton's polygon method so famous, it allows to compute the genus (which is a topological notion) algebraically from the combinatorics of the Newton's polygon:
\bt[Genus from combinatorics]
The genus is the dimension of $\modsp(P)$
\beq
\genus=\dim\modsp(P).
\eeq
If the curve is generic $\Delta\neq 0$, the genus is the number of interior points
\beq
\genus = \# \Newtint.
\eeq
If the curve is not generic, in general we only have $\genus \leq \# \Newtint$.
\et

\bex[Degenerate Weierstrass curve]
$P(x,y)=y^2-x^3+g_2 x+g_3$, with $27g_3^2=4g_2^3$.
We have $\Newtint=\{(0,0)\}$, but $\Newtint^{(\beta)}=\emptyset$ at the nodal point, so that $\modsp(P)=\{0\}$, whose dimension is $\genus=0$.
For the non-degenerate case, there is no nodal point and $\modsp(P)=\Field$, whose dimension is $\genus=1$.
\eex

\subsection{Punctures}


\bd[Canonical local coordinate]
Consider a puncture $\alpha\in\curve$, a pole of either $X$  or $Y$ or both.

$\bullet$ If $X(\alpha)=\infty$, let $a_\alpha=-\deg_\alpha X<0$, and we introduce the canonical local coordinate
\beq
\xi_\alpha = X^{1/a_\alpha}.
\eeq

$\bullet$ If $X(\alpha)\neq\infty$, let $a_\alpha=\operatorname{order}_\alpha (X-X(\alpha))\geq 1$, and we introduce the canonical local coordinate
\beq
\xi_\alpha =(X-X(\alpha))^{1/a_\alpha}.
\eeq
It is convenient to use the convention that $X_\alpha=X(\alpha)$ if $X(\alpha)\neq\infty$ and $X_\alpha=0$ if $X(\alpha)=\infty$, so that \eqref{def:cannonicalzetaalpha} holds in both cases.
\beq\label{def:cannonicalzetaalpha}
\xi_\alpha =(X-X_\alpha)^{1/a_\alpha}.
\eeq

Notice that in both cases $a_\alpha = 1+\operatorname{order}_\alpha dX$, and we have
\beq
dX = a_\alpha \xi_\alpha^{a_\alpha-1} d\xi_\alpha.
\eeq

Notice that $\xi_\alpha$ is defined only up to a $a_\alpha$th root of unity.
Let us denote
\beq
\rho_d = e^{2\pi i /d}.
\eeq
Then $\rho_{a_\alpha}^k \xi_\alpha$ is also a canonical local coordinate for any integer $k$.
The choice of one canonical local coordinate is equivalent to choosing a ray emanating from $\alpha$, on which $\xi_\alpha\in \RR_+$, i.e. the ray $\operatorname{Arg}\xi_\alpha=0$.

In all cases, the canonical local coordinate vanishes linearly at $\alpha$.

\ed

\subsection{Punctures as sides of the Newton's polygon}

{\em 
Punctures can be studied as nodal and branchpoints, by changing $x\to 1/x$ or $y\to 1/y$, and are thus related to the sides of the Newton's polygon.
}

\begin{itemize}
    \item Let $\alpha$ a puncture, with exponents 
\beq
\left\{\begin{array}{l}
x= X_\alpha + \xi_\alpha^{a_\alpha} \cr
y\sim Y_\alpha + \eta_\alpha \xi_\alpha^{b_\alpha} (1+O(\xi_\alpha)) 
\qquad \eta_\alpha\neq 0
\end{array}\right. .
\eeq
(where we take the convention that $X_\alpha$ or $Y_\alpha$ is zero if $\alpha$ is a pole).
A puncture is such that either $a_\alpha<0$ or $b_\alpha<0$ or both.

\item Let
\beq
m_\alpha = \max_{   (i,j)\in \Newt}  a_\alpha i+b_\alpha j.
\eeq
\item Let $D_\alpha$ the line in the $(i,j)$--plane of slope $-a_\alpha/b_\alpha$, of equation
\beq
D_\alpha = \{(i,j)\in \RR\times \RR \ | \ a_\alpha i + b_\alpha j = m_\alpha\}
\eeq
The maximality of $m_\alpha$ means that the line $D_\alpha$ touches the Newton's polygon at at least one summit, and is such that the Newton's polygon lies entirely to one side of $D_\alpha$, the same side as the normal vector $(a_\alpha,b_\alpha)$.

\begin{center}
\begin{tikzpicture}
\begin{scope}[scale = 0.45]
\grid{5}{5};
\fill [red, opacity=0.2] (0,4) --  (3,3) -- (4,2) -- (5,0) -- (0,0)  ;
\draw [red, thick] (0,4) --  (3,3) -- (4,2) -- (5,0) -- (0,0) -- cycle ;
\draw[red] (0,4) node {$\bullet$};
\draw[red] (3,3) node {$\bullet$};
\draw[red] (4,2) node {$\bullet$};
\draw[red] (5,0) node {$\bullet$};
\draw[red] (0,0) node {$\bullet$};

\draw [blue, thick] (0,5) --  (3,3) -- (6,1)  ;
\draw [->,black, thick] (3,3)  -- (2,1.5)  ;

\end{scope}
\end{tikzpicture}
\end{center}

\item Let us denote
\beq
P_\alpha(\eta) = \sum_{(i,j)\in \Newt\cap D_\alpha} P_{i,j} \eta^j.
\eeq
Asymptotically near $\alpha$ we have
\beq
0  = P(x,y) = \xi_\alpha^{m_\alpha} P_\alpha(\eta_\alpha) (1+O(\xi_\alpha)).
\eeq
This implies that 
\beq
P_\alpha(\eta_\alpha)=0.
\eeq
Since $\eta_\alpha\neq 0$, this implies that $P_\alpha$ can't be a monomial, it must have at least to terms, i.e. we must have
\beq
\# (D_\alpha\cap \Newt ) \geq 2.
\eeq

\item The line  $D_\alpha$ must pass through at least 2 distinct points of $\Newt$, and the fact that $m_\alpha$ is maximal, implies that all points of $\Newt$ must lie on one side (that of the normal vector $(a_\alpha,b_\alpha)$).
Therefore $D_\alpha$ must be a tangent to the Newton's polygon.

\begin{center}
\begin{tikzpicture}
\begin{scope}[scale = 0.45]
\grid{5}{5};
\fill [red, opacity=0.2] (0,4) --  (3,3) -- (4,2) -- (5,0) -- (0,0)  ;
\draw [red, thick] (0,4) --  (3,3) -- (4,2) -- (5,0) -- (0,0) -- cycle ;
\draw[red] (0,4) node {$\bullet$};
\draw[red] (3,3) node {$\bullet$};
\draw[red] (4,2) node {$\bullet$};
\draw[red] (5,0) node {$\bullet$};
\draw[red] (0,0) node {$\bullet$};

\draw [blue, thick] (1,5)  -- (5,1)  ;

\draw [->,black, thick] (3,3)  -- (2,2)  ;

\end{scope}
\end{tikzpicture}
\end{center}

\item If $D_\alpha$ contains more than 2 lattice points of $\overline{\Newt}$, then $P_\alpha(\eta)=0 $ possesses as many non-zero solutions as the number of irreducible segments in $D_\alpha\cap \Newt$.
There is a 1-1 correspondence between irreducible boundary segments and punctures.

\begin{center}
\begin{tikzpicture}
\begin{scope}[scale = 0.45]
\grid{5}{5};
\fill [red, opacity=0.2] (0,4) --  (2,3) -- (3,2) -- (4,1) -- (4,0) -- (0,0)  ;
\draw [red, thick] (0,4) --  (2,3) -- (3,2) -- (4,1) -- (4,0) -- (0,0) -- cycle ;
\draw[red] (0,4) node {$\bullet$};
\draw[red] (2,3) node {$\bullet$};
\draw[red] (3,2) node {$\bullet$};
\draw[red] (4,1) node {$\bullet$};
\draw[red] (4,0) node {$\bullet$};
\draw[red] (0,0) node {$\bullet$};

\draw [blue, thick] (0,5)  -- (5,0)  ;

\end{scope}
\end{tikzpicture}
\end{center}
\end{itemize}

The following lemma is an obvious computation
\bl
We have near $\alpha$
\beq
P_y(x,y) \sim \xi_\alpha^{-b_\alpha+m_\alpha} P'_\alpha(\eta_\alpha) (1+O(\xi_\alpha)).
\eeq

\el

\subsubsection{Case of a puncture associated to a horizontal side}
    
    Let $P_d(x)$ the coefficient of $y^d$ in $P(x,y)$, the term of highest degree in $y$. 
    Horizontal punctures are associated to zeros of $P_d(x)$.
    They are points at which $x$ is finite and $y$ is infinite.

\begin{itemize}
    
    \item let $X_\alpha$ a zero of $P_d(x)$, and consider the shifted Newton's polygon $\Newtint^{(X_\alpha,0)}$. 
    Generically, its top-left corner is truncated by a $1\times 1$ triangle.
    
\begin{center}
\begin{tikzpicture}
\begin{scope}[scale = 0.35]
\grid{4}{4};
\fill [red, opacity=0.2] (0,-0.25) -- (0,2) -- (1,3) -- (3,3) -- (4.5,2.25) -- (4.5,-0.25) -- cycle ;
\draw [red, thick] (0,-0.25) -- (0,2) -- (1,3) -- (3,3) -- (4.5,2.25);
\draw[red] (0,2) node {$\bullet$};
\draw[red] (1,3) node {$\bullet$};
\draw[red] (2,3) node {$\bullet$};
\draw[red] (3,3) node {$\bullet$};

\end{scope}
\end{tikzpicture}
\end{center}

    \item However, if $X_\alpha$ is a zero of $P_d$ of higher degree, and possibly a zero of $P_{d-1},P_{d-2},\dots$, the top-left corner can be truncated by a larger area
\begin{center}
\begin{tikzpicture}
\begin{scope}[scale = 0.35]
\grid{4}{4};
\fill [red, opacity=0.2] (0,-0.25) -- (0,0) -- (1,2) -- (2,3) -- (4,3) -- (4.5,2.75) -- (4.5,-0.25) -- cycle ;
\draw [red, thick] (0,-0.25) -- (0,0) -- (1,2) -- (2,3) -- (4,3) -- (4.5,2.75);
\draw[red] (0,0) node {$\bullet$};
\draw[red] (1,2) node {$\bullet$};
\draw[red] (2,3) node {$\bullet$};
\draw[red] (3,3) node {$\bullet$};
\draw[red] (4,3) node {$\bullet$};

\end{scope}
\end{tikzpicture}
\end{center}

\item On each side $\alpha$ (an irreducible lattice segment) of the top-left corner of $\Newt^{(X_\alpha,0)}$, with normal vector $(b_\alpha,a_\alpha)$ (with $a_\alpha>0$ and $b_\alpha<0$) we have the local model
\beq
 y \sim \eta_\alpha (x-X_\alpha)^{b_\alpha/a_\alpha}
\eeq    
with $\eta_\alpha$ solution of $P_\alpha(\eta)=\sum_{(i,j)\in \alpha} P^{(X_\alpha,0)}_{i,j} \eta^j =0$.

\end{itemize}

\section{Differential forms}

Every meromorphic 1-form $\omega$ on $\curve$ can be expressed as $\omega=R(x,y)dx$ where $R(x,y)\in  \CC(x,y) $ is a bivariate rational function. 
More precisely, for every $\omega$, there exists at least one (and in fact infinitely many since we can add multiples of $P(x,y)$) bivariate rational function such that $\omega= R(x,y)dx$ on some open set of $\curve$.

\subsection{Combinatorial 1-forms}
\label{sec:forms}

Let $(i,j)\in \ZZ_+\times \ZZ_+$.

\bp[Combinatorial 1-forms]
Let
\beq\label{def:OmegaComb}
\Omega_{i,j} = \frac{x^i y^j}{P_y(x,y)}dx.
\eeq
\begin{itemize}
\item if $(i,j)\in \Newt''$ then $\Omega_{i,j}$ has poles of degree $m_\alpha-(i'+1)a_{\alpha}-(j+1)b_\alpha+1$ at all punctures $\alpha$ such that $(i+1,j+1)$ lies to the left of the line $D_\alpha $. Notice that this degree  is equal to 1+distance between $D_\alpha$ and the point $(i+1,j+1)$:
$$
m_\alpha-(i+1)a_{\alpha}-(j+1)b_\alpha+1
= 1 + \sqrt{a_\alpha^2+b_\alpha^2} \ \operatorname{dist}(D_\alpha,(i+1,j+1)).
$$

Since $\Omega_{i,j}$ has at least one pole of degree $\geq 2$, it is a 2nd kind differential form.

\item if $(i,j)\in \Newt'''$ then $\Omega_{i,j}$ has simple poles at all punctures $\alpha$ such that $(i+1,j+1)$ lies on the line $D_\alpha $. Notice that points on the boundary always belong to exactly 2 punctures, i.e. $\Omega_{i,j}$ has 2 simple poles. This is called a 3rd kind differential form.
Its residue at the puncture $\alpha$ is
\beq
\Res_{\alpha} \Omega_{i,j} = \frac{-a_\alpha \ \eta_\alpha^j}{P'_\alpha(\eta_\alpha)}.
\eeq

\item Let $\alpha$ a finite puncture such that $Y$ has a pole at $\alpha$, while $X(\alpha)=x_\alpha$ is finite. $\alpha$ corresponds to a horizontal segment of the top boundary of $\Newt$. $x_\alpha$ is a zero of $P_d(x)$ the coefficient of the highest power of $y$ in $P(x,y)$. 
Consider the Newton's polygon shifted by $x_\alpha$, and re-expand  $\Omega_{i,j}(x,y)$ in terms of the shifted $\Omega^{(x_\alpha,0)}_{i,j}(x,y)=\Omega_{i,j}(x+x_\alpha,y)$:
\beq
\Omega_{i,j}(x,y)=\sum_{i'=0}^i \begin{pmatrix}i \cr i'\end{pmatrix} (-x_\alpha)^{i-i'} \ \Omega^{(x_\alpha,0)}_{i',j}(x,y)
\eeq
$\Omega^{(x_\alpha,0)}_{i',j}(x,y)$ has no pole at $\alpha$ iff $(i',j) \in \Newtint^{(x_\alpha,0)}$, otherwise it has a pole of degree $m_\alpha-(i'+1)a_{\alpha}-(j+1)b_\alpha+1$ (and here $a_\alpha<0$).

If $\alpha$ is unramified we have $a_\alpha=-1$, and we have $\Newtint^{(x_\alpha,0)} =\Newtint$, so that $\Omega_{i,j}$ is holomorphic iff $(i,j)\in\Newtint$.

\item if $(i,j)\in \Newtint$ then $\Omega_{i,j}$ has no pole at any puncture.
However it could have poles at zeros of $P_y$ if these are not compensated by zeroes of $dX$, i.e. it can have poles at nodal points.
If there is no nodal point, then it has no pole at all, it is called "holomorphic form", or 1st kind differential form.

This will be studied in further details in the next paragraph.

\end{itemize}

\ep

\proof{}
At a puncture $\alpha$ we have
\beq
\frac{x^i y^j}{P_y(x,y)}dx
\sim a_\alpha\frac{\eta_\alpha^j}{P'_\alpha(\eta_\alpha)} \ \xi_\alpha^{-(i+1)a_\alpha-(j+1)b_\alpha+m_\alpha-1}  (1+O(\xi_\alpha)) d\xi_\alpha 
\eeq
where $\xi_\alpha=(x-x_\alpha)^{1/a_\alpha}$.
$|m_\alpha-(i+1)a_\alpha-(j+1)b_\alpha|$ is the distance of the point $(i+1,j+1)$ to the line $D_\alpha$, times the norm of the normal vector $\sqrt{a_\alpha^2+b_\alpha^2}$:
$$ |m_\alpha-(i+1)a_\alpha-(j+1)b_\alpha| = \sqrt{a_\alpha^2+b_\alpha^2} \operatorname{dist}(D_\alpha,(i+1,j+1)) .$$

\subsection{Holomorphic 1-forms}


Holomorphic 1-forms are those without poles.
The following is a classical theorem
\bt[Riemann]
$\forall \ (i,j)\in \Newtint$, the following differential form
\beq
\Omega_{i,j} = \frac{x^i y^j \ dx}{P_y(x,y)}  
\eeq
is holomorphic at all the punctures. 
Its only poles could be at the zeros of $P_y(x,y)$ if these are not compensated by zeros of $dx$, i.e. these can be only nodal points.

In the generic case, all zeros of $P_y(x,y)$ are simple and are zeros of $dx$, so that this 1-form has no pole at all, it is holomorphic.

$\bullet$ In the generic case we have
\beq
H^1(\Sigma) = \operatorname{Span}\left< \Omega_{i,j} \right>_{(i,j)\in\Newtint}  = \CC[\Newtint] \frac{dx}{P_y(x,y)}
\qquad , \qquad
\genus=\dim H^1(\Sigma)= \#\Newtint.
\eeq

$\bullet$ In the non-generic case we only have
\beq
H^1(\Sigma) = \modsp(P) \frac{dx}{P_y(x,y)}
\qquad , \qquad
\genus=\dim H^1(\Sigma)=\dim \modsp(P) \leq \#\Newtint.
\eeq

In all cases, the holomorphic forms are linear combinations of the $\Omega_{i,j}$, and therefore there exists a rectangular matrix $\hat{\mathcal K}_{k;(i,j)}$ of size $\genus\times \#\Newtint$, such that the normalized holomorphic differentials can be written
\beq
\forall \ k=1,\dots,\genus , \qquad
\omega_k = \sum_{(i,j)\in \Newtint} \hat{\mathcal K}_{k;(i,j)} \frac{x^i y^j \ dx}{P_y(x,y)} = \sum_{(i,j)\in \Newtint} \hat{\mathcal K}_{k;(i,j)} \Omega_{i,j}.
\eeq
Let the $\#\Newtint\times \genus$ rectangular matrix
\beq
{\mathcal {K}}_{(i,j);k} = \oint_{\acycle_k}   \frac{x^i y^j \ dx}{P_y(x,y)} = \oint_{\acycle_k}   \Omega_{i,j} .
\eeq
By definition we have $\hat{\mathcal {K}} {\mathcal {K}} = \text{Id}_{\genus}$, i.e.
\beq
\sum_{(i,j)\in \Newtint} \hat{\mathcal {K}}_{k;(i,j)}   \ {\mathcal {K}}_{(i,j);k}   =\delta_{k,l}.
\eeq
This shows that $\operatorname{rank} \mathcal K  = \operatorname{rank} \hat{\mathcal K} = \genus $.
And this shows that when $\genus=\#\Newtint$, $\mathcal K$ is invertible.

In all cases we have
\beq
\modsp(P) = \hat{\mathcal K}(\CC[\Newtint])
\quad , \quad
H^1(\curve) = \frac{dx}{P_y(x,y)} \hat{\mathcal K}(\CC[\Newtint]) = \frac{dx}{P_y(x,y)} \modsp(P) .
\eeq

\et
\proof{Classical theorem, see \cite{Farkas,fay1973theta,TataLectures,eynardlecturesRS}.}

\medskip
\bex[Weierstrass curve]
$P(x,y)=y^2-x^3+g_2 x+g_3$, for generic $g_2,g_3$. $\curve=\CC/(\ZZ+\tau\ZZ)$ is a torus, with the immersion map $\ii:\curve \to \CC\times \CC$ given by
\bea
X(z) &=& \nu^2 \wp(z,\tau) \cr
Y(z) &=& \frac12\nu^3 \wp'(z,\tau) \cr
\eea
We have
\beq
\Omega_{0,0}=
\frac{dx}{P_y(x,y)} = \frac{dx}{2y} = \frac{\nu^2 \wp'(z,\tau)dz}{\nu^3 \wp'(z,\tau)} = \frac{1}{\nu} dz.
\eeq
$dz$ is indeed a holomorphic form, it has no pole in the parallelogram $(0,1,1+\tau,\tau)$, and it is biperiodic $dz = d(z+1)=d(z+\tau)$.
We choose the Jordan loops $\acycle=[p,p+1]$ and $\bcycle=[p,p+\tau]$ with $p$ a generic point.
The matrix ${\mathcal K}$ is a $1\times 1$ matrix, worth
\beq
{\mathcal K}=\oint_{\acycle} \frac{dx}{P_y(x,y)} = \frac{1}{\nu} \int_p^{p+1} dz = \frac{1}{\nu}.
\eeq
Its inverse is
\beq
\hat{\mathcal K} = \nu.
\eeq
The normalized holomorphic differential is
\beq
\omega= \nu \frac{dx}{P_y(x,y)} = \nu \frac{\nu^2 \wp'(z,\tau)dz}{\nu^3 \wp'(z,\tau)}  = dz .
\eeq
Its $\bcycle$-cycle integral is
\beq
\oint_{\bcycle} \omega = \int_p^{p+\tau} dz = \tau.
\eeq
Riemann's theorem ensures that $\Im\tau>0$.
\eex

\bex[Legendre curve] we have
\beq
{\mathcal K}=\oint_{\acycle} \frac{dx}{P_y(x,y)} = \oint_{\acycle} \frac{dx}{2y}= \frac12 \oint_{\acycle} \frac{dx}{\sqrt{(1-x^2)(1-k^2 x^2)}} = 2K(k)
\eeq
if we choose the $\acycle$-cycle to be the cycle surrounding the cut $[-1,1]$.
\eex

In the degenerate case, the matrix $\hat{\mathcal K}$ can be computed more explicitely from combinatorics of the Newton's polygon, shifted by nodal points $N^{(\beta)}$:

\bp[Degenerate case]

We complete $\mathcal K$ by:
 for $(i',j')\in {\check{\Newtint}}^{(\beta)}$
\beq
\beta\in \mathcal R, \ (i',j')\in {\check{\Newtint}}^{(\beta)}, \quad
\td{\mathcal K}_{(i,j),(\beta,i',j')} = \lim_{r\to 0} \oint_{X^{-1}(x_\beta+r S^1)} (x-x_\beta)^{i'-1}(y-y_\beta)^{j'-1} \  \Omega_{i,j}.
\eeq
which makes  matrix $\td{\mathcal K}$ of size $\#\Newtint \times (\genus+\sum_\beta \ell_\beta \genus_\beta) $, of rank $\#\Newtint$.
Let us in fact denote by $\td{\mathcal K}$, a square invertible submatrix, and then
\beq
\hat{\td{\mathcal K}} = \td{\mathcal K}^{-1}.
\eeq
The matrix $\hat{\mathcal K}$ is then its sub-block of size $\genus\times \#\Newtint$.

\begin{center}
\begin{tikzpicture}
\begin{scope}[scale = 0.35,xshift=-8cm]
\draw[opacity=0.2]  (0,0) -- (5,0) -- (5,4) -- (0,4) -- cycle ;
\draw  (1.5,0) -- (1.5,4) ;
\fill [gray, opacity=0.2] (4,0) -- (4,4) -- (5,4) -- (5,0) -- cycle ;
\draw  (0,0) -- (4,0) -- (4,4) -- (0,4) -- cycle ;
\draw[black, left] (-0.3,2) node {$\td{\mathcal K}=$};
\draw[black] (.8,2) node {${\mathcal K}$};

\draw [decorate,decoration={brace,amplitude=4,mirror,raise=4ex}]
  (0,1.5) -- (1.4,1.5) node[midway,yshift=-3em,font=\tiny]{$\genus$};

\draw [decorate,decoration={brace,amplitude=4,mirror,raise=4ex}]
  (1.6,1.5) -- (5,1.5) node[midway,yshift=-3em,font=\tiny]{$\sum_\beta \ell_\beta\genus_\beta$};

\draw [decorate,decoration={brace,amplitude=4pt},xshift=.4cm,yshift=0pt]
      (5,4) -- (5,0) node [midway,right,xshift=.1cm,font=\tiny] {$\#\Newt$};
      
\draw[<->] (4,4.5) -- (5,4.5);
\node[text width=2cm, above, font=\tiny] at (4.5,4.6) {$not\ independent$};      
      
\end{scope}

\begin{scope}[scale = 0.35,xshift=10cm]

\draw  (0,2.5) -- (4,2.5) ;
\draw  (0,0) -- (4,0) -- (4,4) -- (0,4) -- cycle ;
\draw[black, left] (-0.3,2) node {$, \ \ {\td{\mathcal K}}^{-1}=\hat{\td{\mathcal K}}=$};
\draw[black] (2,3.2) node {$\hat{\mathcal K}$};


\draw [decorate,decoration={brace,amplitude=4,mirror,raise=4ex}]
  (0,1.5) -- (4,1.5) node[midway,yshift=-3em,font=\tiny]{$\#\Newt$};

\draw [decorate,decoration={brace,amplitude=4pt},xshift=.4cm,yshift=0pt]
      (4,4) -- (4,2.5) node [midway,right,xshift=.1cm,font=\tiny] {$\genus$};

\end{scope}

\end{tikzpicture}
\end{center}

\ep

\proof{
$\modsp(P)$ is a subspace of $\Field[\Newtint]$ of dimension $\genus$.
The matrix $\mathcal {\td K}$ codes a morphism $\Field[\Newtint] \to \CC^{\genus+\sum_\beta \ell_\beta\genus\beta} $, such that the image of $\modsp(P)$ should be $\CC^\genus\times 0^{\sum_\beta \ell_\beta\genus\beta}$.
In other words, 
$\hat{\td{\mathcal K}} = \td{\mathcal K}^{-1}$, is such that $\modsp(P) = \hat{\td{\mathcal K}}( \CC^\genus\times 0^{\sum_\beta \ell_\beta\genus\beta} )= \hat{{\mathcal K}} (\CC^\genus)$.
}

\bex[A degenerate case] 
Let $P(x,y) = y^2-(x-c)^2(x^2-a^2)(x^2-b^2)$.
The Newton's polygon has 2 interior points $(1,1),(2,1)$, and thus
\beq
\Newtint=\{(0,0),(1,0)\}.
\eeq
There is one regular nodal point $\beta=(c,0)$, it has $\ell_\beta=2$, $\deg\beta=2$ and
\beq
{\check{\Newtint}}^{(\beta)}=\{(0,0)\}.
\eeq

\begin{center}
\begin{tikzpicture}
\begin{scope}[scale = 0.45,xshift=-6cm]
\grid{6}{3};
\fill [red, opacity=0.2] (0,2) --  (6,0) -- (0,0)  ;
\draw [red, thick] (0,2) --  (6,0) -- (0,0) -- cycle ;
\draw[red] (0,2) node {$\bullet$};
\draw[red] (1,0) node {$\bullet$};
\draw[red] (2,0) node {$\bullet$};
\draw[red] (3,0) node {$\bullet$};
\draw[red] (4,0) node {$\bullet$};
\draw[red] (5,0) node {$\bullet$};
\draw[red] (6,0) node {$\bullet$};
\draw[red] (0,0) node {$\bullet$};
\draw[blue] (1,1) node {$\bullet$};
\draw[blue] (2,1) node {$\bullet$};

\draw[black, left] (-.2,1) node {$\Newtint$};

\end{scope}

\begin{scope}[scale = 0.45,xshift=6cm]
\grid{6}{3};
\fill [red, opacity=0.2] (0,2) --  (6,0) -- (2,0)  ;
\fill [blue, opacity=0.2] (0,2) --  (0,0) -- (2,0)  ;
\draw [red, thick] (0,2) --  (6,0) -- (2,0) -- cycle ;
\draw[red] (0,2) node {$\bullet$};
\draw[red] (2,0) node {$\bullet$};
\draw[red] (3,0) node {$\bullet$};
\draw[red] (4,0) node {$\bullet$};
\draw[red] (5,0) node {$\bullet$};
\draw[red] (6,0) node {$\bullet$};
\draw[blue] (1,1) node {$\bullet$};
\draw[blue] (2,1) node {$\bullet$};

\draw[black, left] (-.2,1) node {$\Newtint^{(\beta)}$};
\end{scope}

\end{tikzpicture}
\end{center}

The genus is $\genus=1$, and $\genus_\beta=1$, so that $\genus+\genus_\beta=2=\#\Newtint$.
The moduli space is
\beq
\modsp(P) = (x-c)\Field
\qquad , \quad
\dim \modsp(P) = \genus=1.
\eeq

We consider the matrix of size $\#\Newtint \times (\genus+2\genus_\beta)$
\beq
\td{\mathcal K}
=\begin{pmatrix}
  \oint_{\acycle} \frac{dx}{2y} & 2\pi\ii \Res_c \frac{dx}{2y}  & -2\pi\ii \Res_c \frac{dx}{2y} \cr
  \oint_{\acycle} \frac{xdx}{2y} & 2\pi\ii \Res_c \frac{xdx}{2y} &  -2\pi\ii \Res_c \frac{xdx}{2y} \cr
\end{pmatrix}
\eeq
of which we select an invertible sub-matrix:
\beq
\td{\mathcal K}
=\begin{pmatrix}
  \oint_{\acycle} \frac{dx}{2y} & 2\pi\ii \Res_c \frac{dx}{2y} \cr
  \oint_{\acycle} \frac{xdx}{2y} & 2\pi\ii \Res_c \frac{xdx}{2y} \cr
\end{pmatrix}
\eeq
Remark that
\beq
\Res_c \frac{dx}{2y} = \frac{1}{2\sqrt{(c^2-a^2)(c^2-b^2)}}
\quad , \quad
\Res_c \frac{xdx}{2y} = \frac{c}{2\sqrt{(c^2-a^2)(c^2-b^2)}}
\eeq
This gives
\beq
{\td{\mathcal K}}^{-1}
= \frac{-2\sqrt{(c^2-a^2)(c^2-b^2)}}{2\pi\ii \oint_{\acycle} \frac{(x-c)dx}{2y} } \begin{pmatrix}
  \frac{2\pi\ii \ c }{2\sqrt{(c^2-a^2)(c^2-b^2)}}  & - \frac{2\pi\ii  }{2\sqrt{(c^2-a^2)(c^2-b^2)}} \cr
  -\oint_{\acycle} \frac{xdx}{2y} & \oint_{\acycle} \frac{dx}{2y}\cr
\end{pmatrix}
\eeq
The matrix ${\hat{\mathcal K}}$ is the first row i.e.
\beq
{\hat{\mathcal K}}
= \frac{1}{\oint_{\acycle} \frac{(x-c)dx}{2y} } \begin{pmatrix}
  -c & 1 
\end{pmatrix}
\eeq
therefore
\beq
\omega_1 = \frac{1}{\oint_{\acycle} \frac{(x-c)dx}{2y} }  \ \frac{(x-c)dx}{2y}.
\eeq
(we indeed have $\omega_1 \in \frac{dx}{2y}\modsp(P)$.)

The second row gives a 1-form
\beq
\omega_2 = \frac{u x + v}{2y}dx 
\eeq
with
\beq
u=\frac{-2\sqrt{(c^2-a^2)(c^2-b^2)}}{2\pi\ii \oint_{\acycle} \frac{(x-c)dx}{2y} } \ \oint_{\acycle} \frac{dx}{2y}
\quad , \quad
v=\frac{2\sqrt{(c^2-a^2)(c^2-b^2)}}{2\pi\ii \oint_{\acycle} \frac{(x-c)dx}{2y} } \ \oint_{\acycle} \frac{xdx}{2y}
\eeq
It has the property that
\beq
\Res_c \omega_2 = \frac{1}{2\pi \ii}
\quad , \quad
\oint_{\acycle} \omega_2 = 0.
\eeq

\eex

\bd[Abel map]\index{Abel map}
\label{def:Abelmap}

Let $o\in\curve$ a generic point, called the "origin" and fixed once for all.
Define:
\beq
F(p)=\{F_i(p)\}_{i=1,\dots,\genus} 
\quad , \quad
F_i(x) = \int_o^p \omega_i
\eeq
where the integral is defined on the universal cover, or in a fundamental domain, such that the integration path is the unique homology chain that doesn't intersect any $\acycle_i$ or $\bcycle_i$ cycles.

\ed
The $\genus$-dimensional vector $F(p)$ generalizes the Legendre elliptic function $F(x,k)$.

\subsection{Fundamental 2nd kind form and Bergman kernel}
\index{Fundamental 2nd kind form}

Riemann introduced the "fundamental 2nd kind form": a symmetric bilinear meromorphic differential on $\curve\times \curve$ having a normalized double pole on the diagonal and no other singularity.

It is not unique because one can add to it any symmetric bilinear combination of 1-forms that have no poles.


\subsubsection{Combinatorial 2nd kind form}

It is easy to build a fundamental 2nd kind form from the combinatorics of the Newton's polygon, purely algebraic.

\bd[Combinatorial Fundamental 2nd kind form \cite{EBergman}]\label{thB}
Let the following $1\otimes 1$ form
\beq\label{defBarithm}
B^{\text{comb}}((x,y);(x',y')) = \left( - \  \frac{P(x,y')P(x',y)}{(x-x')^2(y-y')^2} + Q^{\text{comb}}(x,y;x',y') \right) \ \frac{dx}{P_y(x,y)} \frac{dx'}{P_y(x',y')} 
\eeq
with
\bea\label{defQBarithm}
Q^{\text{comb}}(x,y;x',y')
&=& \text{Sym}\Big( \sum_{(i,j)\in \mathcal N} \sum_{(i',j')\in \mathcal N} P_{i,j} P_{i',j'}
\sum_{(u,v)\in \ZZ^2\cap \text{ triangle }(i,j),(i',j'),(i,j')} \cr
&& |u-i| \ | v-j'|   \delta_{(u,v)\notin \text{interior}}  \  x^{u-1}y^{v-1} x'^{i+i'-u-1}y'^{j+j'-v-1}   \Big) \ . \cr
\eea
Here $\delta_C$ means 1 if the condition $C$ is fulfilled, and 0 if not, and $\text{Sym}$ means that we symmetrize with respect to $(x,y)\leftrightarrow(x',y')$.
\ed

\bex
Consider the following Newton's polygon, and 2 examples of triangles $(i,j),(i',j'),(i,j')$ contributing to $Q^{\text{comb}}$:
\begin{center}
\begin{tikzpicture}
\begin{scope}[scale = 0.45,xshift=-5cm]
\grid{5}{5};
\fill [red, opacity=0.2] (0,5) --  (2,4) -- (4,2) -- (5,0) -- (0,0)  ;
\draw [red, thick] (0,5) --  (2,4) -- (4,2) -- (5,0) -- (0,0) -- cycle ;
\draw[red] (0,0) node (O) {$\bullet$};
\draw[red] (0,5) node (A) {$\bullet$};
\draw[red] (2,4) node {$\bullet$};
\draw[red] (3,3) node {$\bullet$};
\draw[red] (4,2) node {$\bullet$};
\draw[red] (5,0) node (B) {$\bullet$};

\draw[red] (1,3) node {$\bullet$};
\draw[red] (2,1) node {$\bullet$};

\fill [yellow, opacity=0.6] (1,3) --  (5,0) -- (5,3)   ;

\draw[blue] (3,2) node {$\circ$};
\draw[blue] (4,2) node {$\times$};
\draw[blue] (4,1) node {$\circ$};

\end{scope}

\begin{scope}[scale = 0.45, xshift=5cm]
\grid{5}{5};
\fill [red, opacity=0.2] (0,5) --  (2,4) -- (4,2) -- (5,0) -- (0,0)  ;
\draw [red, thick] (0,5) --  (2,4) -- (4,2) -- (5,0) -- (0,0) -- cycle ;
\draw[red] (0,0) node (O) {$\bullet$};
\draw[red] (0,5) node (A) {$\bullet$};
\draw[red] (2,4) node {$\bullet$};
\draw[red] (3,3) node {$\bullet$};
\draw[red] (4,2) node {$\bullet$};
\draw[red] (5,0) node (B) {$\bullet$};

\draw[red] (1,3) node {$\bullet$};
\draw[red] (2,1) node {$\bullet$};

\fill [yellow, opacity=0.6] (2,4) --  (5,0) -- (5,4)   ;

\draw[blue] (4,2) node {$\times$};
\draw[blue] (4,3) node {$\times$};
\draw[blue] (3,3) node {$\times$};

\end{scope}
\end{tikzpicture}

\end{center}
In the first triangle, the only term contributing to $Q^{\text{comb}}$ is
$P_{1,3}P_{5,0} \Big(  x_1^3y_1 x_2 +x_2^3y_2x_1 \Big)$.
In the second triangle, 
the 3 terms contributing to $Q^{\text{comb}}$ are
$P_{2,4}P_{5,0} \Big( (x_1^2x_2^3y_2^2+x_2^2x_1^3y_1^2) + 2(x_1^2y_1x_2^3y_2+x_2^2y_2x_1^3y_1)+2(x_1^3x_2^2y_2^2+x_2^3x_1^2y_1^2) \Big)$.

\eex

\bt
$B^{\text{comb}}$ has a normalized double pole at $(x,y)=(x',y')$, and no pole at the punctures.

At short distance $p\to p'$ it has the behavior
\bea
B^{\text{comb}}(p,p')
& \sim &
\Big( \frac{1}{(x-x')^{2}} + \frac{P_{x}^{2} P_{yy}^{2}}{4 P_{y}^{4}} - \frac{P_{x}^{2} P_{yyy}}{6 P_{y}^{3}} - \frac{P_{xy} P_{x} P_{yy}}{2 P_{y}^{3}} + \frac{P_{xyy} P_{x}}{2 P_{y}^{2}} \cr
&& + \frac{Q^{\text{comb}}(x,y,x,y)}{P_y(x,y)^2} + o(1) \Big) dx\otimes dx' .
\eea

If $P$ is generic, then $B^{\text{comb}}$ is a fundamental second kind form.
If $P$ is not generic, it may have singularities at the zeros of $P_y$, and it is possible (see \cite{EBergman}) to add to it an element of $\CC[\Newtint]\overset{\text{sym}}{\otimes} \CC[\Newtint] \frac{dx}{P_y(x,y)} \frac{dx'}{P_y(x',y')}$ that makes it a fundamental second kind form.
\et
\proof{In \cite{EBergman}.}

\bex[Legendre curve]
$P(x,y) = y^2-(1-x^2)(1-k^2 x^2)$.
\begin{center}
\begin{tikzpicture}
\begin{scope}[scale = 0.55]
\grid{4}{2};
\fill [red, opacity=0.2] (0,0) -- (0,2) -- (2,1) -- (4,0)  ;

\draw [yellow, thick] (0,2) -- (4,2) --  (4,0)  ;
\fill [yellow, opacity=0.2] (4,2) -- (0,2) --  (4,0)  ;

\draw [red, thick] (0,0)  -- (0,2)  -- (4,0)  -- cycle ;
\draw [blue] (3,1) node {$\times$};
\draw [blue] (2,1) node {$\times$};
\def\N{(0,0), (2,0),(4,0),(0,2)};
\drawpolytope[red]\N;
\end{scope}
\end{tikzpicture}
\end{center}
The only triangle with corners $(i,j),(i',j')$ whose interior contains non-interior points of $\Newt$, is the triangle with corners $(0,2),(4,0)$. It contains 2 interior points: $(3,1)$ and $(2,1)$, therefore we have 
\bea
Q^{\text{comb}}(x_1,y_1,x_2,y_2)
&=&  P_{0,2}P_{4,0} ((x_1^2+x_2^2) + 2 x_1 x_2 ) \cr
&=&  -k^2 (x_1^2 + 2 x_1 x_2 + x_2^2) \cr
&=&  -k^2 (x_1+x_2)^2.
\eea
This gives
\bea
&& B^{\text{comb}}(x_1,y_1,x_2,y_2) \cr
&=& \frac{dx_1}{2y_1}\frac{dx_2}{2y_2} \left(-\frac{(y_1^2-y_2^2)(y_2^2-y_1^2)}{(x_1-x_2)^2(y_1-y_2)^2} + Q^{\text{comb}}(x_1,y_1,x_2,y_2) \right) \cr
&=& \frac{dx_1}{2y_1}\frac{dx_2}{2y_2} \left(\frac{(y_1+y_2)^2}{(x_1-x_2)^2} + Q^{\text{comb}}(x_1,y_1,x_2,y_2) \right) \cr
&=& \frac{dx_1 dx_2}{4 y_1 y_2 (x_1-x_2)^2} \left(2y_1y_2+y_1^2+y_2^2 + (x_1-x_2)^2( - k^2(x_1+x_2)^2)  \right) \cr
&=& \frac{dx_1 dx_2}{4 y_1 y_2 (x_1-x_2)^2} \left(2y_1y_2+y_1^2+y_2^2-k^2(x_1^2-x_2^2)^2  \right) \cr
&=& \frac{dx_1 dx_2}{4 y_1 y_2 (x_1-x_2)^2} \left(2y_1y_2+2-(1+k^2)(x_1^2+x_2^2)+2k^2 x_1^2 x_2^2   \right) \cr
\eea

- The numerator is worth $4y_1y_2$ at $(x_1,y_1)=(x_2,y_2)$, and gives the normalized  double pole.

- The numerator is worth $0$ at $(x_1,y_1)=(x_2,-y_2)$, and thus there is no double pole there (and also no simple pole by symmetry).

- The numerator behaves as $O(x_1^2)$ at large $x_1$, and thus $B=O(x_1^{-2}dx_1)$ so that there is no pole at the punctures.

Therefore $B^{\text{comb}}$ is a fundamental second kind form.

At short distance $p'\to p$ it behaves like
\bea
B^{\text{comb}}(p,p')
& \sim & \Big( \frac{1}{(x-x')^{2}} + x^2\frac{((1+k^2)-2k^2 x^2)^{2} }{4 y^4}   - k^2 \frac{x^2}{y^2} + o(1) \Big) dx\otimes dx' \cr
& \sim & \Big( \frac{1}{(x-x')^{2}} + x^2\frac{((1+k^2)-2k^2 x^2)^{2} - 4 k^2 y^2 }{4 y^4}   + o(1) \Big) dx\otimes dx' \cr
& \sim & \Big( \frac{1}{(x-x')^{2}} + \frac{(1-k^2)^2 x^2}{4 y^4}   + o(1) \Big) dx\otimes dx' \cr
\eea

\eex

\subsubsection{Bergman kernel}
\index{Bergman kernel}

\bd

Having made a choice of a Torelli marking, the Bergman kernel is the unique fundamental 2nd kind form, whose $\acycle$-cycle integrals vanish. 
It can be written
\bea
B(p_1,p_2) 
&=& B^{\text{comb}}(x(p_1),y(p_1),x(p_2),y(p_2)) \cr
&& +  \sum_{(i,j),(i',j')\in \Newtint\times \Newtint}
S_{(i,j),(i',j')} \Omega_{i,j}(p_1)\otimes \Omega_{i',j'}(p_2)
\eea
or written symbolically
\beq
B=B^{\text{comb}}+S .\Omega\otimes \Omega
\eeq
where $S\in \CC[\Newtint]\overset{\text{sym}}{\otimes}\CC[\Newtint]$ is  uniquely determined by the fact that the $\acycle$-cycle integrals of $B$ vanish.

In general the coefficients of $S$ are not algebraic functions of the $P_{i,j}$s, they are  transcendantal.

\ed

Here we admit that indeed $S$ exists and is unique.
It is unique because the difference of 2 such forms would be bi-holomorphic, it would be decomposable on the basis $\omega_1,\dots,\omega_\genus$, and it would vanish since all its $\acycle$-cycle integrals vanish.

$S$ is not an algebraic function of the coefficients $P_{i,j}$, it is genuinely transcendantal, and is also not an algebraic function of $\mathcal K$, it is an algebraically independent building block.

It plays a role similar to the Legendre elliptic function $E(k)$.

It can be related to other quantities, for example to derivatives of $\mathcal K$, or to $\Theta$ functions, or to Eisenstein series.
Although this is not necessary for computing, we review the link between $S$ and Theta-functions in section \ref{sec:theta} below.

\subsubsection{Rauch variational formula}

{\em Rauch variational formula \cite{Rauch} is the formula for deformations of the Bergman kernel.
In particular it implies that $S$ is related to derivatives of $\mathcal K$. Let us see how.}

Let us consider a deformation of $P$:
\beq
\delta P(x,y) = 
\sum_{(i,j)\in \Newt} x^i y^j \ \delta P_{i,j}.
\eeq
The branchpoints and nodal points are such that $P(x_a,y_a)=0$ and $P_y(x_a,y_a)=0$.
This implies $0 = \delta P(x_a,y_a) + \delta x_a P_x(x_a,y_a) $. 
For a generic branchpoint $P_x(x_a,y_a)\neq 0$, so we have
\beq\label{eq:deltaxadeltaP}
\delta x_a = - \frac{\delta P(x_a,y_a)}{ P_x(x_a,y_a)}.
\eeq
For non-generic branch or nodal points, we need to consider higher derivatives.

The Rauch variational formula can be written as follows:

\bt[Rauch variational formula \cite{Rauch}]
\label{thm:RauchB}
Under a variation $\delta P(x,y)$ the variation of $B(x_1,y_1;x_2,y_2)$ at fixed $x_1$ and $x_2$ is given by
\bea
\delta B(p_1;p_2)
&=&  \sum_{\beta\in \mathcal R} \sum_{a\in \ii^{-1}(\beta)} 
\Res_{a} \frac{B(x,y;p_1)B(x,y;p_2) \delta P(x,y)}{P_y(x,y)dy}  \cr
&=& - \sum_{\beta\in \mathcal R} \sum_{a\in \ii^{-1}(\beta)} 
\Res_{a} \frac{B(x,y;p_1)B(x,y;p_2) \delta P(x,y)}{P_x(x,y)dx}  \cr
\eea
\et

\proof{Rauch \cite{Rauch} proved that
\beq
\delta B(p_1,p_2)
= \sum_a \delta x_a \ \Res_a \frac{B(x,y;p_1)B(x,y;p_2)}{dx} .
\eeq
For generic branch points, using \eqref{eq:deltaxadeltaP}, this can be rewritten
\beq
\delta B(p_1,p_2)
= - \sum_a  \Res_a \frac{B(x,y;p_1)B(x,y;p_2) \delta P(x,y)}{P_x(x,y) \ dx} .
\eeq
We also notice that on $\td\curve$ we have $P_x dx = -P_y dy$, which gives the other expression.

We admit here that the formula remains valid if the branch points are non-generic (this is easy but tedious to prove).
}

\bt
\label{thm:Rauchomega}
\beq
\delta \omega_{k}(x_1,y_1) = -\sum_{(i,j)\in \Newtint }
\hat{\mathcal K}_{k;(i,j)}
\sum_{a} \Res_{a} \delta P(x,y) \  \ \frac{x^{i}y^{j}}{P_x(x,y)P_y(x,y)} B(x,y;x_1,y_1)
\eeq
\et
\proof{This is a corollary of theorem \ref{thm:RauchB}.
Indeed we have that $\oint_{\bcycle_k} B = 2\pi\ii \omega_k$. Taking the $\bcycle_k$ integral on both sides gives the result.}

\bt
\beq
 \delta \tau_{k,l} = -2\pi\ii \sum_{(i,j),(i',j')\in \Newtint^2 }
\hat{\mathcal K}_{k;(i,j)}\hat{\mathcal K}_{l;(i',j')}
\sum_{a} \Res_{a} \frac{\delta P(x,y) }{P_x(x,y)} \ \frac{x^{i+i'}y^{j+j'}}{P_y(x,y)^2} \ dx
\eeq
\et
\proof{This is a corollary of theorem \ref{thm:Rauchomega}.
Indeed we have that $\oint_{\bcycle_k} B = 2\pi\ii \omega_k$. Taking the $\bcycle_k$ integral on both sides gives the result.}

\bt[$S$ as derivative of $\mathcal K$]
\label{thm:RauchKtoS}
The variation $\delta\mathcal K$ of $\mathcal K$ satisfies
\bea
\sum_{(i',j')\in\Newtint} (\delta {\mathcal K} .\;  \hat{\mathcal K})_{(i,j);(i',j')} x^{i'} y^{j'}
&=& -\frac{x^i y^j\delta P_y(x,y)}{P_y(x,y)} \cr
&& - \left( \frac{j x^i y^{j-1}}{P_y(x,y)} -  \frac{ x^i y^{j}P_{yy}(x,y)}{P_y(x,y)^2}\right) \delta P(x,y) \cr
&& + 2 \sum_{a} \Res_{a} \frac{\delta P(x',y')\hat B(x',y';x,y)x'^iy'^j}{P_x(x',y')P_y(x',y')^2}dx' \cr
\eea
where
$\hat B(x',y';x,y) = B(x',y';x,y) \frac{P_y(x,y)}{dx} \frac{P_y(x',y')}{dx'} $. 
The right hand side is a polynomial $\in \modsp(P)$, its coefficients give the left hand side.

This implies that $\delta \mathcal K . \hat{\mathcal K}$ is an affine function of $S$ with coefficients algebraic in the coefficients of $P$, i.e. in some extension of $\Field$.

Vice-versa $S$ is related to   $\delta\mathcal K$ times $\hat{\mathcal K}$, with algebraic coefficients:
\bea
&& 2\sum_{(,',j'),(i'',j'')} x^{i'}y^{j'} S_{(i',j'),(i'',j'')} \sum_{a} \Res_{a} \frac{\delta P(x',y')x'^{i+i''}y'^{j+j''}}{P_x(x',y')P_y(x',y')^2} dx'  \cr
&=&  \frac{x^i y^j\delta P_y(x,y)}{P_y(x,y)} + \left( \frac{j x^i y^{j-1}}{P_y(x,y)} -  \frac{ x^i y^{j}P_{yy}(x,y)}{P_y(x,y)^2}\right) \delta P(x,y) \cr
&& -2  \sum_{a} \Res_{a} \frac{\delta P(x',y')\hat B^{\text{comb}}(x',y';x,y)x'^iy'^j}{P_x(x',y')P_y(x',y')^2} dx'\cr
&& + \sum_{(i',j')\in\Newtint} (\delta {\mathcal K} \ \hat{\mathcal K})_{(i,j);(i',j')} x^{i'} y^{j'}.
\eea

\et

\proof{
We have
\bea
\delta (\mathcal K.\omega)
&=& \delta \Omega \cr
&=& \delta\left(\frac{x^i y^j}{P_y(x,y)}dx\right) \cr
&=& -\frac{x^i y^j \delta P_y(x,y)}{P_y(x,y)^2}dx - \frac{\delta P(x,y)}{P_y(x,y)^2} \left( j x^i y^{j-1} - x^i y^j \frac{P_{yy}(x,y)}{P_y(x,y)} \right) \cr
\eea
On the other hand we have
\bea
\delta (\mathcal K.\omega)
&=& (\delta \mathcal K)\omega + \mathcal K \delta \omega \cr
&=& (\delta \mathcal K)\hat{\mathcal K}\Omega + \mathcal K \delta \omega .
\eea
Together this implies for every $(i,j)\in\Newtint$:
\bea
&& \sum_{(i',j')\in \Newtint}\sum_{k=1}^\genus
 (\delta \mathcal K_{(i,j),k})\hat{\mathcal K}_{k,(i',j')} \frac{x^{i'}y^{j'}}{P_y(x,y)} dx \cr
&=& \delta (\mathcal K.\omega) - \mathcal K  \delta \omega \cr
&=& -\frac{x^i y^j \delta P_y(x,y)}{P_y(x,y)^2}dx - \frac{\delta P(x,y)}{P_y(x,y)^2} \left( j x^i y^{j-1} - x^i y^j \frac{P_{yy}(x,y)}{P_y(x,y)} \right)dx \cr
&& +  \sum_{a} \Res_{a} \delta P(x',y') \  \ \frac{x'^{i}y'^{j}}{P_x(x',y')P_y(x',y')} B(x,y;x',y') \cr
\eea
Multiplying by $P_y/dx$ we get
\bea
&& \sum_{(i',j')\in \Newtint}\sum_{k=1}^\genus
 (\delta \mathcal K_{(i,j),k})\hat{\mathcal K}_{k,(i',j')} x^{i'}y^{j'} \cr
&=& -\frac{x^i y^j \delta P_y(x,y)}{P_y(x,y)} - \frac{\delta P(x,y)}{P_y(x,y)} \left( j x^i y^{j-1} - x^i y^j \frac{P_{yy}(x,y)}{P_y(x,y)} \right) \cr
&& +  \sum_{a} \Res_{a} \delta P(x',y') \  \ \frac{x'^{i}y'^{j}}{P_x(x',y')P_y(x',y')} B(x,y;x',y') \frac{P_y(x,y)}{dx} \cr
&=& -\frac{x^i y^j \delta P_y(x,y)}{P_y(x,y)} - \frac{\delta P(x,y)}{P_y(x,y)} \left( j x^i y^{j-1} - x^i y^j \frac{P_{yy}(x,y)}{P_y(x,y)} \right) \cr
&& +  \sum_{a} \Res_{a} \delta P(x',y') \  \ \frac{x'^{i}y'^{j}}{P_x(x',y')P_y(x',y')} B^{\text{comb}}(x,y;x',y') \frac{P_y(x,y)}{dx} \cr
&& +  \sum_{a} \Res_{a} \delta P(x',y') \  \ \frac{x'^{i}y'^{j}}{P_x(x',y')P_y(x',y')} S(x,y;x',y') \frac{dx'}{P_y(x',y')} \cr
\eea
The right hand side is a rational fraction of $x,y$, and it seems that it can have poles at the zeros of $P_y$.
However, the residues at the branchpoints, because of the $1/(x-x')^2$ in $B$ also produce poles at the zeros of $P_y$, and in the end the right hand side can't have such poles, it is in fact a polynomial, and moreover, all its monomials are in $\Newtint$, and can be matched to the left hand side.

We have
\bea
&& \sum_{(i',j')\in \Newtint}\sum_{k=1}^\genus
 (\delta \mathcal K_{(i,j),k})\hat{\mathcal K}_{k,(i',j')} x^{i'}y^{j'} \cr
&=& -\frac{x^i y^j \delta P_y(x,y)}{P_y(x,y)} - \frac{\delta P(x,y)}{P_y(x,y)} \left( j x^i y^{j-1} - x^i y^j \frac{P_{yy}(x,y)}{P_y(x,y)} \right) \cr
&& -  \sum_{a} \Res_{a} \delta P(x',y') \  \ \frac{x'^{i}y'^{j}}{P_x(x',y')P_y(x',y')} \frac{P(x,y')(P(x',y)}{(x-x')^2(y-y')^2} \frac{dx'}{P_y(x',y')} \cr
&& +  \sum_{a} \Res_{a} \delta P(x',y') \  \ \frac{x'^{i}y'^{j}}{P_x(x',y')P_y(x',y')} \big(Q^{\text{comb}}(x,y;x',y')\cr
&& \qquad +S(x,y;x',y') \big) \frac{dx'}{P_y(x',y')} \cr
\eea

In the end, we get that $\delta \mathcal K . \hat{\mathcal K}$ is an affine function of $S$ with coefficients in some extension of $\Field$.
}

\bex[Legendre curve]
$P(x,y) = y^2-(1-x^2)(1-k^2x^2)$.
The only parameter is $k$, and we have
\beq
\delta P(x,y) = x^2(1-x^2) 2k \delta k.
\eeq
Since $\hat{\mathcal K}=1/\mathcal K$, we have $\delta \mathcal K.\hat{\mathcal K}=\delta\ln{\mathcal K}$, and 
\bea
\delta \ln\mathcal K
&=&   \frac{\delta P(x,y)}{2y^2}   +  \sum_{a=\pm 1,\pm 1/k} \Res_{a}  \frac{\delta P(x',y')}{ P_x(x',y')} \frac{(y+y')^2}{(x-x')^2} \frac{dx'}{4y'^2} \cr
&& +  \sum_{a=\pm 1,\pm 1/k} \Res_{a}  \frac{\delta P(x',y')}{ P_x(x',y')} (-k^2(x+x')^2+S) \frac{dx'}{4y'^2} .
\eea
This gives
\bea
\frac{\delta \ln\mathcal K}{2k \delta k}
&=&   \frac{x^2(1-x^2)}{2y^2}  \cr
&&  + \sum_{a=\pm 1,\pm 1/k} \Res_{a}  \frac{x'^2(1-x'^2)}{ P_x(x',y')} \frac{(y+y')^2}{(x-x')^2} \frac{dx'}{4y'^2} \cr
&& +  \sum_{a=\pm 1,\pm 1/k} \Res_{a}  \frac{x'^2(1-x'^2)}{ P_x(x',y')} (S-k^2(x+x')^2) \frac{dx'}{4y'^2} \cr
&=&   \frac{x^2}{2(1-k^2 x^2)}  \cr
&& +  \sum_{a=\pm 1,\pm 1/k} \Res_{a}  \frac{x'^2}{ P_x(x',y')} \frac{(y+y')^2}{(x-x')^2} \frac{dx'}{4(1-k^2 x'^2)} \cr
&& +  \sum_{a=\pm 1,\pm 1/k} \Res_{a}  \frac{x'^2}{ P_x(x',y')} (S-k^2(x+x')^2) \frac{dx'}{4(1-k^2 x'^2)} \cr
&=&   \frac{x^2}{2(1-k^2 x^2)}  \cr
&& - \frac{1}{4k^2}  \sum_{a=\pm 1/k}   \frac{a}{ P_x(a,0)} \frac{y^2}{(x-a)^2}  \cr
&& - \frac{1}{4k^2}  \sum_{a=\pm 1/k}  \frac{a}{ P_x(a,0)} (S-k^2(x+a)^2)  \cr
&=&   \frac{x^2}{2(1-k^2 x^2)}  \cr
&& - \frac{1}{4k^2}  \sum_{a=\pm 1/k}   \frac{a}{ 2a(1+k^2-2k^2 a^2)} \frac{y^2}{(x-a)^2}  \cr
&& - \frac{1}{4k^2}  \sum_{a=\pm 1/k}  \frac{a}{ 2a(1+k^2-2k^2 a^2)} (S-k^2(x+a)^2)  \cr
&=&   \frac{x^2}{2(1-k^2 x^2)}  \cr
&& + \frac{1}{8k^2}    \frac{y^2}{ 1-k^2} \left(\frac{1}{(x-1/k)^2} +\frac{1}{(x+1/k)^2}\right) \cr
&& + \frac{1}{4k^2}    \frac{1}{1-k^2} (S-k^2(x^2+k^{-2}))  \cr
&=&   \frac{x^2}{2(1-k^2 x^2)}  \cr
&& + \frac{1}{4}    \frac{y^2}{ 1-k^2}  \frac{1+k^2x^2}{(1-k^2x^2)^2}  \cr
&& + \frac{1}{4k^2}    \frac{1}{1-k^2} (S-1-k^2x^2)  \cr
&=&   \frac{x^2}{2(1-k^2 x^2)}   + \frac{1}{4}    \frac{1-x^2}{ 1-k^2}  \frac{1+k^2x^2}{1-k^2x^2}  
+ \frac{1}{4}    \frac{1-x^2}{1-k^2}   + \frac{S-1-k^2}{4k^2(1-k^2)}   \cr
&=&   \frac{x^2}{2(1-k^2 x^2)}   
+ \frac{1}{2}    \frac{1-x^2}{ 1-k^2}  \frac{1}{1-k^2x^2}  
+ \frac{S-1-k^2}{4k^2(1-k^2)}   \cr
&=&   \frac{1-k^2x^2}{2(1-k^2)(1-k^2 x^2)}   
+ \frac{S-1-k^2}{4k^2(1-k^2)}   \cr
&=&   \frac{1}{2(1-k^2)}   
+ \frac{S-1-k^2}{4k^2(1-k^2)}   \cr
\eea
This gives
\beq
\frac{\delta \ln\mathcal K}{2k \delta k}
=   \frac{S-1+k^2}{4k^2(1-k^2)} =\frac{S}{4k^2(1-k^2)}-\frac{1}{4 k^2} =\frac{2E}{4k^2(1-k^2)K}-\frac{1}{2 k^2}   
\eeq
or equivalently, we recover the well konwn formula for Legendre functions:
\beq
\frac{dK}{dk} = \frac{E}{k(1-k^2)} -\frac{K}{k}.
\eeq

\eex







\subsection{Third kind differentials}

{\it{3rd kind differentials  generalize the $\Pi(u,k)$ from the Legendre elliptic case, they also generalize the Weierstrass zeta-function.}}

3rd kind differentials are differential forms with at most simple poles.
Since the sum of residues must vanish, a 3rd kind differential can never have only 1-pole, it has either no pole, or at least 2 poles.

\bd[Combinatorial 3rd kind form]
Let $p_1=(x_1,y_1)$ and $p_2=(x_2,y_2)$ on $\td\curve$.
Let 
\beq
dS^{\text{comb}}_{p_1,p_2}(x,y)
= \frac{dx}{2P_y(x,y)}  \ \Big( \frac{P(x_1,y)-P(x,y_1)}{(x-x_1)(y-y_1)}  -  \frac{P(x_2,y)-P(x,y_2)}{(x-x_2)(y-y_2)} \Big)
\eeq
\ed

\bp

$dS^{\text{comb}}_{p_1,p_2}(x,y)$ has a simple pole at $(x,y)\to p_1$, with residue $+1$ and a simple pole at $(x,y)\to p_2$, with residue $-1$, and no pole at any puncture.

Near $(x,y)\to (x_1,y_1)$ it behaves like
\bea
dS^{\text{comb}}_{p_1,p_2}(x,y)
& \sim & \frac{dx}{x-x_1} - \frac{dx}{2P_y(x_1,y_1)} \Big( P_{xy}(x_1,y_1)-\frac{P_x(x_1,y_1) P_{yy}(x_1,y_1)}{P_y(x_1,y_1)}  \cr 
&& +  \frac{P(x_2,y_1)-P(x_1,y_2)}{(x_1-x_2)(y_1-y_2)} +o(1) \Big) 
\eea
\ep
\proof{Easy computation.}




\bd[Normalized 3rd kind form]
\beq
dS_{p_1,p_2}(p)
= dS^{\text{comb}}_{p_1,p_2}(x) +
    \frac{dx}{P_y(x,y)} \sum_{(i,j)\in \Newtint} \Big( \zeta_{(i,j)}(p_1)-\zeta_{(i,j)}(p_2) \Big) x^{i}y^{j}  .
\eeq
The coefficients $\zeta_{(i,j)}(p)$ can be chosen so that $dS_{p_1,p_2}$ has a simple pole at $p_1$ with residue $+1$, a simple pole at $p_2$ with residue $-1$ and no other pole, and has vanishing $\acycle$-cycle integrals
($\zeta(p)$ is of course not unique since one can add a constant independent of $p$.)

\ed

({\it{$\zeta(p)$ is in fact a generalization of the Weierstrass zeta function.}})

\bp
\label{prop:dzeta}
The derivatives of $\zeta_{i,j}$ must be such that
\beq
d\zeta_{i,j}(p_1)
= \left( [x^iy^j]S(x,y;x_1,y_1) +  C_{i,j}(x_1,y_1) \right) \frac{dx_1}{P_y(x_1,y_1)} 
\eeq
where $C_{i,j}(x_1,y_1)\in \Field[x_1,y_1]$ is a bivariate polynomial of $x_1,y_1$ with coefficients in $\Field$, given by
\bea
&& \sum_{(i,j)\in \Newtint} C_{i,j}(x_1,y_1) x^{i}y^{j} \cr
&=& Q^{\text{comb}}(x,y;x_1,y_1) \cr
&& - \frac{P(x,y_1)-(x-x_1)P_x(x_1,y_1)}{(x-x_1)^2} \  \frac{P(x_1,y)-(y-y_1)P_y(x_1,y_1)}{(y-y_1)^2} \cr
&& - \frac{P(x_1,y)+P(x,y_1) + (x-x_1)(P_x(x_1,y)-P_x(x_1,y_1))}{2(x-x_1)^2(y-y_1)} \ P_y(x_1,y_1) \cr
&& - \frac{P(x_1,y)+P(x,y_1)+(y-y_1)(P_y(x,y_1)-P_y(x_1,y_1))}{2(x-x_1)(y-y_1)^2}\ P_x(x_1,y_1)  \cr
&& \in \Field[x,y,x_1,y_1].
\eea
The RHS is  a polynomial in all 4 variables $x,y,x_1,y_1$, with coefficients algebraic  in the coefficients of $P$.

\ep

\proof{Use that $d_{p_1} dS_{p_1,p_2}(p)=B(p,p_1) $.
}

\bex[Legendre curve]

\bea
dS^{\text{comb}}_{p_1,p_2}(x,y)
&=& \frac{dx}{2P_y(x,y)}  \ \Big( \frac{P(x_1,y)-P(x,y_1)}{(x-x_1)(y-y_1)}  -  \frac{P(x_2,y)-P(x,y_2)}{(x-x_2)(y-y_2)} \Big) \cr
&=& \frac{dx}{4y}  \ \Big( \frac{2(y^2-y_1^2)}{(x-x_1)(y-y_1)}  -  \frac{2(y^2-y_2^2)}{(x-x_2)(y-y_2)} \Big) \cr
&=& \frac{dx}{2y}  \ \Big( \frac{y+y_1}{x-x_1}  -\frac{y+y_2}{x-x_2} \Big) \cr
\eea
One easily verifies that it has simple pole at $p=p_1$ with residue $+1$, simple pole at $p=p_2$ with residue $-1$, and at the punctures  $x\to\infty$ it behaves as $O(1/x^2)dx$, therefore has no pole at the punctures, and no other pole.
Therefore:
\beq
dS_{p_1,p_2}(x,y)
= \frac{dx}{2y}  \ \Big( \frac{y+y_1}{x-x_1}  -\frac{y+y_2}{x-x_2} +\zeta(p_1)-\zeta(p_2)\Big) ,
\eeq
where $d\zeta(p_1) = (S+C(p_1)) dx_1/2y_1$ with
\bea
&& C(x_1,y_1)  \cr
&=&  - k^2 (x+x_1)^2 \cr
&& - \frac{P(x,y_1)-(x-x_1)P_x(x_1,y_1)}{(x-x_1)^2} \  \frac{P(x_1,y)-(y-y_1)P_y(x_1,y_1)}{(y-y_1)^2} \cr
&=&  - k^2 (x+x_1)^2 \cr
&& - \frac{1}{(x-x_1)^2(y-y_1)^2} \left(P(x,y_1)-(x-x_1)P_x(x_1,y_1)\right)\left(y^2-y_1^2-2(y-y_1)y_1\right) \cr
&=&  - k^2 (x+x_1)^2 \cr
&& - \frac{1}{(x-x_1)^2} \left(P(x,y_1)-(x-x_1)P_x(x_1,y_1)\right) \cr
&=&  - k^2 (x+x_1)^2 \cr
&& - \frac{1}{(x-x_1)^2} \left((1+k^2)(x^2-x_1^2)-k^2(x^4-x_1^4)-2x_1(x-x_1)(1+k^2-2k^2 x_1^2)\right) \cr
&=&  - k^2 (x+x_1)^2 \cr
&& - \frac{1}{(x-x_1)} \left((1+k^2)(x+x_1)-k^2(x^3+x^2x_1+xx_1^2+x_1^3)-2x_1(1+k^2-2k^2 x_1^2)\right) \cr
&=&  - k^2 (x+x_1)^2 \cr
&& - \frac{1}{(x-x_1)} \left((1+k^2)(x-x_1)-k^2(x^3+x^2x_1+xx_1^2-3x_1^3)\right) \cr
&=&  - k^2 (x+x_1)^2 - (1+k^2) +k^2(x^2+2 x x_1+3x_1^2)\cr
&=&   - (1+k^2) +2k^2 x_1^2
\eea

This implies
\beq
d\zeta(p_1) = \frac{S  - (1+k^2) +2k^2 x_1^2}{2y_1} dx_1.
\eeq
Using $dz = dx/(2\mathcal K y) $ this gives
\beq\label{eq:Legendredzeta}
\mathcal K\frac{d}{dz}\zeta(p_1) = \mathcal K^2 (S  - (1+k^2) +2k^2 x_1^2)
\eeq
At $\infty_\pm$ we have $y\sim \pm k x^2$, therefore $dz/dx \sim \frac{\pm 1}{2 k \mathcal K x^2}$, and thus $z-z_{\infty_\pm} \sim - \frac{\pm 1}{2 k \mathcal K x}$, or $x\sim \mp \frac{1}{2k\mathcal K (z-z_{\infty_\pm})}$.

The right hand side has a double pole at the punctures, and behaves as $x_1\to\infty_\pm$ as 
\beq
\mathcal K\frac{d}{dz}\zeta(p_1) = \frac1{2(z_1-z_{\infty_\pm})^2} \left( 1+ \frac83 k^2(1+k^2)\mathcal K^2 (z-z_{\infty_\pm})^2+O(x_1^{-4}) \right)
\eeq
We recognize the behavior of the Weierstrass function 
\beq
\mathcal K\frac{d}{dz}\zeta(p_1) =
\frac12 \wp(z_1-z_{\infty_\pm}) +  
\frac43 k^2(1+k^2)\mathcal K^2 + O(x_1^{-2})
\eeq
and thus that
\bea\label{eq:Legendrezetatoriemann}
\zeta(p_1) 
&=& -\frac{1}{2\mathcal K}  \zeta_{\text{Weierstrass}}(z_1-z_{\infty_+}) -\frac{1}{2\mathcal K} \zeta_{\text{Weierstrass}}(z_1-z_{\infty_-}) \cr
&& + z_1\left(\frac43 k^2(1+k^2)\mathcal K^2 
- \frac{1}{2\mathcal K}  \wp(z_{\infty_+}-z_{\infty_-}) \right)
.
\eea

In other words, our $\zeta$ is closely related to $\zeta_{\text{Weierstrass}}$.

\eex







\section{Theta}
\label{sec:theta}

{\em This section is only to understand how $S$ is related to other  geometric notions, like Theta-functions or Eisenstein series.
The reader who only wants to use the formulas, can consider $S$ as an independent building block (as $E(k)$ in the Legendre case) and can skip this entire section \ref{sec:theta}.}

\subsection{Theta functions}

\bd[Riemann Theta function]
\index{Theta function}
Since $\tau$ is a Siegel matrix ($\tau=\tau^T$ and $\Im\tau>0$) the following sum is absolutely convergent for all $u\in \CC^\genus$, and it defines an analytic entire function of $u$ on $\CC^\genus$:
\beq
\Theta(u,\tau) = \sum_{n\in \ZZ^\genus} e^{2\pi i (u,n)} \ e^{i\pi (n,\tau n)} .
\eeq
Since most of the time $\tau$ will be considered fixed, we shall omit it from the notation and consider the function of $u$ at fixed $\tau$, and often write
\beq
\Theta(u) = \Theta(u,\tau).
\eeq
\ed

\bd[half-characteristics]
An half--integer characteristic $\chi\in \CC^\genus  $ is :
\beq
\chi = \frac12 \alpha + \frac12 \tau \beta
\qquad \alpha \in \ZZ^\genus , \ \ \beta\in\ZZ^\genus .
\eeq
The characteristic is called odd (resp. even) if
\beq
(\alpha,\beta) = \sum_{k=1}^\genus \alpha_k\beta_k = \text{odd} \quad \text{(resp. even)}.
\eeq
We define the Theta function with characteristic $\chi$:
\beq
\Theta_\chi(u,\tau) = \Theta(u+\chi,\tau) e^{\pi i (\beta,u)}.
\eeq

\ed

\bl
It satisfies:
\beq
\Theta_\chi(u+n) = e^{\pi i (\beta,n)} \Theta_\chi(u) \qquad \forall n\in \ZZ^\genus
\eeq
\beq
\Theta_\chi(u+\tau n) = e^{-2\pi i  (n,u)} e^{-\pi i (n,\alpha)} e^{-\pi i (n,\tau n)} \Theta_\chi(u) \qquad \forall n\in \ZZ^\genus
\eeq
\beq
\Theta_\chi(-u) = (-1)^{(\alpha,\beta)} \Theta_\chi(u).
\eeq
It follows that if $\chi$ is an odd characteristic, then $\Theta_\chi$ is an odd function and
\beq
\chi=\text{odd} \quad \implies \quad \Theta_\chi(0)=0.
\eeq

The number of odd characteristics is
\beq
2^{\genus-1}(2^\genus-1).
\eeq

\el
\proof{Easy proof found in any textbooks about Theta function, like \cite{TataLectures,SpecF,Abram}.}

\bd[Regular odd half-characteristics]
an odd characteristic $\chi$ is called regular iff the vector of derivatives $({\Theta_\chi}'_1(0),\dots,{\Theta_\chi}'_\genus(0))$ is not vanishing:
\beq
\exists i \ \quad
\sum_{n\in \ZZ^\genus} (n_i+\frac12 \beta_i) \ e^{\pi i (n,\alpha)} e^{\pi i (n,\tau \beta )} \ e^{i\pi (n,\tau n)} \neq 0.
\eeq
It is known that there always exists at least one regular odd half-characteristic.
Choose one for the rest of this article.

\ed

\bd[Derivatives]
We define the  vector of dimension $\genus$:
\beq
\Theta_\chi' = ({\Theta_\chi}'_1,\dots,{\Theta_\chi}'_\genus)
\eeq
evaluated at $u=0$, 
with
\beq
{\Theta_\chi}'_i = 2\pi i \sum_{n\in \ZZ^\genus} (n_i+\frac12 \beta_i) \ e^{\pi i (n,\alpha)} e^{\pi i (n,\tau \beta )} \ e^{i\pi (n,\tau n)} .
\eeq
And we define the tensor of 3rd derivatives $\Theta_\chi'''$ as
\beq
{\Theta_\chi}'''_{i,j,k} = (2\pi i)^3 \sum_{n\in \ZZ^\genus} (n_i+\frac12 \beta_i)(n_j+\frac12 \beta_j)(n_k+\frac12 \beta_k) \ e^{\pi i (n,\alpha)} e^{\pi i (n,\tau \beta )} \ e^{i\pi (n,\tau n)} .
\eeq
(notice that because of the parity property, all even derivatives at 0 must vanish).

\ed

\bd
We also define the same tensors multiplied by the matrix $\hat{\mathcal  K}$, i.e. in the combinatorial basis rather than the normalized basis of $H^1(\curve)$:
\beq
{{\hat\Theta}_{\chi}\ }'_{(i,j)} = \sum_{k=1}^\genus
{\Theta_{\chi}}'_k  \hat{\mathcal  K}_{k;(i,j)} .
\eeq
\beq
{{\hat\Theta}_{\chi} \ }'''_{(i_1,j_1),(i_2,j_2),(i_3,j_3)} = \sum_{k_1,k_2,k_3=1}^\genus
{\Theta_{\chi}}'''_{k_1,k_2,k_3} \prod_{l=1}^3  \hat{\mathcal  K}_{k_l;(i_l,j_l)} .
\eeq
\ed

\bl
We have
\beq\label{eq:thetatiercefromthetaprime}
{\Theta_\chi}'''_{i,j,k} =  \left( 4\pi i \frac{d}{d\tau_{j,k}} -\pi^2 \beta_j\beta_k \right) .{\Theta_\chi}'_{i} 
\eeq

\el
\proof{Easy computation}.

\bex{Genus 1}

$\tau$ is a $1\times 1$ matrix, i.e. a complex number such that $\Im\tau>0$.
We denote
\beq
q=e^{\pi i \tau} \quad , \qquad |q|<1.
\eeq
The unique (mod $\ZZ+\tau\ZZ$) odd-half integer characteristic is $\chi = \frac12 + \frac12 \tau$.
We have
\bea
\Theta(u,\tau) 
&=& \sum_{n\in \ZZ} e^{2\pi \ii n u}e^{\pi \ii \tau n^2} \cr
&=& \prod_{n\geq 1} (1+e^{2\pi\ii u} q^{2n-1})(1+e^{-2\pi\ii u}q^{2n-1})(1-q^{2n}).
\eea
\bea\label{eq:Thetaprimeqsumprod}
\Theta_\chi(u,\tau) 
&=& e^{\pi\ii u}\sum_{n\in \ZZ} (-1)^n e^{2\pi \ii n u} e^{\pi \ii \tau (n^2+n)} \cr
&=& e^{\pi\ii u}(1-e^{-2\pi\ii u})\prod_{n\geq 1} (1-e^{2\pi\ii u} q^{2n})(1-e^{-2\pi\ii u}q^{2n})(1-q^{2n}).
\eea
Computing the derivative at $u=0$ gives
\bea
\Theta'_\chi 
&=& 2\pi i\sum_{n\in \ZZ} (n+\frac12) (-1)^n q^{ n(n+1)} \cr
&=& 2\pi i\sum_{n=0}^\infty (2n+1) (-1)^n q^{ n(n+1)} \cr
&=& 2\pi\ii \prod_{n\geq 1} (1-q^{2n})^3
\eea
and
\beq
\Theta'''_\chi = (2\pi i)^3 \sum_{n\in \ZZ} (n+\frac12)^3 (-1)^n q^{ n(n+1)}  = \frac12 (\pi i)^3\sum_{n=0}^\infty (2n+1)^3 (-1)^n q^{ n(n+1)}.
\eeq
Using \eqref{eq:thetatiercefromthetaprime} and \eqref{eq:Thetaprimeqsumprod} we get
\beq
\frac{\Theta'''_\chi}{\Theta'_\chi}
= -\pi^2 + 4\pi\ii \frac{d}{d\tau} \ln \Theta'_\chi
= - \pi^2 \left( 1- 24  \sum_{n=1}^\infty \frac{n q^{2n}}{1-q^{2n}}\right) = -\pi^2 E_2(\tau) = -3 G_2(\tau).
\eeq
where $E_2$ and $G_2$ are the Eisenstein quasi-modular series.

\eex

\subsection{Canonical divisor}

The following is a classical theorem
\bp
The following differential form
\beq
\nu = \sum_{k=1}^\genus
{\Theta'_{\chi}}_k \omega_k = \sum_{k=1}^\genus\sum_{(i,j)\in \Newtint} 
{\Theta'_{\chi}}_k  \hat{\mathcal  K}_{k;(i,j)} \Omega_{i,j}
= \sum_{(i,j)\in \Newtint}  \hat\Theta'_{\chi \ (i,j)} \Omega_{i,j} = \Omega.\hat\Theta'
\eeq
is a holomorphic 1-form, that has $\genus-1$ double zeros.
We write it
\beq
\nu = \frac{H(x,y)}{P_y(x,y)}dx
\eeq
with $H\in \modsp(P)$:
\beq
H(x,y) = \sum_{(i,j)\in \Newtint}  \hat\Theta'_{\chi \ (i,j)} x^i y^j.
\eeq
The condition of double zeros implies that the common zeros of $H(x,y)$ and $P(x,y)$, must also be zeros of $H_x P_y - H_y P_x$, and there must be $\genus-1$ of them.

\ep

We shall skip the proof of this classical theorem and refer to \cite{Farkas,TataLectures,eynardlecturesRS}.

Let us mention that every holomorphic 1-form has $2\genus-2$ zeros, and this peculiar  1-form has all its zeros double.

Let us reformulate it as:
There exists some coefficients $H_{(i,j)}$ such that
\beq
\nu = \sum_{(i,j)\in \Newtint} H_{(i,j)}   \Omega_{i,j}
= \frac{H(x,y)}{P_y(x,y)} dx
\eeq
is holomorphic and has $\genus-1$ double zeros.

We know that this happens if we choose a regular odd half characteristics $\chi$ and 
\beq
H_{(i,j)} =  \hat\Theta'_{\chi (i,j)} = \sum_{k=1}^\genus {\Theta'_{\chi}}_k  \hat {\mathcal K}_{k;(i,j)}.
\eeq
However, the issue of whether $H\in \CC[\Newtint]$ such that $\frac{H(x,y)}{P_y(x,y)} dx$ is holomorphic and has $\genus-1$ double zeros, is purely algebraic.

\bp
There exists some $H\in \modsp(P)\subset \CC[\Newtint]$ such that 
$\frac{H(x,y)}{P_y(x,y)} dx$ is a holomorphic 1-form with $\genus-1$ double zeros.

Modulo scalar multiplication, there exists at least as many $H$ as the number of regular odd half characteristics.

The coefficients $H_{i,j}$ are such that the 3 following polynomials, have $\genus-1$ common zeros.
\beq
P, H ,P_x H_y -P_y H_x
\eeq

Equivalently, the curve $H(x,y)=0$ is tangent to the curve $P(x,y)=0$ in $\genus-1$ points.

\begin{figure}
    \centering
        \includegraphics[scale=0.25]{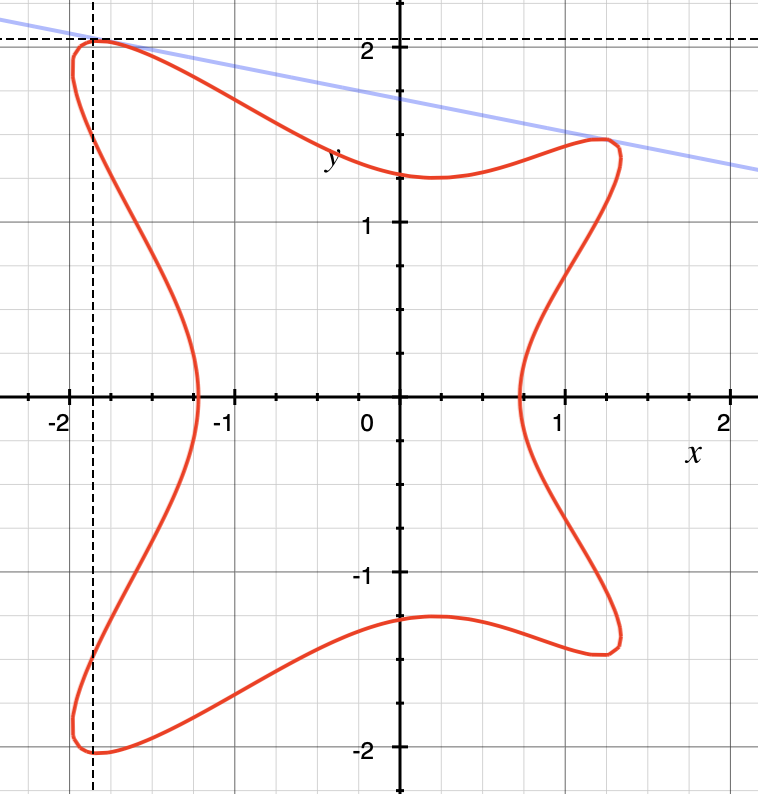}
    \caption{The curve $H(x,y)=0$ is tangent to the curve $P(x,y)=0$ in $\genus-1$ points.}
    \label{fig:my_label}
\end{figure}

%
\ep

\bd[Canonical divisor]
The $\genus-1$ zeros of $H(x,y)=0$ on $\curve$ are called a "canonical divisor".

The following property is not used in this article, but we cite it for completeness:
The Abel map of a canonical divisor  is the Riemann's constant plus a regular  half-integer odd characteristics:
$$
\sum_{i=1}^{\genus-1} F(p_i) = K_{\text{Riemann}}+\chi.
$$

\ed
{\em (we won't explain in this article why it is called caninical divisor and why it plays a very important role for example regarding the Picard group).}

\bex[Hyperelliptic]

Let $n\geq 3$. Let
\beq
P(x,y) = y^2 - \prod_{i=1}^{n} (x-x_i) = y^2-\hat P(x).
\eeq
We have $P_y(x,y)=2y = \pm 2\sqrt{\hat P(x)}$.
We have $\Newtint = \{(0,0),(1,0),\dots,(\lceil n/2\rceil-2)\}$, and $\genus=\lceil n/2\rceil-1$.
$H(x,y)$ must be of the form
\beq
H(x,y) = H(x) = \sum_{i=0}^{\genus-1} H_{i,0}x^i.
\eeq
Therefore $H(x,y)=0$ is a union of $\genus-1$ vertical lines in the $x,y$--plane. 
These can be tangent to the curve $P(x,y)=0$ only where $P(x,y)=0$ has a vertical tangent, i.e. only at the branchpoints.
Said in another way, zeros of $H(x)$ can be double on the curve only if $y=0$, i.e. only if zeros of $H(x)$ are also zeros of $\hat P(x)$.
Therefore $H$ must be a polynomial of degree $\genus-1$ that divides $\hat P$.
\beq
H | \hat P.
\eeq
The number of ways to choose an $H$ is to choose $\genus-1$ among the $n$ roots of $\hat P$:
\beq
I \subset\{x_1,\dots,x_{n}\} \quad , \quad |I|=\genus-1.
\eeq
We get
\beq
\nu = \frac{C}{2} \ \sqrt{\frac{\prod_{i\in I}(x-x_i)}{\prod_{i\notin I} (x-x_i)}} \ dx
\eeq
with a multiplicative constant $C$.

\eex

\subsection{Prime form}
\label{sec:primeform}

{\em This paragraph is not needed for computation of complete integrals. It is needed only for incomplete  integrals of the 3rd kind. It can be skipped in a first approach.}

Fay introduced the following quantity
\bd[Prime form]
\beq
E(p,p') 
= \frac{\Theta_\chi(F(p)-F(p'))}{\sqrt{\nu(p)\nu(p')}} 
= \Theta_\chi(F(p)-F(p')) \  \frac{\sqrt{P_y(x,y)P_y(x',y')}}{\sqrt{H(x,y)H(x',y') dx dx'}}.
\eeq
It is a $(-\frac12) \otimes (-\frac12)$ form (a bi-spinor), it is locally analytic on $\curve\times \curve$, but not globally, it may have monodromies.

It vanishes linearly at $p\to p'$, as 
\beq
E((x,y),(x',y')) \sim \frac{x-x'}{\sqrt{dxdx'}} (1+o(1)),
\eeq
and has no pole and no other zero than coinciding points.
The square-root is locally analytic because $\nu$ has only double zeros.

\ed

It satisfies:
\bea
d_p \ln{(E(p,p_1)/E(p,p_2))}
&=& d_p \ln{(\Theta_\chi(F(p)-F(p_1))/\Theta_\chi(F(p)-F(p_2)))}  \cr
&=& dS_{p_1,p_2}(p) 
\eea
where $dS$ is the 3rd kind form.

As we shall shortly see below, it satisfies:
\beq
d_p\otimes  d_q \ln{E(p,q)} =  d_p\otimes  d_q  \ln{\Theta_\chi(F(p)-F(q))} = B(p,q)
\eeq
where $B$ is the Bergman kernel.

\subsection{Bergman kernel and $S$ from Theta functions}

{\em The goal of this paragraph is to show how $S$ is related to the Theta function. This paragraph is not needed for computation, it is needed only to relate to other formulas. It can be skipped in a first approach.}

An expression of Bergman kernel is known with Theta functions.

\bl
\beq
B(p,q) = d_p\otimes d_q \ln \Theta_\chi\left(\int_q^p \omega , \tau\right)
= d_p\otimes d_q \ln \Theta_\chi\left(F(p)-F(q) , \tau\right)
\eeq
where $F(p)-F(q)=\int_q^p \omega = (\int_q^p \omega_1,\dots,\int_q^p \omega_\genus) \in \CC^\genus $ is  the Abel map (the integral is defined on the universal cover, or in a fundamental domain, such that the integration path is the unique homology chain that doesn't intersect any $\acycle_i$ or $\bcycle_i$ cycles).

Since $B$ is unique, $B$ is independent of the choice of characteristics $\chi$.
\el

From this expression we get that
\bl
Close to the diagonal we have

\beq
B \mathop{\sim}_{x_2\to x_1}  \Big(\frac{1}{(x_1-x_2)^2}   - \frac13 \frac{\Omega\otimes\Omega\otimes\Omega.\hat\Theta'''}{\Omega.\hat \Theta'}  + \frac16 \frac{\Omega''.\hat \Theta' }{\Omega.\hat \Theta'}  - \frac14 \frac{(\Omega'.\hat \Theta')^2}{(\Omega.\hat \Theta')^2} \Big) dx_1\otimes dx_2 
\eeq
More explicitely:
\bea
B & \mathop{\sim}_{x_2\to x_1} & \frac{dx_1\otimes dx_2}{(x_1-x_2)^2}   - \frac{P_y^3}3 \frac{\Omega\otimes\Omega\otimes\Omega.\hat\Theta'''}{P_y^2 H} dx_1 \otimes dx_2 \cr
&& + \Big( \frac{P_{x}^{2} P_{yy}^{2}}{4 P_{y}^{4}}  - \frac{P_{x}^{2} P_{yyy}}{6 P_{y}^{3}} - \frac{P_{x} P_{xy} P_{yy}}{2 P_{y}^{3}} + \frac{P_{x} P_{xyy}}{3 P_{y}^{2}} + \frac{P_{xx} P_{yy}}{6 P_{y}^{2}} - \frac{P_{xxy}}{6 P_{y}} + \frac{P_{xy}^{2}}{12 P_{y}^{2}} \cr
&& - \frac{H_{x} P_{x} P_{yy}}{6 H P_{y}^{2}} + \frac{H_{x} P_{xy}}{6 H P_{y}} + \frac{H_{xx}}{6 H} - \frac{2H_{xy} P_{x}}{6 H P_{y}} + \frac{H_{y} P_{x} P_{xy}}{6 H P_{y}^{2}}  - \frac{H_{y} P_{xx}}{6 H P_{y}} + \frac{H_{yy} P_{x}^{2}}{6 H P_{y}^{2}} \cr
&& - \frac{(H_x P_y-H_{y} P_{x})^{2}}{4 H^{2} P_{y}^{2}}
\Big) dx_1\otimes dx_2.
\eea

\el

\proof{(The expression above is verified by {\em{sympy}} symbolic computation).}

This implies at coinciding point
\bea
\frac{Q^{\text{comb}}+S}{P_y^2} 
&=&  - \frac{P_y^3}3 \frac{\Omega\otimes\Omega\otimes\Omega.\hat\Theta'''}{P_y^2 H}  \cr
&& +     \frac{P_{xx} P_{yy}}{6 P_{y}^{2}} - \frac{P_{xxy}}{6 P_{y}} + \frac{P_{xy}^{2}}{12 P_{y}^{2}}  - \frac{P_{xyy}P_x}{6P_y^2} \cr
&&  - \frac{H_{x} P_{x} P_{yy}}{6 H P_{y}^{2}} + \frac{H_{x} P_{xy}}{6 H P_{y}} + \frac{H_{xx}}{6 H} - \frac{H_{xy} P_{x}}{3 H P_{y}} + \frac{H_{y} P_{x} P_{xy}}{6 H P_{y}^{2}}  - \frac{H_{y} P_{xx}}{6 H P_{y}} + \frac{H_{yy} P_{x}^{2}}{6 H P_{y}^{2}} \cr
&& - \frac{(H_{x}P_y-H_y P_x)^{2}}{4 P_y^2 H^{2}}  .
\eea

\bea
Q^{\text{comb}}+S
&=&   -\frac{P_y^3}{3H} \Omega\otimes\Omega\otimes\Omega.\hat\Theta'''  \cr
&& +  \frac16 \Big(  P_{xx} P_{yy} - P_{xxy}P_y + \frac12 P_{xy}^{2} - P_{xyy}P_x \Big) \cr
&& +\frac{1}{6H}\Big(  - H_{x} P_{x} P_{yy} + H_{x} P_{xy}P_y + H_{xx}P_y^2 - 2H_{xy} P_{x}P_y \cr 
&& + H_{yy} P_{x}^{2} + H_{y} P_{x} P_{xy}  - H_{y} P_{xx}P_y  \Big) \cr
&& - \frac{(H_{x}P_y-H_y P_x)^{2}}{4 H^{2}}  \cr
&=&   -\frac{P_y^3}{3H} \Omega\otimes\Omega\otimes\Omega.\hat\Theta'''  \cr
&& +  \frac16 \Big(  P_{xx} P_{yy} - P_{xxy}P_y + \frac12 P_{xy}^{2} - P_{xyy}P_x \Big) \cr
&& +\frac{1}{6H P_y } (P_{y} P_{xy} - P_x P_{yy})(H_xP_y-P_x H_y)  \cr
&& +\frac{1}{6H}\Big( H_{xx}P_y^2 - 2H_{xy} P_{x}P_y + H_{yy} P_{x}^{2} + 2H_{y} P_{x} P_{xy}  \cr 
&& - H_{y} P_{xx}P_y-H_y P_{yy}P_x^2/P_y  \Big) \cr
&& - \frac{(H_{x}P_y-H_y P_x)^{2}}{4 H^{2}}  .
\eea

Canceling the double pole at the zeros of $H$, we get
\bl
If $\genus\geq 2$, 
at the zeros of $H$ we have
\beq
-P_y^3 \Omega\otimes\Omega\otimes\Omega.\hat\Theta''' 
= \frac12\Big( H_{xx}P_y^2 - 2H_{xy} P_{x}P_y+ H_{yy} P_{x}^{2} + 2H_{y} P_{x} P_{xy}  - H_{y} P_{xx}P_y-H_x P_{yy}P_x  \Big)
\eeq
\el

We have
\bl
There exists a fully symmetric tensor $\hat\Theta'''$ and a symmetric matrix $S$, such that
\bea
&& S(x,y;x,y)+\frac{P_y^3}{3H} \Omega\otimes\Omega\otimes\Omega.\hat\Theta''' \cr
&=&    \frac16 \Big(  P_{xx} P_{yy} - P_{xxy}P_y + \frac12 P_{xy}^{2} - P_{xyy}P_x \Big) -Q^{\text{comb}}(x,y;x,y) \cr
&& +\frac{1}{6H P_y } (P_{y} P_{xy} - P_x P_{yy})(H_xP_y-P_x H_y)  \cr
&& +\frac{1}{6H}\Big( H_{xx}P_y^2 - 2H_{xy} P_{x}P_y + H_{yy} P_{x}^{2} + 2H_{y} P_{x} P_{xy}  - H_{y} P_{xx}P_y-H_y P_{yy}P_x^2/P_y  \Big) \cr
&& - \frac{(H_{x}P_y-H_y P_x)^{2}}{4 H^{2}}  .
\eea
\el










\bex[Legendre curve]

$H=\hat\Theta'_\chi=\Theta'_\chi/\mathcal K$ is a constant.
\beq
\nu  = \frac{Hdx}{P_y} = \Theta'_\chi \omega = \Theta'_\chi \frac{dx}{\mathcal K P_y} = \hat\Theta'_\chi \frac{dx}{P_y}.
\eeq
We have
\bea
 S+\frac{\hat\Theta'''_\chi}{3H}  
&=&    \frac16 \Big(  P_{xx} P_{yy} - P_{xxy}P_y + \frac12 P_{xy}^{2} - P_{xyy}P_x \Big) +k^2(x+x)^2 \cr
&=&    \frac{P_{yy}}6 P_{xx} +k^2(x+x)^2 \cr
&=&    \frac13 (2(1+k^2)-12 k^2 x^2)  +4 k^2x^2 \cr 
&=&    \frac23 (1+k^2) 
\eea
i.e.
\bea
S 
&=&  \frac23 (1+k^2) - \frac13 \frac{\hat\Theta'''_\chi}{\hat\Theta'_\chi}  \cr
&=&  \frac23 (1+k^2) - \frac13 \frac{\mathcal K^{-3}\Theta'''_\chi}{ \mathcal K^{-1} \Theta'_\chi}  \cr
&=&  \frac23 (1+k^2) - \frac13 \frac{\Theta'''_\chi}{ \mathcal K^2 \Theta'_\chi}  \cr
&=&  \frac23 (1+k^2) + \frac{G_2}{\mathcal K^2}
. 
\eea

\eex

\section{Change of basis of cycles}
\label{sec:changecycles}

{\it{We compute integrals in terms of the building blocks, so we never actually need to know the cycles. The $\acycle$ cycles are encoded once for all in $\mathcal K$ and $S$, and the $\bcycle$ cycles are encoded in $\tau$.
However, it is useful to know how $\mathcal K$, $S$, $\tau$  change if we change the cycles.}}

Let us change the basis of cycles:
\bea
\acycle'_i &=& \sum_{j=1}^{\genus} \alpha_{i,j} \acycle_j + \sum_{j=1}^{\genus} \beta_{i,j} \bcycle_j \cr
\bcycle'_i &=& \sum_{j=1}^{\genus} \gamma_{i,j} \acycle_j + \sum_{j=1}^{\genus} \delta_{i,j} \bcycle_j \cr
\eea
where the $2\genus\times 2\genus$ matrix $U=\begin{pmatrix}
  \alpha & \beta \cr \gamma & \delta
\end{pmatrix}\in PSL(2\genus,\ZZ) $ is symplectic, with integer coefficients (which implies $\alpha\beta^T=\beta\alpha^T$, $\gamma\delta^T=\delta\gamma^T$ and $\alpha\delta^T-\beta\gamma^T=\text{Id}$):
\beq
U J U^T  = J
\quad , \quad J=\begin{pmatrix}
  0 & \text{Id} \cr    -\text{Id} & 0\cr
\end{pmatrix}.
\eeq

\bp
We have
\beq
\hat{\mathcal K}'= (\alpha^T+\tau \beta^T)^{-1}\hat{\mathcal K}.
\eeq

\beq\label{eq:SchangeTorelli}
S'= S -2\pi\ii \ \hat{\mathcal K}^T (\alpha+\beta\tau)^{-1}\beta \hat{\mathcal K}.
\eeq

\beq\label{eq:tauchangeTorelli}
\tau' = (\alpha^T+\tau \beta^T)^{-1}(\gamma^T+\tau \delta^T)
= (\gamma+\delta \tau)(\alpha+\beta\tau)^{-1}
.
\eeq

\bea\label{eq:zetachangeTorelli}
\zeta'
&=& \zeta -2\pi\ii \ \Omega^T \hat{\mathcal K}^T (\alpha+\beta\tau)^{-1}\beta \hat{\mathcal K} \cr
&=& \zeta -2\pi\ii \ \omega^T  (\alpha+\beta\tau)^{-1}\beta \hat{\mathcal K} \cr 
&=& \zeta -2\pi\ii \ \omega'^T  \beta \hat{\mathcal K} .
\eea

\ep

\proof{
Let $\omega'_i = \sum_{j=1}^\genus C_{i,j} \omega_j$.
In order to have $\oint_{\acycle'_j}\omega'_i = \delta_{i,j}$ we must have
\beq
\text{Id} = C (\alpha+\beta\tau)^T.
\eeq
Moreover $\omega'_k = \sum_{(i,j)\in\Newtint} \hat{\mathcal K}'_{k,(i,j)} \Omega_{i,j}$ and $\omega_k = \sum_{(i,j)\in\Newtint} \hat{\mathcal K}_{k,(i,j)} \Omega_{i,j}$ imply that
\beq
\hat{\mathcal K}' = C \hat{\mathcal K} = (\alpha^T+\tau\beta^T)^{-1}\hat{\mathcal K}.
\eeq
In order to prove \eqref{eq:SchangeTorelli}, it suffices to prove that $\int_{\acycle'_k} (B+ \Omega^T(S'-S)\Omega) = 0$.
We have

Let us thus compute
\bea
\int_{\acycle'_k} B 
&=& \sum_l \beta_{k,l} \int_{\bcycle_l} B  \cr
&=& 2\pi\ii \sum_l \beta_{k,l} \omega_l  \cr
&=& 2\pi\ii \sum_l \sum_{(i,j)\in\Newtint} \beta_{k,l} \hat {\mathcal K}_{l,(i,j)} \Omega_{i,j}  .
\eea
On the other hand
\bea
\int_{\acycle'} 
\hat{\mathcal K}^T \Omega^T (\alpha+\beta\tau)^{-1}\beta \hat{\mathcal K}
&=& 
\int_{\acycle'} 
\omega^T (\alpha+\beta\tau)^{-1}\beta \hat{\mathcal K} \cr
&=& 
\alpha \int_{\acycle} 
\omega^T (\alpha+\beta\tau)^{-1}\beta \hat{\mathcal K} 
+\beta \int_{\bcycle} 
\omega^T (\alpha+\beta\tau)^{-1}\beta \hat{\mathcal K} \cr
&=& 
(\alpha+\beta\tau)  (\alpha+\beta\tau)^{-1}\beta \hat{\mathcal K}   \cr
&=&  \beta \hat{\mathcal K}   
\eea
which is the same as $\int_{\acycle'} B$.
This concludes the proof of \eqref{eq:SchangeTorelli}.

}

\section{Computation of integrals}
\label{sec:summary}

{\em This section presents the core of the method.}

\medskip

\subsection{Decomposition of a 1-form}

Let $R(x,y)\in \check\Field(x,y)$, where $\check\Field$ is an extension of $\Field$.
Our goal is to compute integrals
 \beq
\int_\gamma R(x,y) \ dx
\eeq
as an algebraic combination (algebraic in the coefficients of $P$ and of the coefficients of $R$) of the elementary blocks $S,\mathcal K,\zeta, F,\tau$.

Let us proceed as follows:

\begin{itemize}

\item Let $\mathcal P_{R}$ the set of poles of $R(x,y)dx$.

\item At each pole $p\in \mathcal P_{R} $, let the times 
\bea
t_{p,k}(R) 
&=& \Res_p \xi_p^k \ R(x,y)dx  \cr
&=& \Res_p (x-x_p)^{\frac{k}{a_p}} \ R(x,y) dx \cr
&=& \Res_{z\to z_p} (X(z)-x_p)^{\frac{k}{a_p}} \ R(X(z),Y(z)) \ dX(z) .
\eea
The times are \textbf{algebraic} in some finite algebraic field extension of $\check\Field$:
\beq
t_{p,k}(R) \in \check\Field[x_p,y_p].
\eeq

\item Let us introduce for $k>0$
\beq
B_{p,k} = \frac{1}{k}\Res_p \xi_p^{-k} B.
\eeq
This is an algebraic combination of coefficients of $B$.
In particular this is polynomial in the coefficients of $S$ of degree 1.

\item For $k=0$, let $o$ a generic reference point chosen once for all (nothing will depend on it afterwards), consider:
\beq
B_{p,0} = dS_{p,o}.
\eeq
This is an algebraic combination of coefficients of $P$ and this is polynomial in $\zeta(p)$ of degree 1. 
(it is also polyonmial in $\zeta(o)$ but $\zeta(o)$ should cancel at the end.)

\item
Let the 1-form
\beq
\td R(x,y)dx = R(x,y)dx - \sum_{p\in \mathcal P_{R}} \sum_{k=1}^{-1+\deg_p R(x,y)dx} t_{p,k}(R) B_{p,k}
 - \sum_{p\in \mathcal P_{R}} t_{p,0}(R) dS_{p,o}.
\eeq
Since we have subtracted all poles,
this 1-form is a holomorphic form, therefore it is a linear combination of the $\Omega_{(i,j)}$ with $(i,j)\in\Newtint$.
In other words:
\beq
P_y(x,y) \td R(x,y) = \sum_{(i,j)\in\Newtint} \td R_{i,j}(R) x^i y^j \in \modsp(P).
\eeq
The coefficients $\td R_{i,j}(R)$ can be computed algebraically as follows:
\beq
\td R_{i,j}(R) = [x^i y^j] \left( P_y(x,y) \td R(x,y) \right).
\eeq
They are polynomials of degree 1 in  $S$ and  $\zeta$.

\item Therefore we have
\bea
R(x,y)dx &=& \sum_{p\in \mathcal P_{R}} \sum_{k=1}^{-1+\deg_p R(x,y)dx} t_{p,k}(R) B_{p,k} \cr
&& + \sum_{p\in \mathcal P_{R}} t_{p,0}(R) dS_{p,o} \cr
&& + \sum_{(i,j)\in\Newtint} \td R_{i,j}(R) \Omega_{(i,j)} \cr
\eea

\end{itemize}

\subsection{Integrals}

It remains to compute the integral on $\gamma$ of each term.

We distinguish 2 cases:
\begin{itemize}
    \item \textbf{Complete integrals:} this is the case when $\gamma$ is a closed Jordan loop. In that case, we decompose $\gamma$ as a combination of the loops $\acycle_i$, $\bcycle_i$, and possibly some small circles $\mathcal C_p$ surrounding some poles $p$ of $R(x,y)dx$:
    \beq
        \gamma = \sum_{i=1}^\genus c_i \acycle_i + \td c_i \bcycle_i + \sum_{p\in \mathcal P_R} c_p \mathcal C_p
    \eeq    
    \item \textbf{Incomplete integrals:} this is the case when $\gamma$ is an open Jordan arc.
    First, up to adding a closed Jordan arc, we can assume that $\gamma$ doesn't cross any $\acycle_i$, $\bcycle_i$ cycles.
    
\end{itemize}





    


\textbf{Summary:}

\begin{itemize}

\item[$\circ$] \textbf{Complete integrals:} 
\begin{itemize}
    \item[$\bullet$] $\oint_{\acycle_i} B_{p,k} = 0  $, $\oint_{\acycle_i} dS_{o,p} = 0$, $\oint_{\acycle_i} \omega_j = \delta_{i,j}  $.
    \item[$\bullet$] $\oint_{\bcycle_i} B_{p,k} = \frac{1}{k}\Res_{p} (x-x_p)^{-k/a_p} \omega_i$, $\oint_{\bcycle_i} dS_{o,p} = F_i(p)-F_i(o) $, $\oint_{\bcycle_i} \omega_j = \tau_{i,j}  $.
    \item[$\bullet$] $\oint_{\mathcal C_p} B_{q,k} = 0$, $\oint_{\mathcal C_p} dS_{o,q} = 2\pi\ii(\delta_{p,q}-\delta_{p,o}) $, $\oint_{\mathcal C_p} \omega_j = 0  $.
\end{itemize}
 
\item[$\circ$] \textbf{Incomplete integrals:}  if $\gamma = \gamma_{p_1\to p_2}$ is an open Jordan arc, we have
\begin{itemize}
        \item[$\bullet$] $\int_{\gamma} B_{p,k} = \frac{1}{k}\Res_{p} (x-x_p)^{-k/a_p} dS_{p_2,p_1}$
        \item[$\bullet$] $\int_{\gamma} \omega_j = F_j(p_2)-F_j(p_1)  $.
        \item[$\bullet$] $\int_{\gamma} dS_{o,p} = \ln{\left({E(p_1,p)E(p_2,o)}/{E(p_1,o)E(p_2,p)}\right)} $
        where $E(p,q)$ is the prime form.
\end{itemize}

\end{itemize}

\subsection{Full example: Legendre elliptic curve}

\beq
P(x,y) = y^2 - (1-x^2)(1-k^2 x^2).
\eeq

\subsubsection{Elementary blocks}

\begin{itemize}

    \item The discriminant is $\Delta(x) = 4(1-x^2)(1-k^2 x^2)$,
    and  $\Delta = 2^{16} k^2 (1-k^2)^4$.

    \item  If $k\notin \{0,1,-1\}$, the curve is non-degenerate, it has 4 regular branch-points at
\beq    
\mathcal R_x=\{1,-1,\frac{1}{k},\frac{-1}{k}\}.
\eeq

\item The Newton's polygon 
\begin{center}
\begin{tikzpicture}
\begin{scope}[scale = 0.55]
\grid{4}{2};
\fill [red, opacity=0.2] (0,0) -- (0,2) -- (2,1) -- (4,0)  ;

\draw [red, thick] (0,0)  -- (0,2)  -- (4,0)  -- cycle ;
\draw [blue] (1,1) node{$\circ$};
\def\N{(0,0), (2,0),(4,0),(0,2)};

\drawpolytope[red]\N;

\draw[blue] (2,1) node {$\bullet$};

\end{scope}
\end{tikzpicture}
\end{center}
has 1 interior point $(1,1)$, and its boundary has 2 integer segments $[(0,2),(2,1)], [(2,1),(4,0)]$, therefore we have
\bea
\Newtint & = & \{(0,0)\} \cr
\Newt''' & = & \{(1,0)\} \cr
\Newt'' & = & \{(2,0),(4,0),(0,2)\} 
\eea

\item There are 2 punctures, denoted $\infty_\pm$, with $x=\infty$, $y=\infty$, associated to the segments $[(0,2),(2,1)], [(2,1),(4,0)]$ both with normal vector $(a,b)=(-1,-2)$, and with local model $y\sim \pm k x^2$.

\item There is a unique interior point to the Newton's polygon of coordinate $(1,1)$, thus there is a unique combinatorial holomorphic 1-form:
\beq
\Omega_{0,0}=\frac{x^{1-1}y^{1-1}dx}{P_y(x,y)}=\frac{dx}{2y}.
\eeq

\item We define the $1\times 1$ matrix (written as a complex number):
\beq
\mathcal K = \oint_{\acycle} \frac{dx}{2y} = \hat{\mathcal K}^{-1}.
\eeq
If we choose $\acycle$ the cycle that surrounds the segment $[-1,1]$, it corresponds to
$
\mathcal K = 2 K(k)
$
in the usual Legendre Elliptic functions.

\item If we choose $\bcycle$ the cycle that surrounds the segment $[1,1/k]$, we have
\bea
\tau 
&=& \frac{1}{\mathcal K}\oint_{\bcycle} \frac{dx}{2y} \cr
&=& \frac{\ii}{2K(k)} \int_{1}^{1/k} \frac{dx}{\sqrt{(x^2-1)(1-k^2 x^2)}} \cr
&=& \frac{\ii K'(k)}{K(k)}.
\eea

(had we chosen $\acycle $ the contour that surrounds $[1,1/k]$, i.e. the exchange $(\acycle',\bcycle')=(\bcycle,-\acycle)$, using section \ref{sec:changecycles} we would have obtained $\mathcal K' = \tau \mathcal K = 2\ii K'(k)$.)

\item The normalized holomorphic 1-form is then
\beq
\omega = \frac{dx}{2\mathcal K y}.
\eeq 
\item Its integral is the Abel map
\beq
z=F(x,k) 
\qquad , \qquad
dz = \omega = \frac{dx}{2\mathcal K y}.
\eeq

\item Bergman kernel. We have
\bea
 Q^{comb}(x_1,y_1,x_2,y_2) 
 &=&  P_{0,2}P_{4,0}(x_1^2+2x_1x_2+x_2^2) \cr
 &=&  - k^2(x_1^2+2x_1x_2+x_2^2) \cr
 &=&  - k^2(x_1+x_2)^2,
\eea
and thus
\bea
&& B^{comb}(x_1,y_1,x_2,y_2) \cr
&=&  \frac{dx_1 \otimes dx_2}{4 y_1 y_2}\Big(-
\frac{(y_2^2-(1-x_1^2)(1-k^2x_1^2))(y_1^2-(1-x_2^2)(1-k^2x_2^2))}{(x_1-x_2)^2(y_1-y_2)^2} +  Q^{comb}  \Big) \cr 
&=&  \frac{dx_1 \otimes dx_2}{4 y_1 y_2}\Big(-
\frac{(y_2^2-y_1^2)(y_1^2-y_2^2)}{(x_1-x_2)^2(y_1-y_2)^2} +  Q^{comb}  \Big) \cr 
&=&  \frac{dx_1 \otimes dx_2}{4 y_1 y_2}\Big(
\frac{(y_1+y_2)^2}{(x_1-x_2)^2} - k^2 (2x_1 x_2+x_1^2+x_2^2)     \Big) \cr 
&=&  \frac{dx_1 \otimes dx_2}{4 y_1 y_2(x_1-x_2)^2}\Big(
2y_1 y_2 + y_1^2+y_2^2 - k^2 (2x_1 x_2+x_1^2+x_2^2)(x_1-x_2)^2    \Big) \cr 
&=&  \frac{dx_1 \otimes dx_2}{4 y_1 y_2(x_1-x_2)^2}\Big(
2y_1 y_2 + y_1^2+y_2^2 - k^2 (x_1^2-x_2^2)^2   \Big) \cr 
&=&  \frac{dx_1 \otimes dx_2}{4 y_1 y_2 (x_1-x_2)^2}
\Big( 2y_1 y_2 + 2k^2x_1^2x_2^2 - (1+k^2)(x_1^2+x_2^2)+2   \Big) 
\eea
and thus
\bea
B(x_1,y_1,x_2,y_2)
&=& \frac{dx_1 \otimes dx_2}{4 y_1 y_2 (x_1-x_2)^2}
\Big( 2y_1 y_2 + 2k^2x_1^2x_2^2 - (1+k^2)(x_1^2+x_2^2)+2  \cr
&& +S(x_1-x_2)^2 \Big) .
\eea

\item To compute $S$, use the small distance behavior
\bea
B(x_1,y_1,x_2,y_2)
&\mathop{\sim}_{p_1\to p_2} & dx_1 \otimes dx_2 \Big( \frac{1}{(x_1-x_2)^2} + \frac{y'^2-4 k^2 x^2}{4y^2} + \frac{S}{4y^2} \cr 
&& + O(x_1-x_2) \Big) \cr
\eea
Change variable to $z_1,z_2$ such that $dx/dz=2\mathcal K y$, and use the Schwarzian derivative (see appendix \ref{App:Schwartzian}), i.e.
\bea
B(x_1,y_1,x_2,y_2)
&\mathop{\sim}_{p_1\to p_2} & dz_1 \otimes dz_2 \Big( \frac{1}{(z_1-z_2)^2} + \frac16\left(\frac{d^2y/dz^2}{y} - \frac32 \frac{(dy/dz)^2}{y^2}\right) \cr
&& + \mathcal K^2(y'^2-4 k^2 x^2+S)  + O(x_1-x_2) \Big) \cr
&\mathop{\sim}_{p_1\to p_2} & dz_1 \otimes dz_2 \Big( \frac{1}{(z_1-z_2)^2} + \frac{4\mathcal K^2 }6\left((y'^2+y y'') - \frac32 y'^2\right) \cr 
&& + \mathcal K^2(y'^2-4 k^2 x^2+S)  + O(x_1-x_2) \Big) \cr
&\mathop{\sim}_{p_1\to p_2} & dz_1 \otimes dz_2 \Big( \frac{1}{(z_1-z_2)^2} + \frac{2\mathcal K^2 }3\left(6 k^2 x^2 - (1+k^2) \right) \cr 
&& + \mathcal K^2(-4 k^2 x^2+S)  + O(x_1-x_2) \Big) \cr
&\mathop{\sim}_{p_1\to p_2} & dz_1 \otimes dz_2 \Big( \frac{1}{(z_1-z_2)^2} + \mathcal K^2\left( S - \frac23(1+k^2) \right)  \cr 
&& + O(x_1-x_2) \Big) \cr
\eea

and compare with the second derivative of the log Theta function, i.e. 
\bea
B 
&=& \left(\wp(z_1-z_2)+G_2(\tau)\right)dz_1 \otimes dz_2 \cr
&=& \left(\frac{1}{(z_1-z_2)^2}+G_2(\tau) + O((z_1-z_2)^2)\right)dz_1 \otimes dz_2
\eea

This gives
\beq
S= \frac{G_2}{\mathcal K^2} + \frac23(1+k^2)
= k^2-1+2\frac{E}{K}.
\eeq

Recall that
\beq
G_2(\tau) = -\frac13 \frac{\theta'''(\chi)}{\theta'(\chi)}
= -\frac{(2\pi\ii)^2}3 \frac{\sum_{n} n^3 (-1)^n q^{\frac12 n(n+1)}}{\sum_{n} n (-1)^n q^{\frac12 n(n+1)}}
\eeq
and
\beq
G_2 = \frac{\pi^2}{3}\left(1-24 \sum_{n\geq 1} \frac{q^n}{(1-q^n)^2}\right)
\eeq

Theorem \ref{thm:RauchKtoS}  gives
\beq
S = (1-k^2) \left(1 + 4 k^2  \frac{\delta \ln\mathcal K}{2k\delta k}\right).
\eeq


\item 3rd kind form
\bea
dS_{p_1,p_2}(x)
&=& \frac{dx}{2y}  \ \Big( \frac{y+y_1}{(x-x_1)}  - \frac{y+y_2}{(x-x_2)}   + \zeta(p_1)-\zeta(p_2) \Big) \cr
\eea
In \eqref{eq:Legendredzeta}, we found that
\beq
\mathcal K\frac{d}{dz}\zeta(p_1) = \mathcal K^2 (S  - (1+k^2) +2k^2 x_1^2)
\eeq
and \eqref{eq:Legendrezetatoriemann}
\bea
\zeta(p_1) 
&=& -\frac{1}{2\mathcal K}  \zeta_{\text{Weierstrass}}(z_1-z_{\infty_+}) -\frac{1}{2\mathcal K} \zeta_{\text{Weierstrass}}(z_1-z_{\infty_-}) \cr
&& + z_1\left(\frac43 k^2(1+k^2)\mathcal K^2 
- \frac{1}{2\mathcal K}  \wp(z_{\infty_+}-z_{\infty_-}) \right)
.
\eea

\item Prime form:
Let $\chi = \frac12(1+\tau)$, $q=e^{\pi\ii \tau}$, recall that
\beq
\theta(z+\chi) = (e^{\pi \ii z}-e^{-\pi \ii z}) \prod_{n=1}^\infty (1-e^{2\pi \ii z} q^{2n})(1-e^{-2\pi \ii z}q^{2n})(1-q^{2n})
\eeq
\beq
E(z_1,z_2) = \frac{\theta(z_1-z_2+\chi)}{\theta'(\chi)\  \sqrt{dz_1 dz_2}}.
\eeq

\end{itemize}

\subsubsection{Examples of integrals}

Let us now consider examples of integrals
\beq
\int_\gamma \Omega \quad , \quad \Omega = R(x,y) \frac{dx}{2y}.
\eeq

\begin{itemize}

\item $R(x,y)=1$, 

\begin{itemize}
\item $\gamma = \acycle$:
\beq
\oint_{\acycle} \frac{dx}{2y} = \mathcal K.
\eeq

\item $\gamma = \bcycle$:
\beq
\oint_{\bcycle} \frac{dx}{2y} = \tau \mathcal K.
\eeq

\item 3rd kind cycle $\gamma = \gamma_{p_1\to p_2}$:
\beq
\int_{p_1}^{p_2} \frac{dx}{2y} = \mathcal K(F(p_2)-F(p_1)).
\eeq

\end{itemize}

\item $R(x,y)=2y^2$, in other words
\beq
\Omega = ydx
\eeq
$\Omega$ has a pole at $x=\infty$, i.e. 2 poles $\infty_\pm$ on $\curve$, at which
\beq
y \sim \pm kx^2(1-\frac{1+k^2}{2k^2x^2} + O(x^{-4}))
= \pm  ( k x^2-\frac{1+k^2}{2k} + O(x^{-2}))
\eeq
This gives
\beq
t_{\infty_\pm,3}(\Omega) = \mp k
\quad , \quad
t_{\infty_\pm,1}(\Omega) = \pm \frac{1+k^2}{2k}
\eeq
and all the other times are zero.

We have
\bea
&& j B_{\infty_\pm,j}(x,y) \cr
&=&  \Res_{\infty_\pm} \td x^j B(\td x,\td y,x,y) \cr
&=&  \Res_{\infty_\pm} \td x^j \frac{dx\otimes d\td x}{4 y \td y(x-\td x)^2} \Big( 2 y \td y + 2 k^2 x^2 \td x^2 - (1+k^2)(x^2+\td x^2)+2 \cr 
&& \qquad +S(x-\td x)^2 \Big) \cr
&=& \frac12 dx 
{\Res_{\infty_\pm}} {\td x}^{j-2} (1-x/\td x)^{-2} d\td x  \cr
&& \pm \frac{dx}{4y}
{\Res_{\infty_\pm}} {\td x}^j 
\frac{d\td x}{k{\td x}^2}(1+\frac{1+k^2}{2k^2}x^{-2}+O({\td x}^{-4}))(1-x/{\td x})^{-2} \cr
&&  \Big(  2 k^2 x^2   - (1+k^2)(1+x^2{\td x}^{-2})+2 \td x^{-2} \Big) \cr
&& \pm S \ \frac{dx}{4y}
{\Res_{\infty_\pm}} {\td x}^j 
\frac{d\td x}{k{\td x}^2}(1+\frac{1+k^2}{2k^2}x^{-2}+O({\td x}^{-4}))  \cr
\eea
If $j=1$ this gives
\bea
B_{\infty_\pm,1}(x,y)
&=& -\frac{dx}{2} \mp \frac{dx}{4ky} \left( 2k^2 x^2-(1+k^2)+S  \right)
\eea

If $j=3$ this gives

\bea
&& B_{\infty_\pm,3}(x,y) \cr
&=& -\frac12 x^2 dx 
\mp \frac{dx}{12ky} \Big( 3x^2 (2k^2 x^2 -(1+k^2))  \cr && + \frac{1+k^2}{2k^2}(2k^2x^2-(1+k^2))  + 2-(1+k^2)x^2 
 +S\frac{1+k^2}{2k^2}\Big) \cr
&=& -\frac12 x^2 dx 
\mp \frac{dx}{12ky} \Big( 6 k^2 x^4 -3(1+k^2)x^2  - \frac{(1-k^2)^2}{2k^2}  
 +S\frac{1+k^2}{2k^2}\Big) \cr
&=& -\frac12 x^2 dx 
\mp \frac{dx}{2ky} \Big(  k^2 x^4 -\frac12 (1+k^2)x^2  - \frac{(1-k^2)^2}{12k^2}  
 +S\frac{1+k^2}{12k^2}\Big) \cr
\eea

This gives that
\bea
\Omega' 
&=& ydx - \sum_{j=1,3}\sum_{\eta=\infty_\pm} t_{\eta,j} B_{\eta,j} \cr
&=& \frac{dx}{y} \frac{1}{6k^2} \left( (1+k^2)S - (1-k^2)^2 \right) .
\eea

In particular this gives
\beq
\oint_{\acycle} ydx = \frac{\mathcal K}{3k^2} \left( (1+k^2)S - (1-k^2)^2 \right)  .
\eeq
Notice that the computation was only algebraic, there was no need to compute integrals.



\item $R(x,y)=2(1-k^2 x^2)$, in other words
\beq
\Omega = \frac{1-k^2 x^2}{\sqrt{(1-x^2)(1-k^2 x^2)}}dx  =  \frac{-k  \sqrt{1-1/k^2 x^2}}{\sqrt{1-1/x^2}}dx.
\eeq
The only pole is at $x=\infty$, which corresponds to 2 punctures $\alpha=\infty_\pm$, at which
the only non-vanishing times are
\beq
t_{\infty_\pm,1} 
= \mp \Res_{\infty_\pm} x^{-1} \frac{k(1-\frac1{2 k^2x^2}+O(x^{-4})}{(1-\frac1{2 x^2}+O(x^{-4})} dx = \pm k.
\eeq
We have
\beq
B_{\infty_\pm,1}(x,y) = \pm \Res_{\infty_\pm} x' B(x',y',x,y)  
= \frac{dx}{2} \mp \frac{dx}{4ky} \left( 2k^2 x^2-(1+k^2)+S  \right).
\eeq
Then:
\bea
\Omega-\sum_{\epsilon=\pm} t_{\infty_\epsilon,1} B_{\infty_\epsilon,1}
&=& \frac{2(1-k^2 x^2)dx}{2y} + 2k  \frac{dx}{4ky} \left( 2k^2 x^2-(1+k^2)+S  \right) \cr
&=&  \frac{dx}{2y} \left( 1-k^2+S  \right) .
\eea
This gives the period
\beq
t_1 = \mathcal K \ (1-k^2 +S).
\eeq
therefore we have the decomposition of $\Omega$:
\beq
\Omega = k B_{\infty_+,1} - k B_{\infty_-,1} +  \mathcal K \ (1-k^2 +S) \omega_1.
\eeq
This implies in particular that
\beq
4 E(k) = \oint_{\acycle} \frac{1-k^2 x^2}{y}dx = \mathcal K \ (1-k^2 +S) = 2 K(1-k^2+S)
\eeq
in other words we recover
\beq
\frac{2 E(k)}{K(k)} = 1-k^2+S.
\eeq

\item $R(x,y)=\frac{1}{x-x_0}$, in other words
\beq
\Omega = \frac{dx}{2(x-x_0)y}
\eeq
$\Omega$ has a pole at $x=x_{0}$, i.e. 2 poles $\pm z_{0} = (x_0,\pm y_0)$ on $\curve$.
We have
\beq
t_{\pm z_0,0} = \Res_{\pm z_0} \Omega = \frac{1}{2 Y(\pm z_0)} = \frac{\pm 1}{2 Y(z_0)}= \frac{\pm 1}{2 y_0}.
\eeq
We have
\bea
\Omega' 
& = & \Omega-\sum_{\pm} t_{\pm z_0,0} dS_{\pm z_0,o} \cr
& = & \frac{dx}{2y(x-x_0)} - t_{+ z_0,0} dS_{z_0,-z_0} \cr
& = & \frac{dx}{2y} \left( \frac{1}{x-x_0} - \frac{1}{2 y_0} \left( \frac{y+y_0}{x-x_0}-\frac{y-y_0}{x-x_0} + \zeta(z_0)-\zeta(-z_0) \right) \right) \cr
& = & -\frac{dx}{2y} \frac{\zeta(z_0)-\zeta(-z_0)}{4y_0} \cr
\eea
i.e.
\beq
\Omega = \sum_{\pm} t_{\pm z_0,0} dS_{\pm z_0,o} - \frac{dx}{2y} \frac{\zeta(z_0)-\zeta(-z_0)}{4y_0} 
\eeq

For example this implies
\beq
\oint_{\acycle} \frac{dx}{2(x-x_0)y}
= - \mathcal K \frac{\zeta(z_0)-\zeta(-z_0)}{4y_0} 
\eeq

\item $R(x,y)=\frac{1}{1-u x^2}$, in other words
\beq
\Omega = \frac{dx}{2(1-u x^2)y}
\eeq
Let $x_0=1/\sqrt{u}$.
We get
\bea
\Pi(u,k) = \mathcal K \frac{\zeta(z_0)-\zeta(-z_0)}{2 u y_0} 
\eea

\end{itemize}









\subsection{Example: even hyperelliptic curve}

\beq
P(x,y) = y^2-\hat P(x).
\eeq
where  $\deg\hat P = 2d$ is even, and worth:
\beq
\hat P(x) = C \prod_{j=1}^{2d}(x-a_j).
\eeq

\begin{center}
\begin{tikzpicture}
\begin{scope}[scale = 0.45]
\grid{9}{2};
\fill [red, opacity=0.2] (0,2) --  (8,0) -- (0,0);
\draw [red, thick] (0,2) --  (8,0) -- (0,0) -- cycle ;
\draw[red] (0,2) node {$\bullet$};
\draw[red] (1,0) node {$\bullet$};
\draw[red] (2,0) node {$\bullet$};
\draw[red] (3,0) node {$\bullet$};
\draw[red] (4,0) node {$\bullet$};
\draw[red] (5,0) node {$\bullet$};
\draw[red] (6,0) node {$\bullet$};
\draw[red] (7,0) node {$\bullet$};
\draw[red] (8,0) node {$\bullet$};
\draw[red] (0,0) node {$\bullet$};
\draw[red] (4,1) node {$\times$};
\draw[blue] (1,1) node {$\circ$};
\draw[blue] (2,1) node {$\circ$};
\draw[blue] (3,1) node {$\circ$};

\end{scope}
\end{tikzpicture}
\end{center}

\beq
\Newtint=\{(i,0) \ | \ i=0,\dots,d-2\}.
\eeq
The genus is generically
\beq
\genus = d-1.
\eeq

Let us define
\beq
A(x) = \left(\sqrt{\hat P(x)}\right)_+
\quad , \quad
B(x) = \hat P(x)-A(x)^2,
\eeq
where $()_+$ means keeping only positive powers of $x$ in the large $x$ expansion.
We have
\beq
\deg A= d
\quad , \quad
\deg B\leq d-1.
\eeq
The Bergman kernel takes the following form
\bea
B(x_1,y_1,x_2,y_2)
&=& \frac{dx_1\otimes dx_2}{4 y_1 y_2 (x_1-x_2)^2}\Big( 2y_1 y_1 + 2 A(x_1)A(x_2)+B(x_1)+B(x_2) \cr 
&& \qquad + (x_1-x_2)^2 s(x_1,x_2)\Big)
\eea
where $s(x_1,x_2)$ is a symmetric polynomial of degree $d-2$ in each variable.

Using theorem \ref{thm:RauchKtoS} we have
\bea
&& \sum_{i'=0}^{d-2} (\delta {\mathcal K} \ \hat{\mathcal K})_{(i,0);(i',0)} x^{i'} \cr
&=& -\frac{x^i \delta P_y(x,y)}{P_y(x,y)}  +  \frac{ x^i P_{yy}(x,y)}{P_y(x,y)^2} \delta P(x,y) \cr
&& + 2 \sum_{a} \Res_{a} \frac{\delta P(x',y')\hat B(x',y';x,y)x'^i }{P_x(x',y')P_y(x',y')^2}dx' \cr
&=& -  \frac{ 2x^i }{4y^2} \delta \hat P(x) \cr
&& + 2 \sum_{a} \Res_{a} \frac{\delta \hat P(x')\hat B(x',y';x,y)x'^i }{4\hat P'(x')y'^2}dx' \cr
&=& -  \frac{ x^i }{2\hat P(x)} \delta \hat P(x) \cr
&& + \frac12 \sum_{a} \Res_{a} \frac{x'^i \delta \hat P(x')(2yy'+2A(x)A(x')+B(x)+B(x')) }{\hat P'(x')\hat P(x') (x-x')^2}dx' \cr
&& + \frac12 \sum_{a} \Res_{a} \frac{x'^i \delta \hat P(x')s(x,x') }{\hat P'(x')\hat P(x') }dx' \cr
&=& -  \frac{ x^i }{2\hat P(x)} \delta \hat P(x) \cr
&& + \frac12 \sum_{a} \frac{a^i \delta \hat P(a)(2A(x)A(a)+B(x)+B(a)) }{\hat P'(a)^2 (x-a)^2} \cr
&& +  \frac12 \sum_{a} \frac{a^i \delta \hat P(a)s(x,a) }{\hat P'(a)^2 } \cr
&=& -  \frac{ x^i }{2\hat P(x)} \delta \hat P(x) \cr
&& + \frac12 \sum_{a} \frac{a^i \delta \hat P(a)(2(A(x)-A(a))A(a)+B(x)-B(a)) }{\hat P'(a)^2 (x-a)^2} 
+ \frac{a^i \delta \hat P(a)s(x,a) }{\hat P'(a)^2 } \cr
&=& - x^i \frac{\delta c}{2c} - \frac12  \sum_{j=1}^{2d} \frac{\delta a_j}{x-a_j} (x^i-a_j^i) +\frac{\delta a_j}{x-a_j} a_j^i  \cr
&& -\frac12 \sum_{j=1}^{2d} \delta a_j \left( \frac{a_j^i  (2(A(x)-A(a_j))A(a_j)+B(x)-B(a_j)) }{\hat P'(a_j) (x-a_j)^2} 
+ \frac{a_j^i s(x,a_j) }{\hat P'(a_j) } \right) \cr
&=& - x^i \frac{\delta c}{2c} - \frac12  \sum_{j=1}^{2d} \delta a_j\frac{x^i-a_j^i}{x-a_j}  \cr
&& -\frac12 \sum_{j=1}^{2d} \frac{a_j^i \delta a_j}{\hat P'(a_j)} \Big( 2\frac{ A(x)-A(a_j)-(x-a_j)A'(a_j))A(a_j) }{ (x-a_j)^2} \cr
&& + \frac{B(x)-B(a_j)-(x-a_j)B'(a_j) }{ (x-a_j)^2}  +  s(x,a_j)  \Big) \cr
&=& - x^i \frac{\delta c}{2c} - \frac12  \sum_{j=1}^{2d} \sum_{i'=0}^{i-1} \delta a_j x^{i'} a_j^{i-1-i'}  \cr
&& -\frac12 \sum_{j=1}^{2d} \frac{a_j^i \delta a_j}{\hat P'(a_j)} \Big( 2A(a_j) \sum_{k=2}^d \frac{A^{(k)}(a_j)}{k!} (x-a_j)^k \cr
&& + \sum_{k=2}^{d-1} \frac{B^{(k)}(a_j)}{k!} (x-a_j)^k  +  s(x,a_j)  \Big) \cr
\eea
This gives
\bea
-2  \sum_{l=1}^\genus \frac{\partial \mathcal K_{i,l}}{\partial a_j} \hat{\mathcal K}_{l,i'} 
&=& a_j^{i-i'-1} + \frac{a_j^i}{\hat P'(a_j)} \ [x^{i'}]s(x,a_j) \cr 
&& + \frac{a_j^i}{\hat P'(a_j)}  \sum_{k=2}^d \frac{2A(a_j) A^{(k)}(a_j)+B^{(k)}(a_j)}{i'! (k-i')!} (-a_j)^{k-i'} \cr
\eea

Third kind form:

\beq
dS_{p_1,p_2} = \frac{dx}{2y}\left( \frac{y+y_1}{x-x_1}-\frac{y+y_2}{x-x_2}+\zeta(x,p_1)-\zeta(x,p_2) \right)
\eeq
We have

\bea
2y_1\frac{d\zeta(x,p_1)}{dx_1}
&=& s(x,x_1) + C(x,x_1)
\eea
where
\bea
C(x,x_1)
&=&  \frac{\hat P(x)-\hat P(x_1) -(x-x_1)\hat P'(x_1)}{(x-x_1)^2}  - \frac{(A(x)-A(x_1))^2}{(x-x_1)^2}  \cr
&=&  2A(x_1)\frac{A(x)-A(x_1)-(x-x_1)A'(x_1)}{(x-x_1)^2}    \cr
&& + \frac{B(x)-B(x_1)-(x-x_1)B'(x_1)}{(x-x_1)^2}   \cr
\eea


\subsection{Example cubic elliptic curve}

\beq
P(x,y)=1+x^3+y^3+t xy=0
\eeq
Its discriminant is $4 t^3 x^3 - 9 (x^3+1)^2$, which is non-degenerate as long as $t\neq 0$ and $ t^3\neq 9$.

Its Newton's polygon has a unique interior point, $\Newtint=\{(0,0)\}$.

\begin{center}
\begin{tikzpicture}
\begin{scope}[scale = 0.45]
\grid{4}{4};
\fill [red, opacity=0.2] (0,3) --  (3,0) -- (0,0);
\draw [red, thick] (0,3) --  (3,0) -- (0,0) -- cycle ;
\draw[red] (0,3) node {$\bullet$};
\draw[red] (3,0) node {$\bullet$};
\draw[red] (1,1) node {$\bullet$};
\draw[red] (0,0) node {$\bullet$};
\draw[blue] (1,1) node {$\circ$};

\end{scope}
\end{tikzpicture}
\end{center}

Bergman kernel. We have
\bea
&& B^{comb}(x_1,y_1,x_2,y_2) \cr
&=& \frac{dx_1 \otimes dx_2}{(3 y_1^2+tx_1)(3y_2^2+tx_2) }
\Big( -\frac{(y_1^3+x_2^3+1+t x_2 y_1)(y_2^3+x_1^3+1+t x_1 y_2)}{(x_1-x_2)^2(y_1-y_2)^2} +Q^{comb}  \Big)  \cr
&=& \frac{dx_1 \otimes dx_2}{(3 y_1^2+tx_1)(3y_2^2+tx_2) }
\Big( -\frac{(y_1^3-y_2^3+t x_2 (y_1-y_2))(y_2^3-y_1^3 t x_1 (y_2-y_1))}{(x_1-x_2)^2(y_1-y_2)^2} +Q^{comb}  \Big)  \cr
&=& \frac{dx_1 \otimes dx_2}{(3 y_1^2+tx_1)(3y_2^2+tx_2) }
\Big( \frac{(y_1^2+y_1y_2+y_2^2+t x_2)(y_1^2+y_1y_2+y_2^2+ t x_1)}{(x_1-x_2)^2} \cr
&& \quad +x_1 y_1 + x_2 y_2 + 2x_1 y_2 +2x_2 y_1   \Big)  .
\eea

\bea
&& dS_{p_1,p_2}(x) \cr
&=& \frac{dx}{2(3y^2+tx)}  \ \Big( \frac{x_1^3+y^3+tx_1y -txy_1-x^3-y_1^3}{(x-x_1)(y-y_1)}  -  \frac{P(x_2,y)-P(x,y_2)}{(x-x_2)(y-y_2)} \cr
&&  +  (\zeta(p_1)-\zeta(p_2))   \Big) \cr
&=& \frac{dx}{2(3y^2+tx)}  \ \Big( \frac{2y^3+txy+tx_1y -txy_1-tx_1y_1-2y_1^3}{(x-x_1)(y-y_1)}  -  \frac{P(x_2,y)-P(x,y_2)}{(x-x_2)(y-y_2)} \cr
&&  +  (\zeta(p_1)-\zeta(p_2))   \Big) \cr
&=& \frac{dx}{2(3y^2+tx)}  \ \Big( \frac{2(y^2+yy_1+y_1^2)+t(x+x_1)}{(x-x_1)}  -  \frac{P(x_2,y)-P(x,y_2)}{(x-x_2)(y-y_2)} \cr
&&  +  (\zeta(p_1)-\zeta(p_2))   \Big) \cr
\eea

and, using prop~\ref{prop:dzeta}:
\beq
d\zeta(p_1)
= \left( S +  C(x_1,y_1) \right) \frac{dx_1}{P_y(x_1,y_1)} 
\eeq
with
\bea
 C(p_1) 
&=& Q^{\text{comb}}(x,y;x_1,y_1) \cr
&& - \frac{P(x,y_1)-(x-x_1)P_x(x_1,y_1)}{(x-x_1)^2} \  \frac{P(x_1,y)-(y-y_1)P_y(x_1,y_1)}{(y-y_1)^2} \cr
&& - \frac{P(x_1,y)+P(x,y_1) + (x-x_1)(P_x(x_1,y)-P_x(x_1,y_1))}{2(x-x_1)^2(y-y_1)} \ P_y(x_1,y_1) \cr
&& - \frac{P(x_1,y)+P(x,y_1)+(y-y_1)(P_y(x,y_1)-P_y(x_1,y_1))}{2(x-x_1)(y-y_1)^2}\ P_x(x_1,y_1)  \cr
&=& x y + x_1 y_1 + 2x y_1 +2x_1 y  \cr
&& - \frac{x^3+y_1^3+1+txy_1-(x-x_1)(3x_1^2+ty_1)}{(x-x_1)^2} \cr 
&& \qquad \frac{x_1^3+y^3+t x_1 y+1 -(y-y_1)(3y_1^2+t x_1)}{(y-y_1)^2} \cr
&& - \frac{t (x_1y+xy_1-xy-x_1y_1) + (x-x_1)t(y-y_1))}{2(x-x_1)^2(y-y_1)} \ P_y(x_1,y_1) \cr
&& - \frac{t (x_1y+xy_1-xy-x_1y_1)+(y-y_1)t(x-x_1)}{2(x-x_1)(y-y_1)^2}\ P_x(x_1,y_1)  \cr
&=& x y + x_1 y_1 + 2x y_1 +2x_1 y  \cr
&& - \frac{x^3-x_1^3-tx_1y_1+txy_1-(x-x_1)(3x_1^2+ty_1)}{(x-x_1)^2} \cr 
&& \qquad \frac{y^3-y_1^3-tx_1y_1+t x_1 y -(y-y_1)(3y_1^2+t x_1)}{(y-y_1)^2} \cr
&=& x y + x_1 y_1 + 2x y_1 +2x_1 y  \cr
&& - \frac{x^2+xx_1+x_1^2+ty_1-(3x_1^2+ty_1)}{(x-x_1)} \  \frac{y^2+yy_1+y_1^2+t x_1 -(3y_1^2+t x_1)}{(y-y_1)} \cr
&=& x y + x_1 y_1 + 2x y_1 +2x_1 y  
- \frac{x^2+xx_1-2x_1^2}{(x-x_1)} \  \frac{y^2+yy_1-2y_1^2}{(y-y_1)} \cr
&=& x y + x_1 y_1 + 2x y_1 +2x_1 y  
- (x+2x_1) \ (y+2y_1) \cr
&=& -3 x_1 y_1
\eea
i.e.
\beq
P_y(p_1)\frac{d\zeta_1}{dx_1} = S-3x_1 y_1.
\eeq
Notice that the right hand side has double pole at the punctures. The derivative of $\zeta$ is thus an affine combination of Weierstrass functions, which means that $\zeta$ is an affine combination of $\zeta_{\text{Weierstrass}}$.







\section{Conclusions}

This is a summary of the Newton's polygon's method to compute integrals.
The method is rather automatic, and should be implemented in math softwares.

\section*{Aknowledgements}

This work is supported by the ERC-SyG project, Recursive and Exact New Quantum Theory (ReNewQuantum) which received funding from the European Research Council (ERC) under the European Union's Horizon 2020 research and innovation programme under grant agreement No 810573.
We thank F. Zerbini for discussions on this topic.

\bigskip

\hrule

\medskip

\appendix

\section{Discriminant}
\label{App:Discr}

Let $P(y) = \sum_{j=0}^{d} P_j y^j \in \Field[y] $ a polynomial of degree $d=\deg P$.

\bd[Discriminant]
Then the discriminant is the following determinant of size $2d-1$
\beq\label{eqApp:discrmatrix}
\Delta(x) = \det
\begin{pmatrix}
  P_d    & P_{d-1} & \dots   & P_0 & 0      &       \dots 0         \cr
   0      & \ddots  &         &     & \ddots & \vdots    \cr
   0\dots      &    0    &  P_d    &     & \dots  & P_0       \cr
  dP_d    & (d-1)P_{d-1} & \dots   & P_1 & 0      &      \dots 0          \cr
   0      & \ddots  &         &     & \ddots & \vdots     \cr
   0\dots      &    0    &  d P_d    &     & \dots  & P_1           \cr
\end{pmatrix}
\eeq
\ed

Example for $P(x,y) = ay^2+by+c$, we have
\beq
\Delta(x) = \det\begin{pmatrix}
a   & b      & c        \cr
2 a & b      & 0          \cr
 0    & 2 a    & b        \cr
\end{pmatrix}
= - a ( b^2 - 4 ac) .
\eeq

\section{Schwartz derivative}
\label{App:Schwartzian}

\bl
Let $x=f(z)$
We have
\beq
\frac{dx_1 dx_2}{(x_1-x_2)^2} \sim_{x_1\to x_2} \left( \frac{1}{(z_1-z_2)^2} + \frac16 \{ f,z_1\} +O(z_1-z_2) \right) dz_1 dz_2
\eeq
where
\beq
\{ f,z\}=\frac{f'''}{f'} - \frac32 \frac{f''^2}{f'^2} = \left(\frac{f''}{f'}\right)' - \frac12 \frac{f''^2}{f'^2}
\eeq
\el

\section{Short distance expansions}
\label{App:shortdistance}

Let $x_2=x_1+\epsilon$ and $y_2=y_1+\eta$, such that $P(x_1,y_1)=P(x_2,y_2)=0$.
In the limit $\epsilon\to 0$ we have
\bea
\eta
& = & - \frac{P_{x}}{P_{y}} \epsilon \cr
&& +\left( - \frac{P_{x}^{2} P_{yy}}{2 P_{y}^{3}} + \frac{P_{x} P_{xy}}{P_{y}^{2}} - \frac{P_{xx}}{2 P_{y}} \right) \epsilon^{2}  \cr
&& + \Big( \frac{P_{x}^{3} P_{yyy}}{6 P_{y}^{4}} - \frac{P_{x}^{3} P_{yy}^{2}}{2 P_{y}^{5}} + \frac{3 P_{x}^{2} P_{xy} P_{yy}}{2 P_{y}^{4}} - \frac{P_{x}^{2} P_{xyy}}{2 P_{y}^{3}} \cr
&& - \frac{P_{x} P_{xx} P_{yy}}{2 P_{y}^{3}} + \frac{P_{x} P_{xxy}}{2 P_{y}^{2}} - \frac{P_{x} P_{xy}^{2}}{P_{y}^{3}} + \frac{P_{xx} P_{xy}}{2 P_{y}^{2}} - \frac{P_{xxx}}{6 P_{y}} \Big) \epsilon^{3} \cr
&&  + O\left(\epsilon^{4}\right)
\eea


\section{Elliptic functions}
\label{App:Elliptic}

First kind complete Elliptic function
\bea
K(k) 
&=& \frac{\pi}{2} \left( \sum_{n=0}^\infty \left(\frac{(2n-1)!!}{(2n)!!}\right)^2 k^{2n} \right)
= \frac{\pi}{2} \left( 1+\frac{k^2}{4}+\frac{9k^4}{64} + \dots \right) \cr
&=& 2\pi \left( \frac14 +  \sum_{n=1}^\infty \frac{q^n}{1+q^{2n}} \right)
= \frac{\pi}{2} \left( 1 +  4q+4q^2+ \dots \right) \cr
&=& \frac{\pi}{2} \theta_3(q)^2  = \frac{\pi}{2} \left( \sum_{n\in \ZZ} q^{n^2} \right)^2
= \frac{\pi}{2} \left( 1 +  4q+4q^2+ \dots \right)
\eea
This gives
\beq
q = \frac{k^2}{16} + \frac{k^4}{32}  + \dots
\eeq

Second kind complete Elliptic function
\beq
E(k) = \frac{\pi}{2} \left( 1-\sum_{n=1}^\infty \left(\frac{(2n-1)!!}{(2n)!!}\right)^2 \frac{k^{2n}}{2n-1} \right)
= \frac{\pi}{2} \left( 1-\frac{k^2}{4}-\frac{3k^4}{64} + \dots \right)
\eeq

Second Eisenstein series
\beq
G_2 = \frac{\pi^2}{3}\left(1-24 \sum_{n=1}^\infty \frac{nq^n}{1-q^n} \right)
= \frac{\pi^2}{3}\left(1-24 (q+3q^2+\dots) \right)
\eeq
\beq
G_2 
= \pi^2\left( \frac{1}{3} - 8q-24 q^2 + \dots \right)
= \pi^2\left( \frac{1}{3} - \frac{k^2}{2}-\frac{11}{32} k^4 + \dots \right)
\eeq

\beq
\frac{G_2}{4 K^2} = \frac{2E}{K} -\frac53+\frac{1}{3}k^2 .
\eeq

\printbibliography



\end{document}